\documentclass[a4paper,12pt]{article} \pdfoutput=1

\usepackage{amsmath,amssymb,amsfonts,amsthm}
\usepackage{todonotes}
\usepackage[caption=false]{subfig}
\usepackage{microtype}
\usepackage{booktabs}
\usepackage{authblk}
\usepackage{grffile} 

\usepackage[sorting=none,maxnames=5,firstinits=true,doi=true,url=true]{biblatex}
\usepackage{braket}
\usepackage{hyperref}
\hypersetup{pdfauthor={Lisa Glaser},pdftitle={}}

\newcommand{\RomanNumeralCaps}[1]
    {\MakeUppercase{\romannumeral #1}}

\newcommand{\I}{region \RomanNumeralCaps{1} }
\newcommand{\II}{region \RomanNumeralCaps{2} }
\newcommand{\III}{region \RomanNumeralCaps{3} }

\newcommand{\av}[1]{\langle #1 \rangle}
\newcommand{\pd}[2]{\frac{\partial #1}{\partial #2}}
\newcommand{\mc}[1]{\mathcal{#1}}
\newcommand{\C}{\mathcal{C}}

\newcommand{\brOne}{\beta_{\mathrm{red},\mathrm{mag}}}
\newcommand{\brTwo}{\beta_{\mathrm{red}, \mathrm{geo}}}

\begin{document}
\title{Phase transitions in $2$d orders coupled to the Ising model}
\author[1]{Lisa Glaser}
\affil[1]{Faculty of Physics, University of Vienna}
\date{\today}

\maketitle

\begin{abstract}
  The $2$d orders are a sub class of causal sets, which is especially amenable to computer simulations.
  Past work has shown that the $2$d orders have a first order phase transition between a random and a crystalline phase.
  When coupling the $2$d orders to the Ising model, this phase transition coincides with the transition of the Ising model.
  The coupled system also shows a new phase, at negative $\beta$, where the Ising model induces the geometric transition.
  In this article we examine the phase transitions of the coupled system, to determine their order, as well as how they scale when the system size is changed.
  We find that the transition at positive $\beta$ seems to be of mixed order, while the two transitions at negative $\beta$ appear continous/ first order for the Ising model/ the geometry respectively.
  The scaling of the observables with the system size on the other hand is fairly simple, and does, where applicable, agree with that found for the pure $2$d orders.
  We find that the location of these transitions has fractional scaling in the system size.
\end{abstract}

Causal set theory replaces the smooth manifolds of general relativity with discrete, partially ordered sets~\cite{Bombelli_Lee_Meyer_Sorkin_1987}.
The partial order encodes the causal structure of the manifold, while the discreteness encodes the volume of space-time regions.
Causal sets thus promise to introduce a regularization of the path integral over geometries, while retaining the Lorentzian structure of space-time.
However including realistic matter beyond scalar fields proves very difficult, due to the lack of a concept of tangent spaces.

To calculate the path integral over causal sets one has to define a measure on the space of causal sets.
The simple counting measure on the space of all causal sets is disfavoured due to the entropic dominance of the so-called Kleitmann-Rothschild orders~\cite{Kleitman_Rothschild_1975}, however there are ideas about how to define a measure using a growth process for causal sets~\cite{Rideout_Sorkin_2000,Dowker_Zalel_2017,Zalel_2020}.
Another simple measure, using a subset of causal sets called the $2$d orders, was proposed in~\cite{Brightwell_Henson_Surya_2008}.
The advantage of the $2$d orders is that they always embed into $2$d space-time, and that a random $2$d order will, with high likely-hood, be a sprinkling into $2$d Minkowski space.
At the same time, they still include plenty of non-manifoldlike partial orders, hence giving us a good starting point for our investigations.
They are also an ideal model system, since they can easily be explored on the computer.
In~\cite{Surya_2012} and~\cite{Glaser:2017sbe} it was shown that path integral over the $2$d orders using the Benincasa Dowker action $S_c$~\cite{Benincasa_Dowker_2010}, has a first order phase transition and that this system shows very clear scaling behaviour with the number of elements in the $2$d order.

In~\cite{Glaser:2018jss} Ising spins were coupled to the $2$d orders, with an Ising action $S_I$ coupling spins only along links, leading to an overall action $S=S_c+S_I$.
This system has three coupling constants, an inverse temperature $\beta$, the non-locality parameter $\epsilon$ of the Benincasa Dowker action, and the coupling of the Ising model $j$.
Since~\cite{Surya_2012,Glaser:2017sbe} showed the system to be stable over a wide range of $\epsilon$, we fixed $\epsilon=0.21$ in ~\cite{Glaser:2018jss}.
In the phase diagram of the two remaining couplings a number of transitions arise.
The most interesting of these is at negative $\beta$ and positive $j$, where the Ising model becomes completely magnetized, and then pushes the geometry into a crystalline state.
Energetically this crystalline state would be disfavoured for the pure $2$d orders at negative $\beta$, however this geometric configuration allows for a maximisation of the Ising action, by increasing the number of available connections.
Hence the matter changes the geometry.
In~\cite{Glaser:2018jss} these phases were observed at fixed size, and the overall phase diagram was explored.
Here we follow up and explore the phase transitions in more detail, and for varying system size.

The questions explored in this paper are:
\begin{itemize}
  \item How do the phase transitions in this system behave?
  We know there is a transition of geometry that is forced by the coupled matter.
  Does it remain a first order transition or does it change? What are the indicators for this transition?
  \item The $2$d orders show simple power-law scalings with size which indicates some very consistent behaviour with increasing $N$, can the same be said for the $2$d orders coupled to this simple matter?
  If even matter as simple as the Ising model throws the system into chaotic behaviour this might spell trouble for causal sets and matter, since it is unlikely that more complicated systems behave more regular than simple systems.
  \item What are the critical exponents for this model, do they compare to the ordinary Ising model or to an Ising model on random graphs in any meaningful way? In the ordinary Ising model nearest neighbours would be space-like, while here the neighbours are time-like, and the system undergoing equilibration could be considered as time free. How does this affect the outcome?
\end{itemize}

To examine these questions we pick two lines at fixed $j$ and vary $\beta$ to cross the phase transition.
The main results we find for $j=-1$ and $\beta>0$ are:
\begin{itemize}
  \item The phase transition for matter and geometry happens concurrently, it is likely of first order in the geometry, but continuous in the matter.
  \item The phase transition point $\beta_c$ scales like $N^{-0.72 \pm 0.01}$ with varying system size.
  \item The actions $S,S_I,S_c$ all scale like $N$ for $\beta<\beta_c$ and like $N^2$ for $\beta>\beta_c$.
\end{itemize}
For $j=1$ and $\beta<0$ we find two phase transitions, one that only changes the Ising spins:
 \begin{itemize}
   \item Judging by the fourth oder cumulant and the histograms of observables this transition is likely continous.
  \item The transition point scales like $\beta_{c,mag} \sim  N^{-0.41 \pm 0.04}$.
  \item And the actions $S,S_I,S_c$ scale like $N$ for $\beta<\beta_{c,mag}$.
\end{itemize}
And a second transition point at which the geometry changes:
\begin{itemize}
  \item Considering the quantities we examined this point is likely of first order.
  \item The transition point scales like $\beta_{c,geo}\sim N^{-0.77 \pm 0.01}$.
  \item The actions $S,S_I,S_c$  scale like $N^2$ for $\beta>\beta_{c,geo}$.
  \item No clear scalings were determined for the region between the transitions.
\end{itemize}

In Section~\ref{sec:intro} we introduce causal sets, and review the results of~\cite{Glaser:2017sbe} and~\cite{Glaser:2018jss} about scaling in $2$d orders and the Ising model coupled to $2$d orders.
We then explore the location of the phase transitions and how they scale with the system size in Section~\ref{sec:PTloc}, and examine their order in Section~\ref{sec:PTorder}.
The last item to explore is the scaling of the observables, and an attempt to determine critical exponents, which we do in~\ref{sec:scaling}, followed by a discussion in Section~\ref{sec:discussion} before we finish with the traditional summary and outlook in Section~\ref{sec:sum}.

\section{Introduction}\label{sec:intro}

\subsection{Some details about causal sets}
Causal set theory is based on Malament's theorem~\cite{Malament_1977} which shows that the causal structure of a space-time is sufficient to encode its data up to a conformal factor.
A discretised form of the causal structure is then encoded in the discrete partial order, the causal set $\C$.
For two elements $a,b \in \C$ we write $a \prec b$ if $a$ is to the past of $b$.
A causal set is defined as a partial order that is
\begin{itemize}
\item {\bf reflexive} $a \prec a$ for any $a \in \C$
\item {\bf transitive} for all $a,b,c \in \C$, if $a \prec b$ and $b \prec c$ then $a \prec c$
\item {\bf antisymmetric} if $a,b \in \C$ then $a \prec b \prec a$ only if $a=b$ so there are no closed time-like curves
\item {\bf locally finite} for all $a,b \in \C$  $| I(a,b)| < \infty $, the number of elements in between any pair of elements is always finite, this condition ensures discreteness.
\end{itemize}
If $a \prec b$ and there is no $c$ such that $a \prec c \prec b$ we say $a \prec_l b$, the elements are linked.
A chain is a set of elements $\{a_0, \ldots , a_n\}$ so that $a_0 \prec a_1 \prec \ldots \prec a_n$, if all relations in a chain are links  $a_0 \prec_l a_1, a_1 \prec_l a_2 ,\ldots , a_{n-1} \prec_l a_n$ it is called a path.
All elements $c$ such that $a\prec c \prec b$ form the causal interval, sometimes also called Alexandrov interval $I(a,b)$.
The number of causal intervals of size $i$ is denoted as $N_i$ and the values of $N_i$ counted in a causal set can be used to measure the manifold likeness and dimension for a causal set~\cite{Glaser:2013pca}.
A causal set can also be written down using the adjacency matrix $A_{ik}$ and the link matrix $L_{ik}$.
Using a Kronecker-delta $\delta_{i\prec j}$ which is $1$ if the relation exists, and $0$ otherwise, we can write these as upper triangular matrices
\begin{align}
  A_{ik}&= \delta_{i\prec k} & L_{ik}&= \delta_{i\prec_l k} \;,
\end{align}
where we assume that the elements are labelled in a way that if $i<k$ then $i\prec k$ or $i$ is not related to $k$, a labelling of this type is also called a natural labelling.
The same causal set can have different natural labelings, however causal set observables are constructed to be label invariant.

The $2$d orders are a class $\Omega_{2d}$ of partial orders whose interest for causal set theory was first investigated in~\cite{Brightwell_Henson_Surya_2008}.
They can be defined as the intersection of two total orders.
For a given set $S = \{1, . . . , N \}$ let $U = (u_1 , u_2 , . . . ,u_N )$ and $ V = (v_1 , v_2 , . . ., v_N )$, such that $u_i,v_i\in S$, with $u_i=u_k \Rightarrow i=k$, and $v_i=v_k\Rightarrow i=k$.
Then total orders on $U$ and $V$ are induced by the order on the integers, and the $2$d order $C \equiv U \cap V$ is a partial order with elements $e_i\equiv (u_i,v_i)\in C $ where the ordering is induced as $e_i \prec e_k$ in $C$ iff $u_i < u_k$ and $v_i < v_k$~\cite{El-Zahar_Sauer_1988,Winkler_1990,Brightwell_Henson_Surya_2008}.
\begin{figure}
  \includegraphics[width=\textwidth]{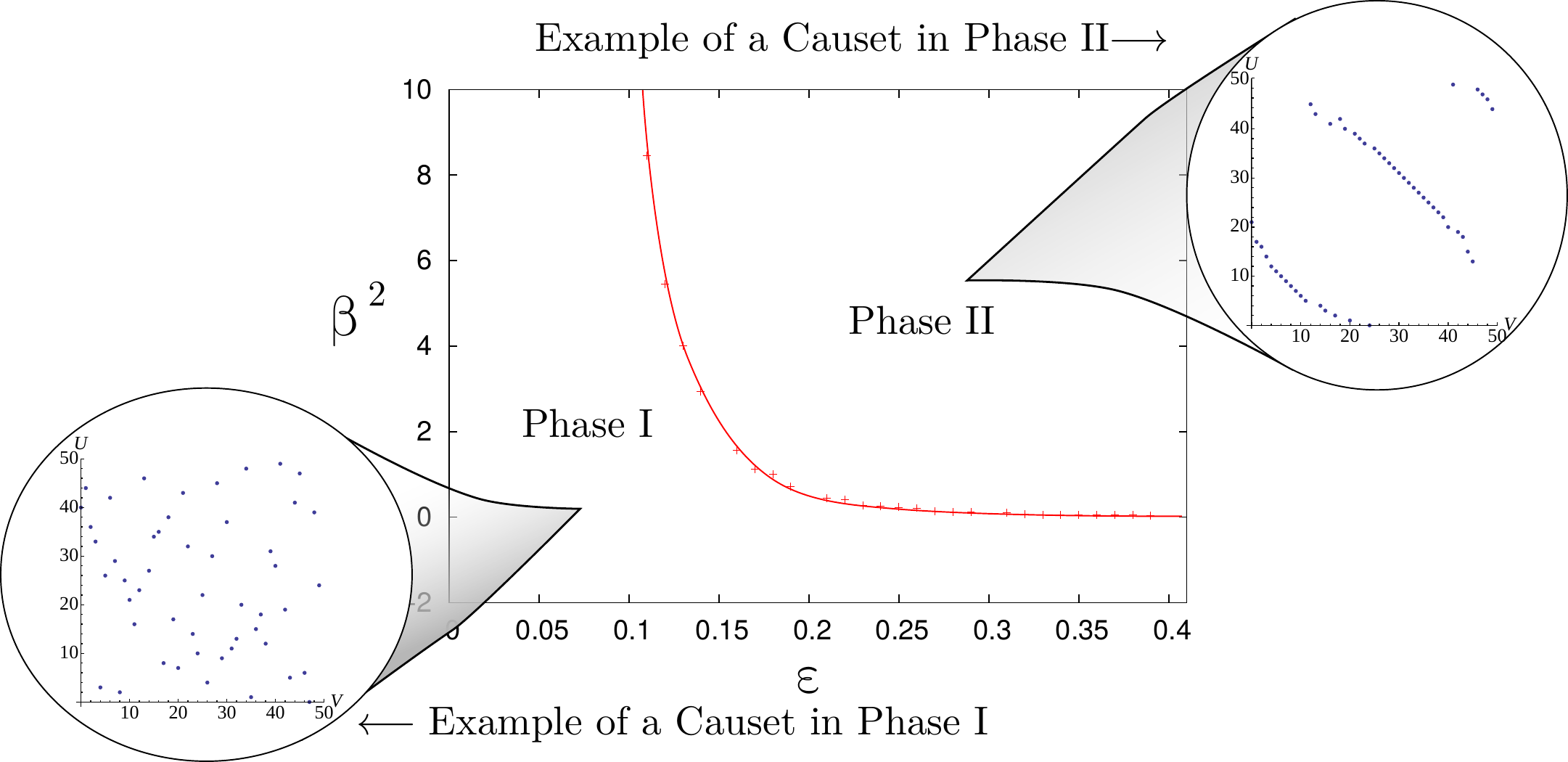}
  \caption{Phase diagram of the $2$d orders, phase I are the random $2$d orders, while phase II are the crystalline ones. This figure is a modification of a figure from~\cite{Surya_2012}.}\label{fig:2dPhases}
\end{figure}

In~\cite{Surya_2012} Surya first explored the path integral over the $2$d orders, studying

\begin{align}
  \mathcal{Z}= \sum_{p \in \Omega_{2d}} e^{- \beta S_c(p,\epsilon)}
\end{align}
using Markov Chain Monte Carlo simulations.
The configurations were weighted by the Benincasa-Dowker action~\cite{Benincasa_Dowker_2010,Dowker_Glaser_2013,Glaser_2014}, which for a causal set $p$ with $N$ elements and interval abundances $N_i$ is
\begin{align}
 S_{c}(p,\epsilon) &=4 \epsilon \left( N- \sum_{n} N_n f(n,\epsilon) \right) \\
  f(n,\epsilon)&= (1-\epsilon)^{n} \left(1- \frac{2 n \epsilon}{1-\epsilon}+ \frac{n (n-1)\epsilon^2}{2 (1-\epsilon)^2} \right)\;.
\end{align}
Here $\epsilon$ is an intermediate non-locality scale, as introduced in~\cite{Sorkin_2007} to dampen the fluctuations.
The simulations in~\cite{Surya_2012} revealed two different phases, a random phase, dominated by $2$d orders that were manifold like, and a crystalline phase, in which the $2$d order froze into an energetically optimal configuration.
The resulting phase diagram, with examples of the two phases, is shown in Figure~\ref{fig:2dPhases}.

The transition between these two phases is dependent on the non-locality scale $\epsilon$, and the system size $N$.
The order of this phase transition and the scaling of the observables for the $2$d orders in $N$ and $\epsilon$ was studied in~\cite{Glaser:2017sbe}.
There we found that the phase transition is of first order, and shows very clear scaling with $N$ and $\epsilon$, which indicates that the transition will persist for large $N$.
The phase transition scales like
\begin{align}
\beta_c = \frac{1.66}{\epsilon^2 N} + O(\frac{1}{N^2})
\end{align}
to very good precision, and even the sub-leading $N^{-2}$ term could be determined.
The action scales like $N$ for small $\beta$ and like $N^2$ for large $\beta$, the large $\beta$ scaling is explained by the fact that, to first order, the action scales like the number of links, and the number of links in the crystalline causal sets scales like $N^2$.

Away from the phase transition, the scaling of observables can be determined using the free energy $\beta F = -\ln{Z}$, and hence
\begin{align}
  \av{S}&= \pd{(\beta F)}{\beta}  &  \mathrm{Var}(S)&= \left( \av{S -\av{S}}\right)^2=- \pd{^2(\beta F)}{\beta^2}
\end{align}
which means that the scaling of any one of $S,F,\mathrm{Var}(S)$ should determine the scaling of all others.
Assuming that $\beta_c \sim N^{\lambda}$  and $\av{S} \sim N^{\nu}$, one then expects $F\sim N^{\nu}$ and $\mathrm{Var}(S)\sim N^{\nu-\lambda}$.
Our notation here differs from that in~\cite{Glaser:2017sbe} by some factors of $\beta$, since the scaling of the phase transition points $\beta_c$ with the system size is more complicated for the coupled system.

While including the Ising model might change the scaling with $N$, and could introduce new scalings with $j$,
there is no a priori reason why the $\epsilon$ dependence of the system should be changed through coupling to the Ising model.
While it is possible that this might have an influence through the change of the crystalline structure with $\epsilon$, we chose
to only examine the single value $\epsilon=0.21$ in this work.

\subsection{Ising model and $2$d orders}
In~\cite{Glaser:2018jss} we examined the $2$d orders coupled to the Ising model.
We did this using Markov Chain Monte Carlo simulations that included steps for the Ising model to equilibrate on the causal sets.

The action for the Ising model on the system, with spins $s_i \in \{ \pm 1 \}$  is defined as
\begin{align}
  S_{I}=j \sum_{i,k} s_i s_k L_{i k}\;,
\end{align}
which corresponds to spins only interacting along links.
The overall action for the system, used in the simulations, is a linear combination of this and the Benincasa Dowker action defined above
\begin{align}
  S&= S_{c}+ S_{I} \;. \\
\intertext{The full simulated action is then:}
 \beta S&=   \beta  4 \epsilon \left( N- \sum_{n} N_n f(n,\epsilon) \right) + \beta j \sum_{i,k} s_i s_k L_{i k}  \;,
\end{align}
In defining this overall action we had the choice to leave $\beta$ as an overall prefactor or to use $\beta$ only as the prefactor of $S_c$ and have a prefactor $\tilde{j}= \beta j$ in front of $S_I$.
The decision to use $\beta$ as an overall prefactor was made, since then the relative strength and signs of the $S_c$ and $S_I$ remain constant along lines of fixed $j$.
This is particularly interesting for us since, as explained in more detail in~\cite{Sorkin_Lorentzian} and in~\cite{Surya_2012} in the context of simulations of the $2$d orders, $\beta$ here should be considered as a Wick rotation parameter, which would need to be analytically continued to imaginary values to obtain a quantum theory of geometry.
This origin of $\beta$ also motivates the study of negative $\beta$, which would be questionable if we were to consider $\beta$ only as an inverse temperature.

We will use these three actions as observables in our attempt to understand the system.
We also define two other observables, to better examine the spin system, one of which is the absolute magnetisation of the system
\begin{align}
  |M|=|\frac{1}{N} \sum_i s_i \;|.
\end{align}
We chose to work with the absolute value $|M|$ to remove the global symmetry induced when all spins are multiplied by $-1$.

\begin{figure}
  \centering
  \includegraphics[width=0.6\textwidth]{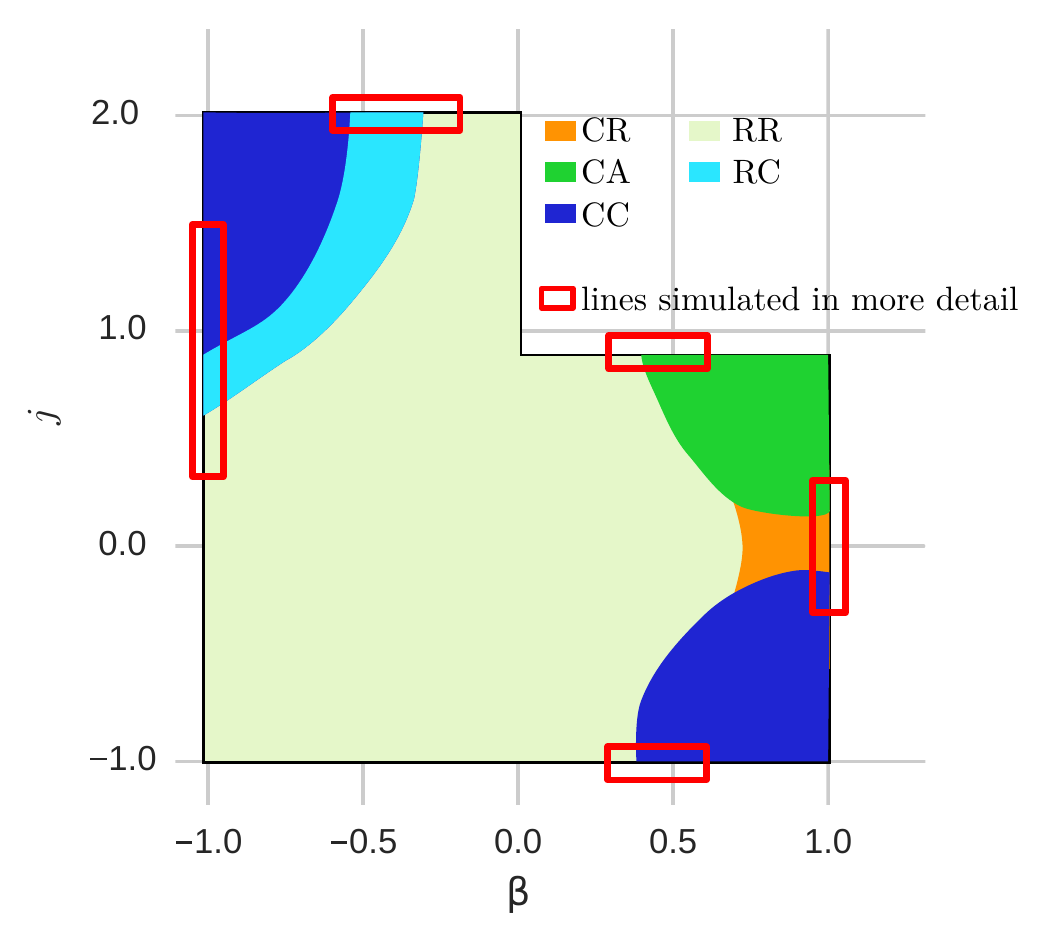}
  \caption{\label{fig:phases}This figure is from~\cite{Glaser:2018jss} and shows the structure of phase transitions found there, the red boxes mark the lines that were studied in more detail there. The different phases are explained in the legend, with the first letter pertaining to the state of the causal set random (R) or crystalline (C) and the second letter pertaining to the state of the Ising model random (R), correlated (C) or anticorrelated (A).}
\end{figure}

The last is the relation correlation, which was defined in~\cite{Glaser:2018jss}\footnote{In~\cite{Glaser:2018jss} a factor of $1/N$ in the definition of $R$ was omitted in the manuscript, however since the simulations there were done at fixed $N$ this does not change any of the conclusions reached there.}
\begin{align}
  R=\frac{1}{N}  \sum_{i,k} s_i s_k A_{i k} \;.
\end{align}
In~\cite{Glaser:2018jss}, before starting to explore the full dynamical system, we explored the Ising model on fixed $2$d orders.
For this we picked $20$ causal sets from the random phase at $\beta=0$ and $20$ causal sets in the crystalline phase for $\beta=1.5$ and $\epsilon=0.21$.
The phases we found there agreed well with those arising in the dynamical simulations.
There are two distinct phases for the geometry, random, and crystalline, which correspond to those found for the random $2$d orders\footnote{One would expect that even the pure random orders at very negative $\beta$ might show a new phase, which we did explore too, however this phase does appear in the data coupled to the Ising model for the parameter region explored.}.
The Ising spins can be found in random, correlated and anti-correlated states, where the anti-correlated state only happens for positive $\beta, j$, and only on crystalline causal sets.
Scanning the phase space at fixed $N$ we find the phase diagram shown in Figure~\ref{fig:phases}.
In~\cite{Glaser:2018jss} we chose to label the five different phases by the state of the Ising model, random, correlated or anticorrelated, and the state of the causal set, random or crystalline.
In Figure~\ref{fig:2dPhases} the completely random state is marked in light green, the correlated, crystalline state is marked dark blue, the state of random causal sets with correlated spins is marked in light blue, the state of crystalline causal sets with random spins is orange, and the state of crystalline causal sets and anti-correlated spins is marked in green
Not all possible permutations of the possible states arise, and it is in particularly glaring that no new phase arises in the lower left corner where both $\beta$ and $j$ take large negative values.
One possible reason for this, discussed in~\cite{Glaser:2018jss} is that we do not have a suitable observable to find anti-correlated states on random causal sets.

Of particular interest here is the region in the upper left corner, where the Ising spins seem to push the geometry into the crystalline state.
In the present work we will use the same observables to explore the scaling of the system with size.

\section{Location of the phase transitions}\label{sec:PTloc}
For this article we look in detail at two lines in the phase space, a line from the completely random state to the correlated crystalline state at positive $\beta$, $j=-1, \beta \in [0,0.8]$ and a line at negative $\beta$ from the completely random state through the random causal sets with correlated Ising spins to the correlated crystalline state, $j=1, \beta\in [-1.4,0]$.
The first is chosen since it allows for easy comparison with the scaling results for the pure $2$d orders in~\cite{Glaser:2017sbe}, to see how the inclusion of the Ising model changes the phase transition.
The second line is chosen to cross the region in which the matter most drastically changes the geometry.
We chose to use fixed $j$ so that the relative strengths of the geometric and the magnetic action remains fixed along the lines explored.
The choice of $j=\pm 1$ was made for symmetry reasons, and without considering the choices made in~\cite{Glaser:2018jss}.

In considering our results we can draw some intuition from considering that negative $\beta$ should lead the value of $S$ to be maximized, which will require a balance between states that maximize $S_c$ and those that maximize $S_I$, as we will see, for negative beta, the geometric states that maximize $S_I$ will minimize $S_c$, leading the states to be in conflict with each other.

We use the same code as in~\cite{Glaser:2018jss} which is a modification of an earlier version of the causal set generator code~\cite{Cunningham_Krioukov_2017} and is available online at~\url{https://github.com/LisaGlaser/IsingCauset}.
The data generated for this article is uploaded, with some examples of analysis code, here~\cite{IsingCausetData}.

The duration of the simulations and amount of data generated here makes it impractical to use the gold standard test for thermalisation of starting the simulations for the same parameter values from different initial conditions and waiting for their convergences.
Instead we applied different heuristics that work reasonably well for our situation.
First we use very long simulations, with more than $500$k sweeps (how many exactly varies with the parameter value $(N,\beta,j)$), which corresponds to at least $500\, 000 N^2$ attempted moves for the causal set elements and $500\, 000 N$ attempted moves for the Ising spins, and only use the last $100$k of these in our analysis.
The first heuristic for thermalisation is a simple consistency check, neighbouring parameter values of $\beta$ (we keep $j$ fixed) should have similar values for the observables, unless they are at the phase transition, hence points that were not at the phase transition but showed different averages were simulated longer.
Next, the autocorrelation of distant states of the simulations can be used as a test of thermalisation.
In an equilibrium state the autocorrelation should fall off exponentially with the number of sweeps between states, hence parameter values that do not show this behavior are examined closer again, and simulated longer if necessary.
Third we visually inspect how the action changes with each sweep.
If this shows a visible slope, or a jump to a different state where it remains, within the last $100$k sweeps we simulate the point longer.
Combining these measures then leads us to be reasonably certain that our data is thermalized.

\begin{figure}
\subfloat[$N=30$]{
\includegraphics[width=0.5\textwidth]{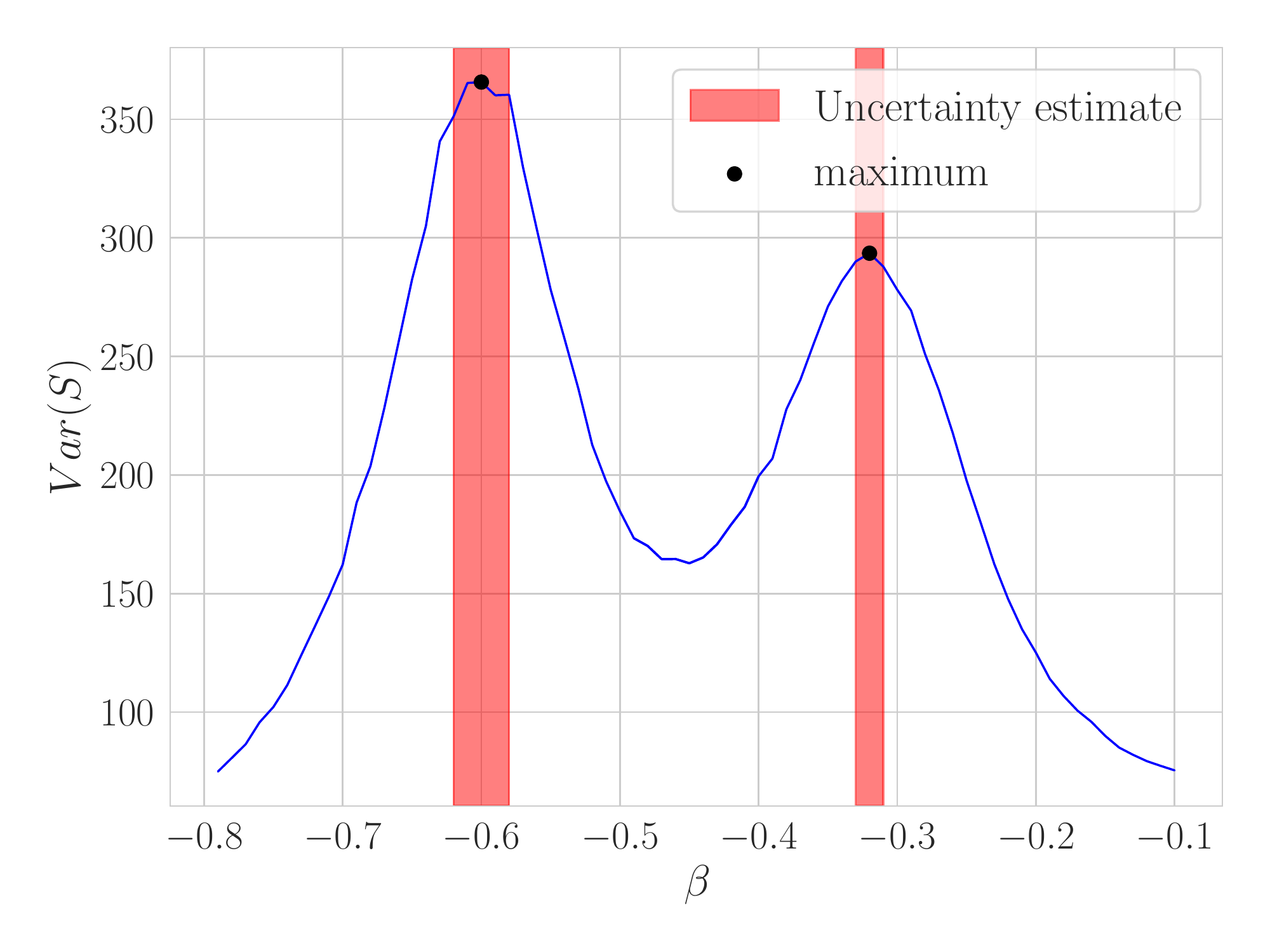}}
\subfloat[$N=70$]{
  \includegraphics[width=0.5\textwidth]{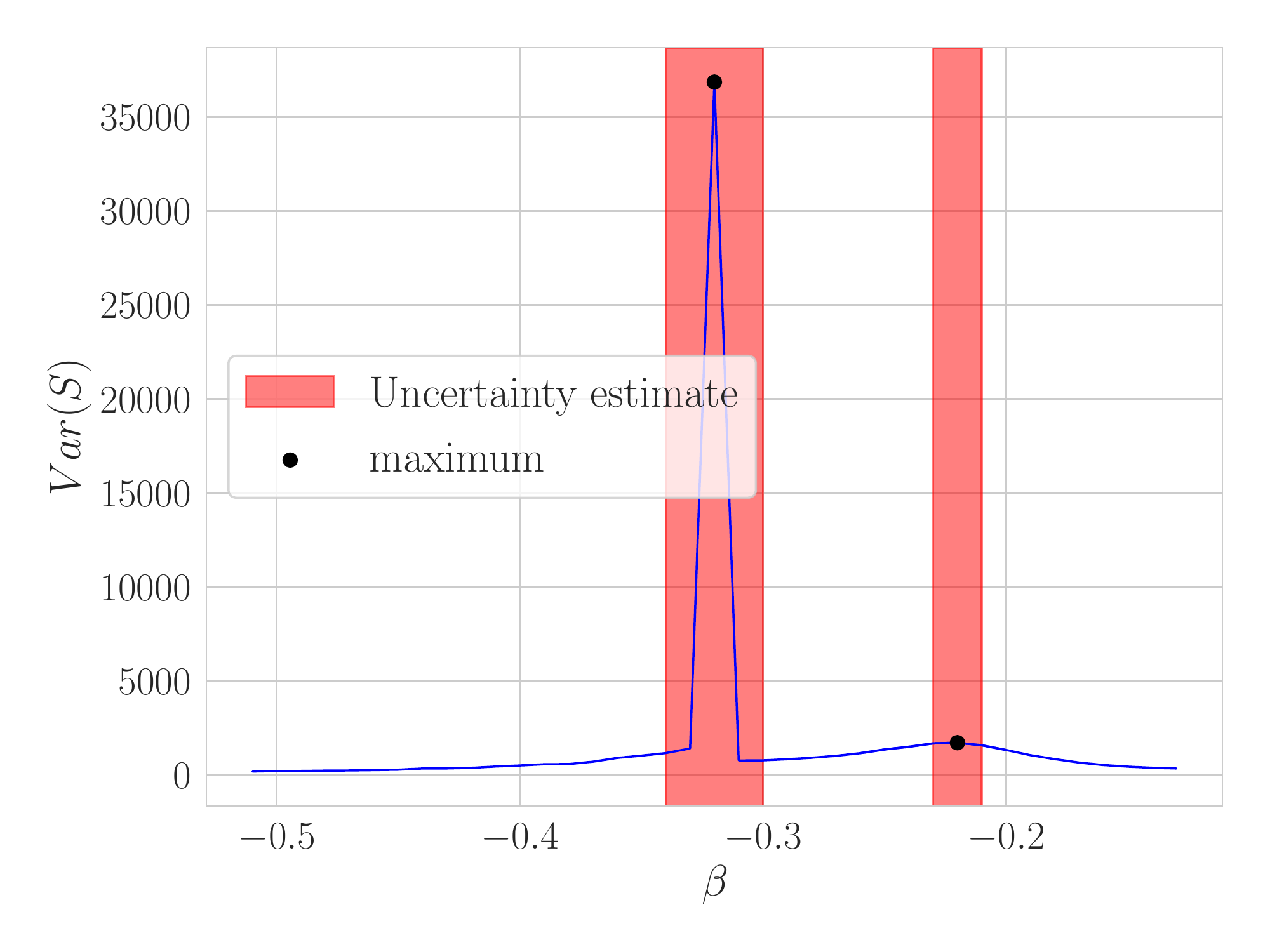}}
  \caption{\label{fig:peakfind} Phase transition points, determined from $\mathrm{Var}(R)$ and $\mathrm{Var}(S)$, with their associated uncertainty marked as red shading along the $j=1$ line for $N=30,70$}
\end{figure}

To determine the location of the phase transitions we use a simple peak finder on the variance of our observables.
The peak finder works based on a guess of the phase transition location $\mu_g$ and its width $\sigma_g$, using this input it finds the maximum value in the region of $\mu_g \pm 5 \sigma_g$ the location of which approximates the phase transition point.
We estimate the uncertainty in the peak position as the next to next value of $\beta$ simulated, which might be an over estimate for large system sizes.
Similar results could be achieved by manually reading off the results, however this is impractical for the amount of data generated here.
We show some examples of the peaks analysed, for $N=30,70$ and $j=1$ in Figure~\ref{fig:peakfind}.

\subsection{Scaling of the phase transition along the $j=-1$ line}

\begin{figure}
\subfloat[$\beta=0$]{ \includegraphics[width=0.33\textwidth]{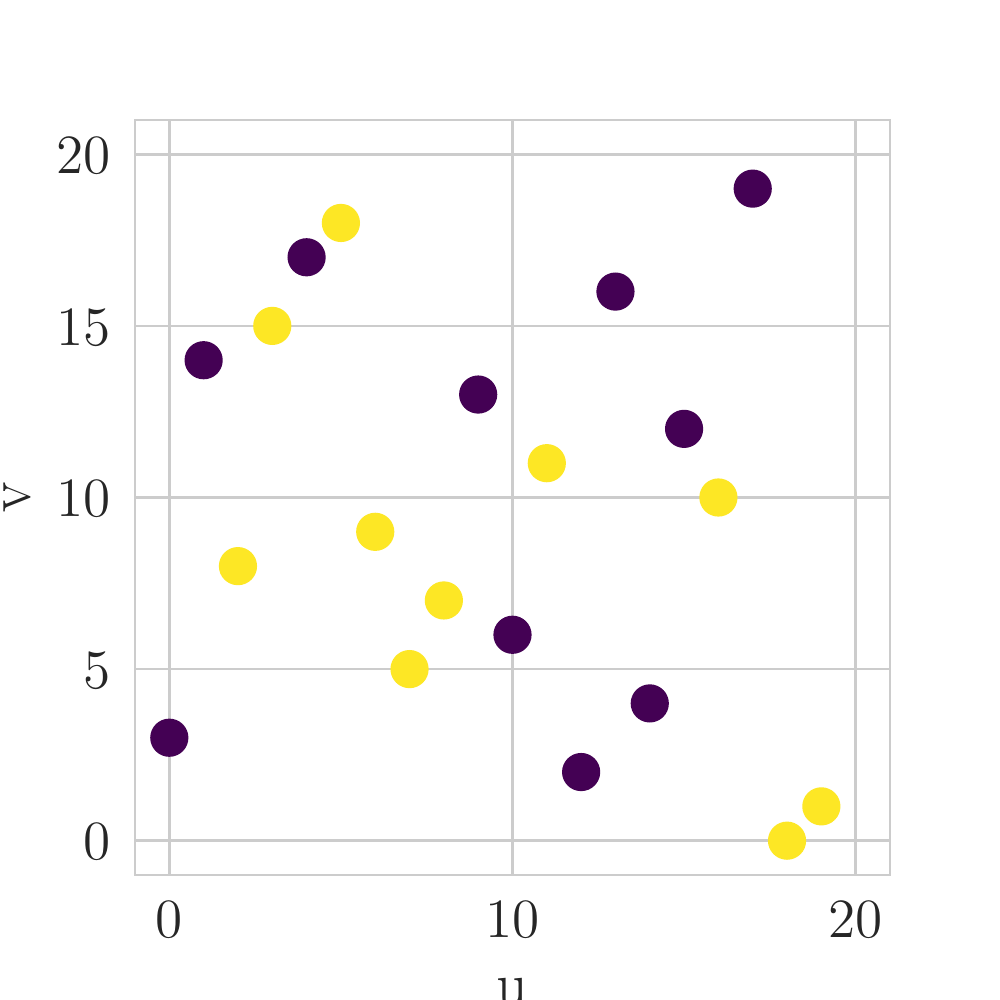}}
\subfloat[$\beta=0.4$]{\includegraphics[width=0.33\textwidth]{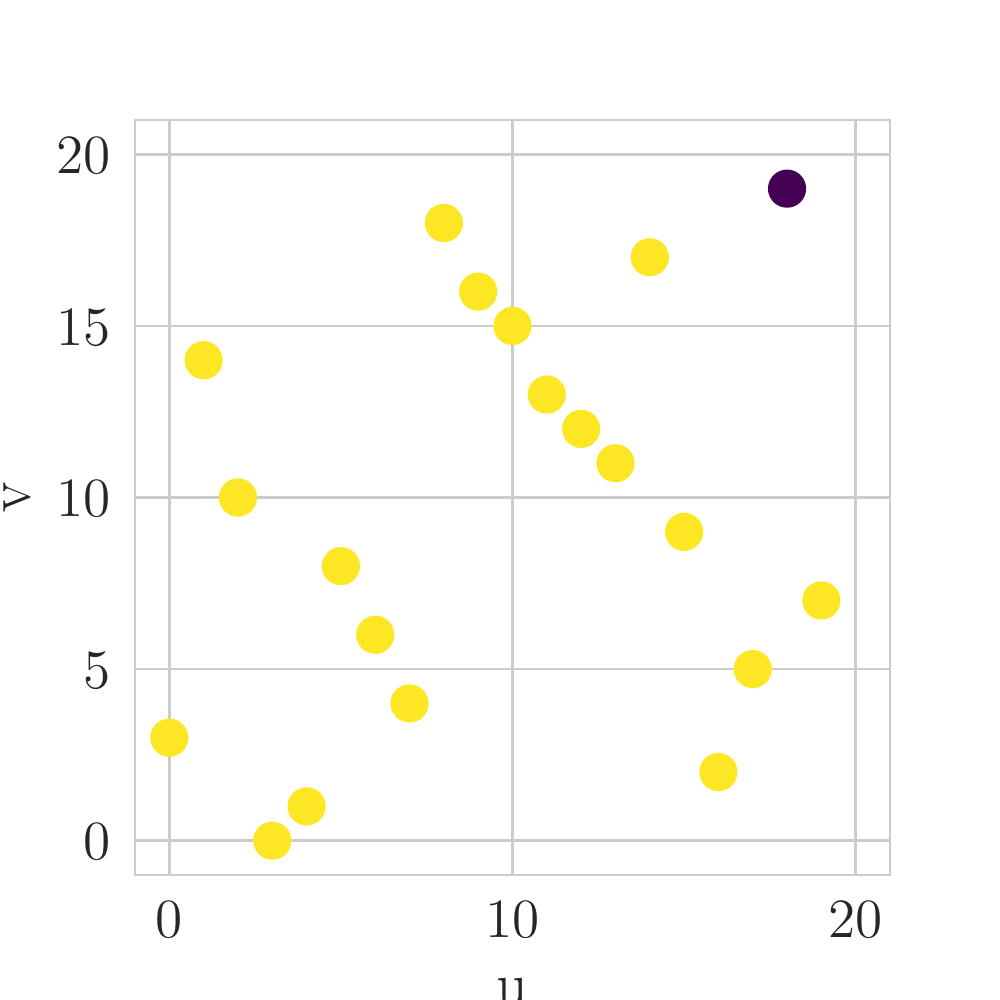}}
\subfloat[$\beta=0.8$]{  \includegraphics[width=0.33\textwidth]{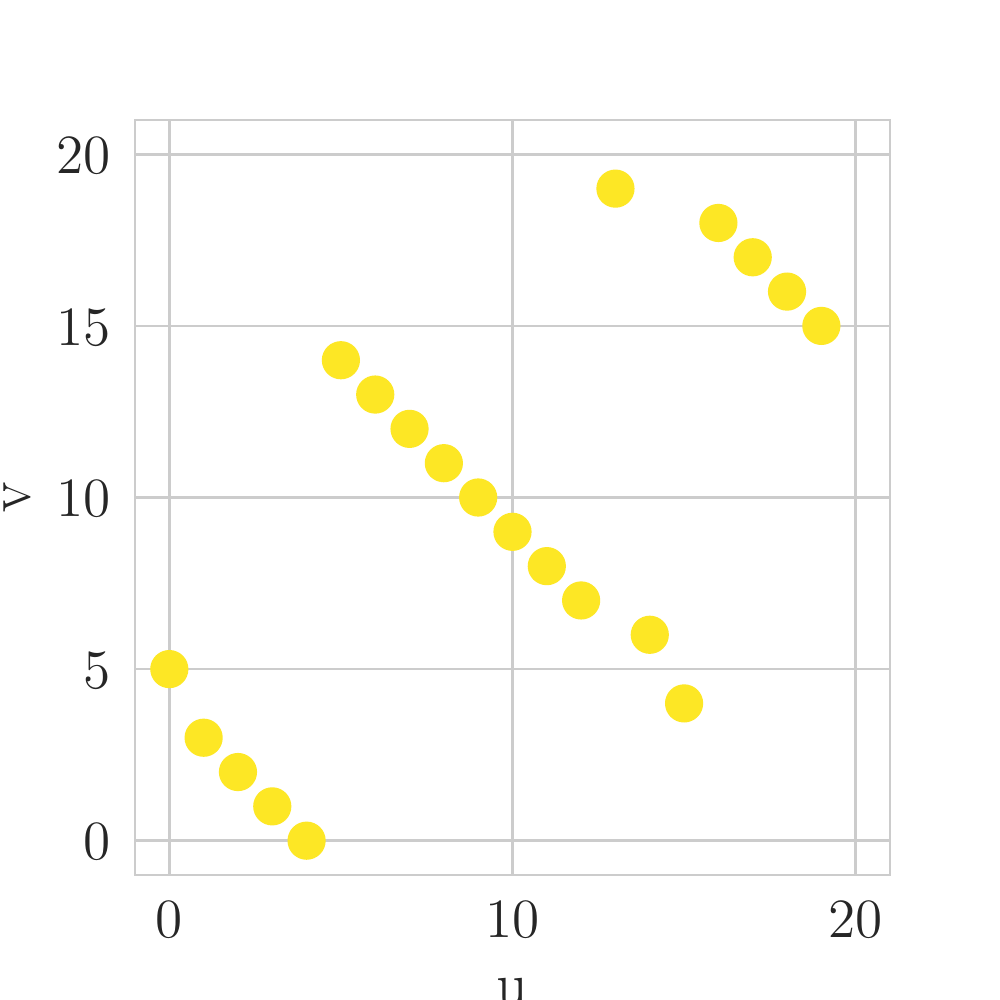}}
  \caption{\label{fig:N20phases}For $N=20$  and $j=-1$ we can find 3 different phases, where at intermediate size the spins are oriented, but the causal set elements are not fully crystallized yet.}
\end{figure}

Examining the location of the phase transition at different sizes reveals that for $N<40$ the phase transition in the spins happens at slightly lower $\beta$ than for the geometry, however these two transition happen concurrently for larger $N$.
We show the seemingly three phases for $N=20$ in Figure~\ref{fig:N20phases}, the middle one of these disappears for large $N$.
The separate phase transitions thus seem to be a finite size effect, hence we only include $N\geq 40$ to find the scaling of the transition point.

\begin{figure}
  \includegraphics[width=\textwidth]{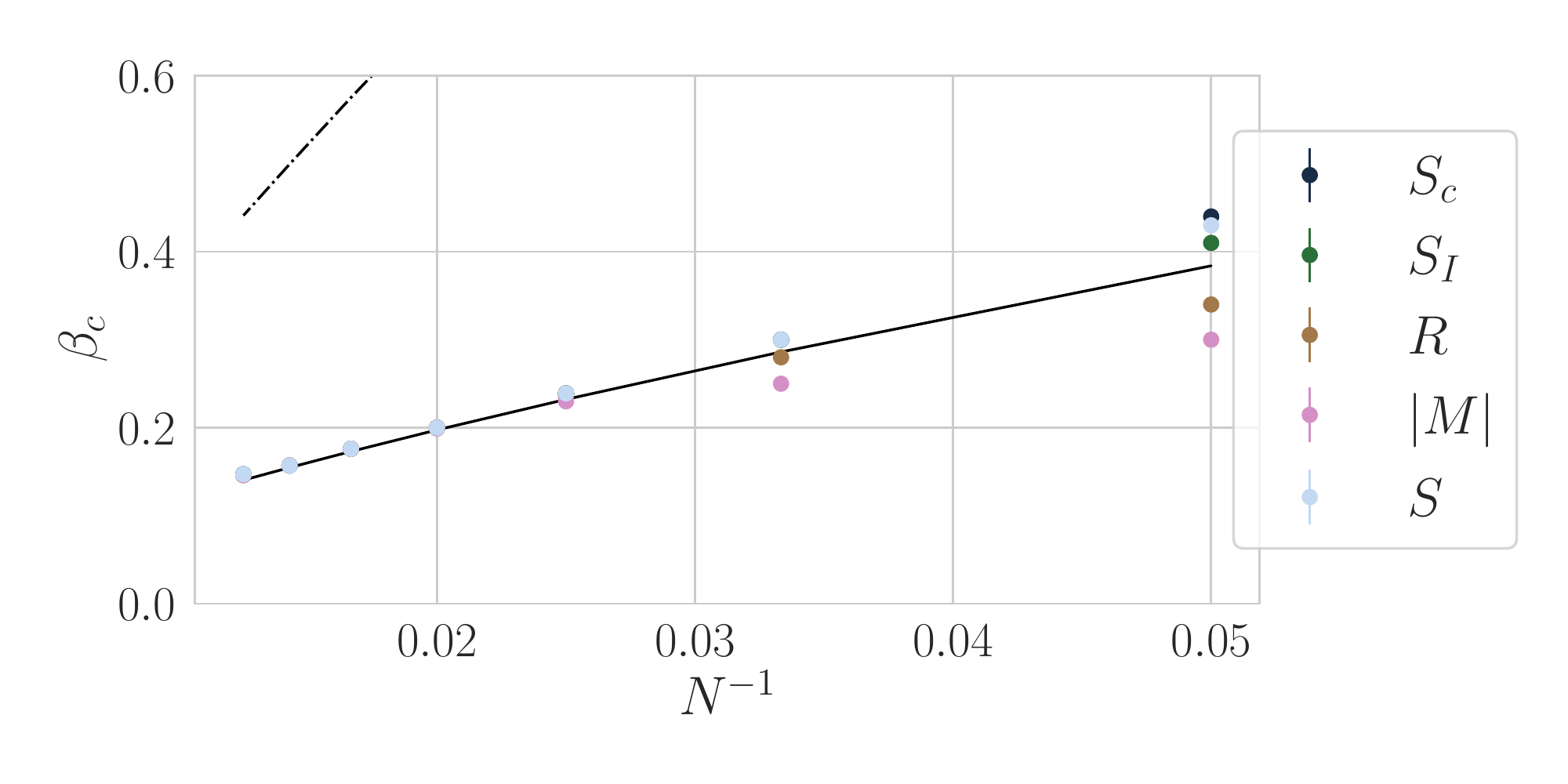}
  \caption{Scaling of the phase transition point $\beta_c$ vs $N$, for different $N$ and all observables. For the line of $j=-1,\beta\in[0,0.8]$.
The solid black curve is the best fit, $\beta_c(N)= (3.35 \pm 0.15 )\cdot N^{-0.72 \pm  0.01}$, and the dash-dotted black curve in the upper left corner is the best fit from~\cite{Glaser:2017sbe} which for $\epsilon=0.21$ is $ 37.64 N^{-1} - 278101.10 N^{-2}$. }\label{fig:betac_jm1}
\end{figure}

As we can see in Figure~\ref{fig:betac_jm1}, we are well justified to combine the phase transition locations for all different observables to determine the scaling of the transition.
We fit the phase transitions with $\beta_c=a N^{b}$, and thus find that $\beta_c \propto N^{-0.72}$.
The entire fit data is given in Table~\ref{tab:fits}, and the fit of the curve is shown in Figure~\ref{fig:betac_jm1}.

This fractional scaling is surprising, since in the pure causal set model the phase transition location scales like $N^{-1}$.
The best explanation is that this arises through an interaction between the phase transition of the causal set with the $N^{-1}$ scaling, and the Ising model, for which the critical temperature is not size dependent, so $N^0$.
The value of $N^{-0.72}$ then suggests that the causal set structure, at least at $j=-1$, has a stronger influence on the phase transition than the Ising model.
Some, limited, data gathered at $j=-0.5$ indicates a scaling with $N^{-0.71}$, however with a much larger uncertainty than that obtained for $j=-1$, so it can neither confirm nor refute the hypothesis that lower $j$ might lead to stronger size dependence in the scaling.
It is likely that with additional data we could also determine the scaling with $j$, however this is left for future study.

The dashed curve in Figure~\ref{fig:betac_jm1} shows the best fit for $\beta_c$ from~\cite{Glaser:2017sbe} for the given value of $\epsilon$, it is thus clear that the phase transition happens at lower $\beta_c$ when the Ising model is added.
The two systems are clearly influencing each other, since the causal set coupled to the Ising model transitions at lower $\beta$ compared to the pure model, while at the same time the Ising model on a fixed crystalline causal set transitions at lower $\beta$ than for a fixed random causal set.
When studying the behaviour of the Ising model on fixed causal sets as a precursor to the full system in~\cite{Glaser:2018jss}, we found that, for $\beta=0.1$ the phase transition on random causal sets happen around $j\simeq -5.5$, while for the crystalline orders it happens around $j\simeq -0.2$.
So crystalline orders lead to  phase transitions at significantly lower $\beta$.

\subsection{Scaling of the phase transition along the $j=1$ line}
For the line at $j=1, \beta\in [-1.4,0]$ we find two phase transitions, one of the spins and one of the geometry which is induced through the spins.
The fact that the separate phase transitions for the positive $\beta$ line are a finite size effect means we need to at least consider this possibility here as well.
To take this into account we add one point at $N=120$, to improve the confidence in our extrapolation.
Large $N$ requires longer computation time, and considerably slows down the thermalisation, which makes generating data here harder.
Hence we simulated fewer points far from the phase transition than for lower $N$, using a first estimate of the scaling relations as derived for the lower $N$ data, to focus on the interesting region.
An additional difficulty is that the phase transitions become sharper at larger $N$, hence requiring a finer grained net to clearly resolve.
To allow for more simulations at larger $N$ one could implement parallel tempering algorithms, however this is outside the scope of the present work.

\begin{figure}
\subfloat[$\beta=0$]{ \includegraphics[width=0.33\textwidth]{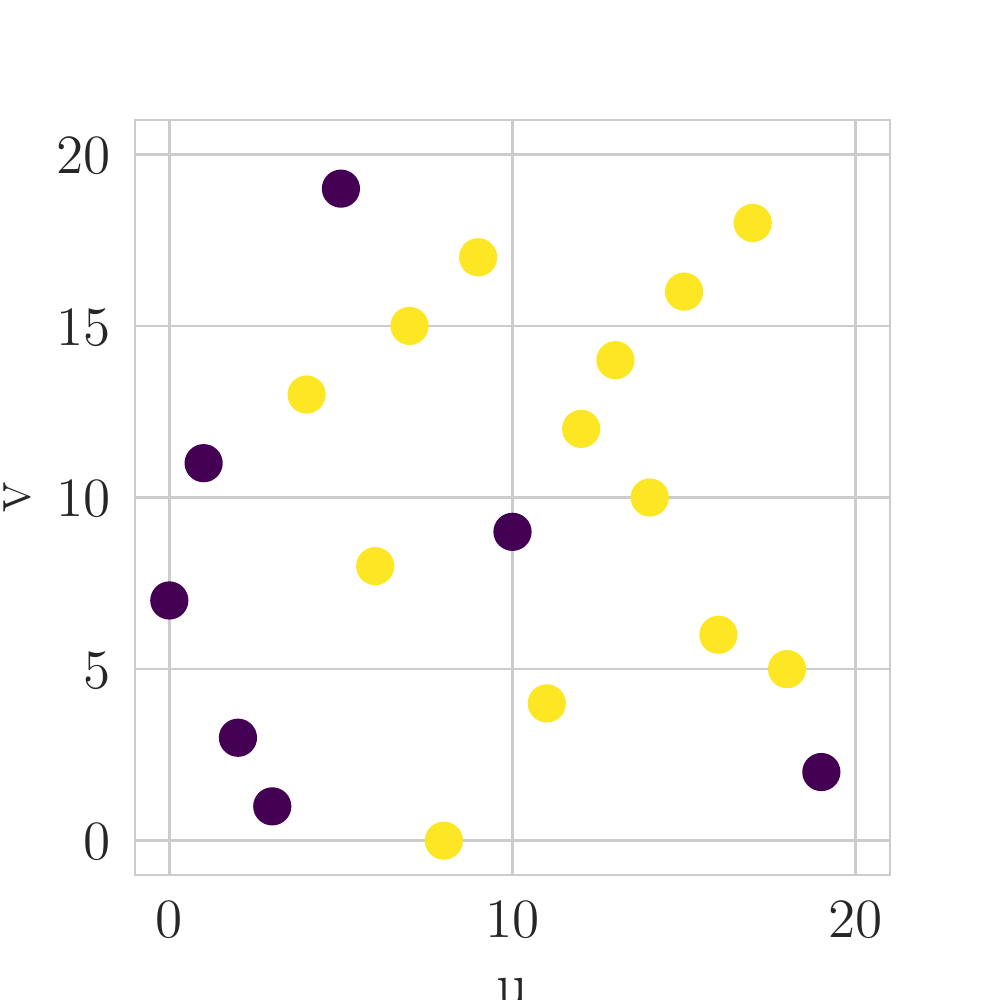}}
\subfloat[$\beta=-0.55$]{\includegraphics[width=0.33\textwidth]{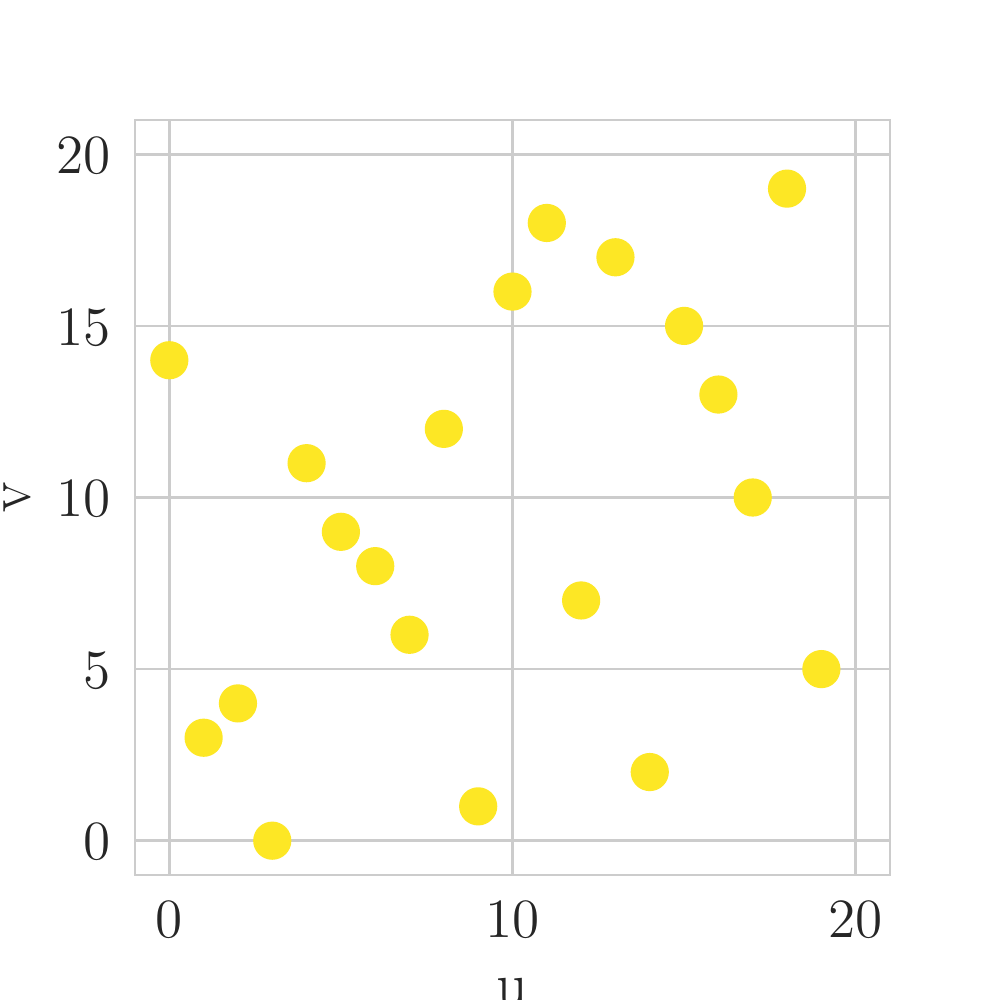}}
\subfloat[$\beta=-1.0$]{  \includegraphics[width=0.33\textwidth]{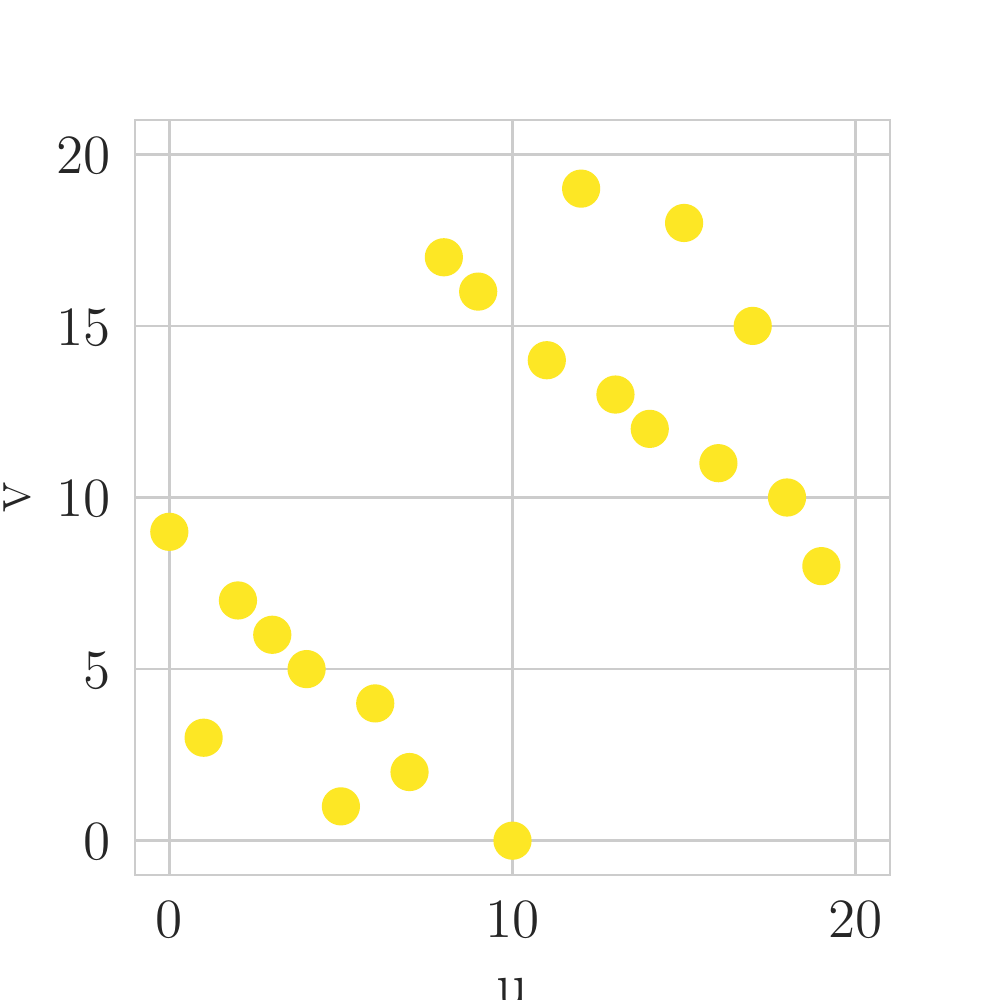}}
  \caption{\label{fig:N20phasesj1}The three different phases at negative $\beta$ and positive $j=1$ for $N=20$, where at intermediate size the spins are oriented, but the causal set elements are not fully crystallized yet.}
\end{figure}

The three phases along this line look very similar to those seen in Figure~\ref{fig:N20phases} above, they are shown in Figure~\ref{fig:N20phasesj1}

\begin{table}
\begin{tabular}{c l l l l }
\toprule
$j=1$    & magnetic  & & geometry & \\
\midrule
          & $a$ & $b$ & $a$ & $b$ \\
\midrule
  & $-1.22 \pm 0.20$ & $-0.41 \pm 0.04$ & $-8.58 \pm 0.34$ &  $-0.77 \pm 0.01$\\
\midrule
\midrule
$j=-1$   & a &b & & \\
\midrule
     & $ 3.35 \pm0.15$ & $-0.72 \pm  0.01$& & \\
\midrule
$j=-0.5$   & a &b & & \\
\midrule
     & $ 5.26 \pm1.50$ & $-0.71 \pm  0.07$& & \\
\bottomrule
\end{tabular}
  \caption{Fit parameters for $\beta_c = a N^{b}$}\label{tab:fits}
\end{table}

\begin{figure}
  \includegraphics[width=\textwidth]{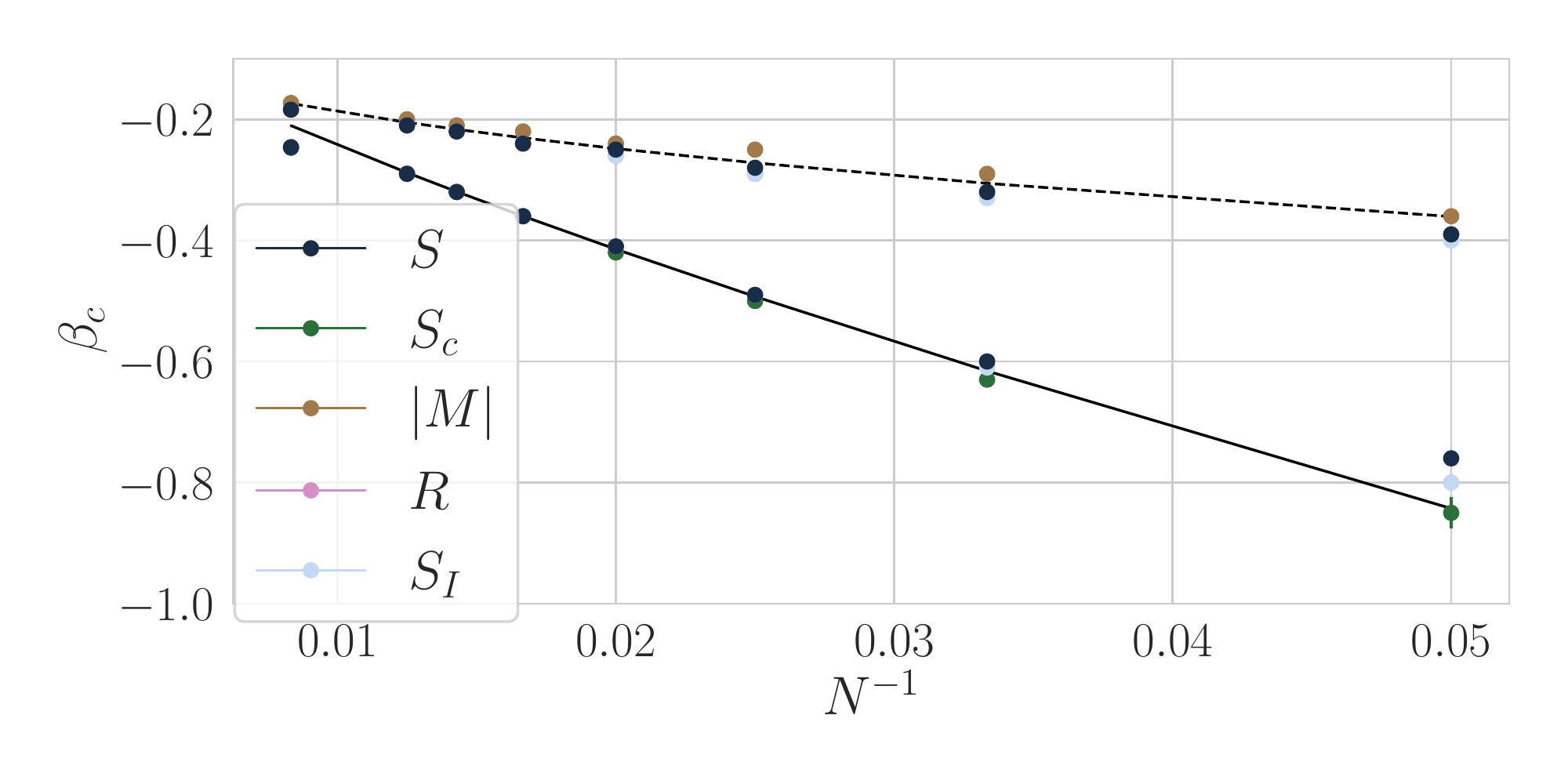}
  \caption{Scaling of the phase transition point $\beta_c$ vs $N$, for different $N$ and all observables. For the line of $j=1,\beta\in[-1.4,0]$.
The dashed black curve is the best fit for the magnetic phase transition as $\beta_{c,mag} = (-1.22 \pm 0.20) \cdot N^{-0.41 \pm 0.04}$ and the solid curve is the best fit for the geometric phase transition at $\beta_{c,geo}=  (-8.58 \pm 0.34) \cdot N^{-0.77 \pm 0.01}$
  }\label{fig:betac_j1}
\end{figure}

To determine the scaling of the phase transitions we again fit the power-law $\beta_c = a N^{b}$, and again exclude the points for $N=20,30$ from the fits for their strong finite size effects.
The best fit parameters with their errors are given in Table~\ref{tab:fits}, and the best fit is plotted in Figure~\ref{fig:betac_j1}.
With the pre-factors we find, the two phase transitions would cross each other and exchange position around $N\approx 500$, or using the uncertainty estimates, latest by $N\approx1000$.
We know that physically this is not possible, since the transition in the geometry is forced by the transition of the spins.
Looking closer at the curve for the geometric phase transition  however, the $N=120$ point is below the best fit, so one might speculate about a levelling off of the transition curve.
To investigate this further would however need more data which can not be generated with the means of production currently available.

\section{Order of the phase transitions}\label{sec:PTorder}

To better understand which order the phase transitions have we will use two methods.
First we will examine the distribution of values for our observables, and how it changes with size.
Usually for a first order transition the system at the transition point is jumping between the two phases.
This is visible if we plot histograms of the observables, and shows as  two distinct peaks of roughly equal size.
As the system size increases these peaks wander further apart, it also becomes more computationally challenging to measure them both since at larger system size the system will take longer to switch between the two phases.

At a second order transition on the other hand we expect the system to take on a new type of state, unique to the transition point.
This leads to a single peak in the histogram of observables at the point of the transition.

The other method we will use is the fourth order cumulant, often also called Binder cumulant.
\begin{align}\label{eq:binder}
  B_{\mc{O}}=1-\frac{\av{\mc{O}^4}}{3 \av{\mc{O}^2}^2} \;.
\end{align}
This cumulant can be calculated for any observable $\mc{O}$, but we will focus on it for the three action observables, $S,S_I,S_c$, and the magnetisation $M$.
The fourth order cumulant is commonly used to determine the order of a phase transition, as explained in~\cite{Binder_1981,Challa_Landau_Binder_1986}.
The cumulant takes on values of $2/3$ for random, states where the distribution of $\mc{O}$ is well approximated as a fairly wide gaussian, while it goes to $0$ for states that have a strongly peaked distribution of $\mc{O}$.
At a continous phase transition the system can be approximated through a single gaussian the whole time, hence the cumulant will take on one of these values, or smoothly change between them, depending on the distribution of states.
At a first order phase transition however, the distribution of $\mc{O}$ would be better described by a double gaussian, hence the value of $B_{\mc{O}}$ becomes negative.
One could then determine the order of the phase transition by just plotting the value of $B_{\mc{O}}$ at the phase transition point.
For first order transitions one could also locate the transition point by looking for the minimum of $B_{\mc{O}}$.
This does not work well in our case, since the value of $B_{\mc{O}}$, is influenced by how close to the phase transition the measurement is taken.
Instead we decided to plot $B_{\mc{O}}$ against $\beta$, which allows us to use the entire curve to guide our interpretation, instead of relying on a single point.

A similar problem also arises when we try to determine the order of the phase transition by directly looking at the histograms of our observables at $\beta_c$.
As the phase transitions becomes sharper with increasing system size they need to be measured very close to the exact location of the phase transition,

to show the double peaks.
So if we have not run simulations for a value $\beta$ sufficently close to $\beta_c$ this can mask the system behaviour.
For our causal set system, an increase in system size to size $N$, increases the number of possible links to $N^2$, and since the BD-action $S_{BD}$ scales to first order like the number of links, its extremal value scales like $N^2$.
This means that as the system size increases, the suppression of the  random, high entropy states, relative to the ordered, low entropy states, gets stronger.
This has two effects, on the one hand we need to be closer to the phase transition to observe a flipping between the two phases, while on the other hand these flips become rarer, requiring us to take more measurements.
Thus creating more measurements closer to the phase transition at larger $N$ becomes very expensive in computing time.
This makes it impractical to add sufficiently many points at large $N$ to ensure optimal resolution.
Instead we will disregard larger $N$ for our conclusion when this limitation appears.
This is an example of critical slowing down at a phase transition, which is usually introduced in higher order transitions, but can also appear at first order transitions.
It seems likely that it appears here due to the non-local character of the $2$d orders.

\subsection{Order of the single phase transition at positive $\beta$}
\begin{figure}
  \centering
  \includegraphics[width=0.9\textwidth]{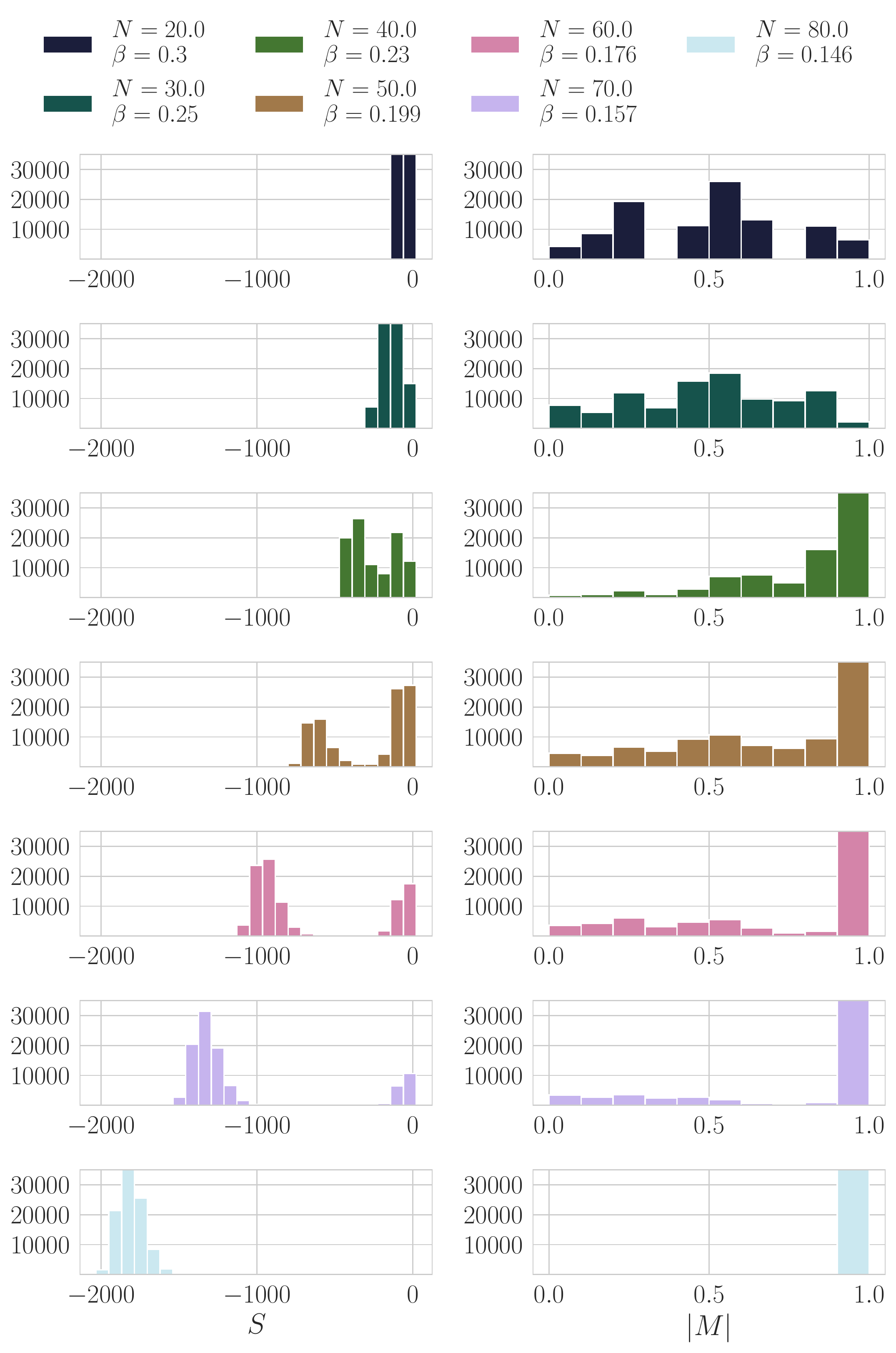}
  \caption{Histograms for the action and absolute magnetisation at the critical $\beta$ for $j=-1$. The actions shows two peaks of increasing separation up to $N=80$, where the peak around $S=0$ is no longer probed. The magnetisation shows the same behaviour, just without a clear peak at $|M|=0$ instead being almost flat there. }\label{fig:jm1Hist}
\end{figure}

Looking at the histogram of the action at this phase transition we see a double peak structure, the two peaks become first visible for $N=40$ and then move away from each other as the system size increases\footnote{As discussed above we do not see the double peak in $N=80$ since there the point closest to the phase transition we measured does not probe both phases.}.
For $|M|$ we see a clear peak at $|M|\approx1$ and a wider distribution starting from $|M|=0$ (again, excepting $N=80$ where the distribution only probes $|M|=1$).
These two effects are illustrated in Figure~\ref{fig:jm1Hist}.

\begin{figure}
\subfloat[$S$]{\includegraphics[width=0.5\textwidth]{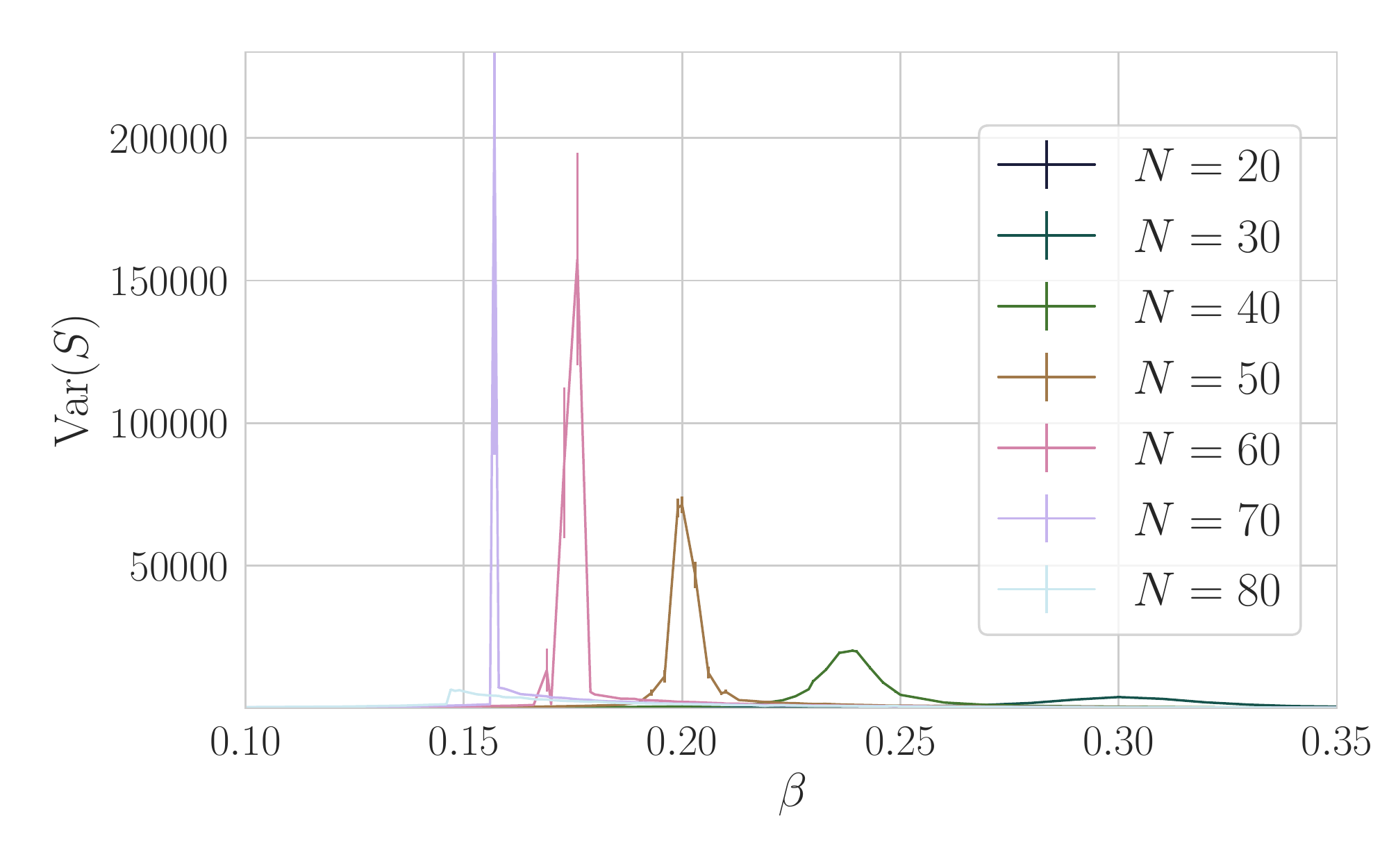}}
\subfloat[$|M|$]{\includegraphics[width=0.5\textwidth]{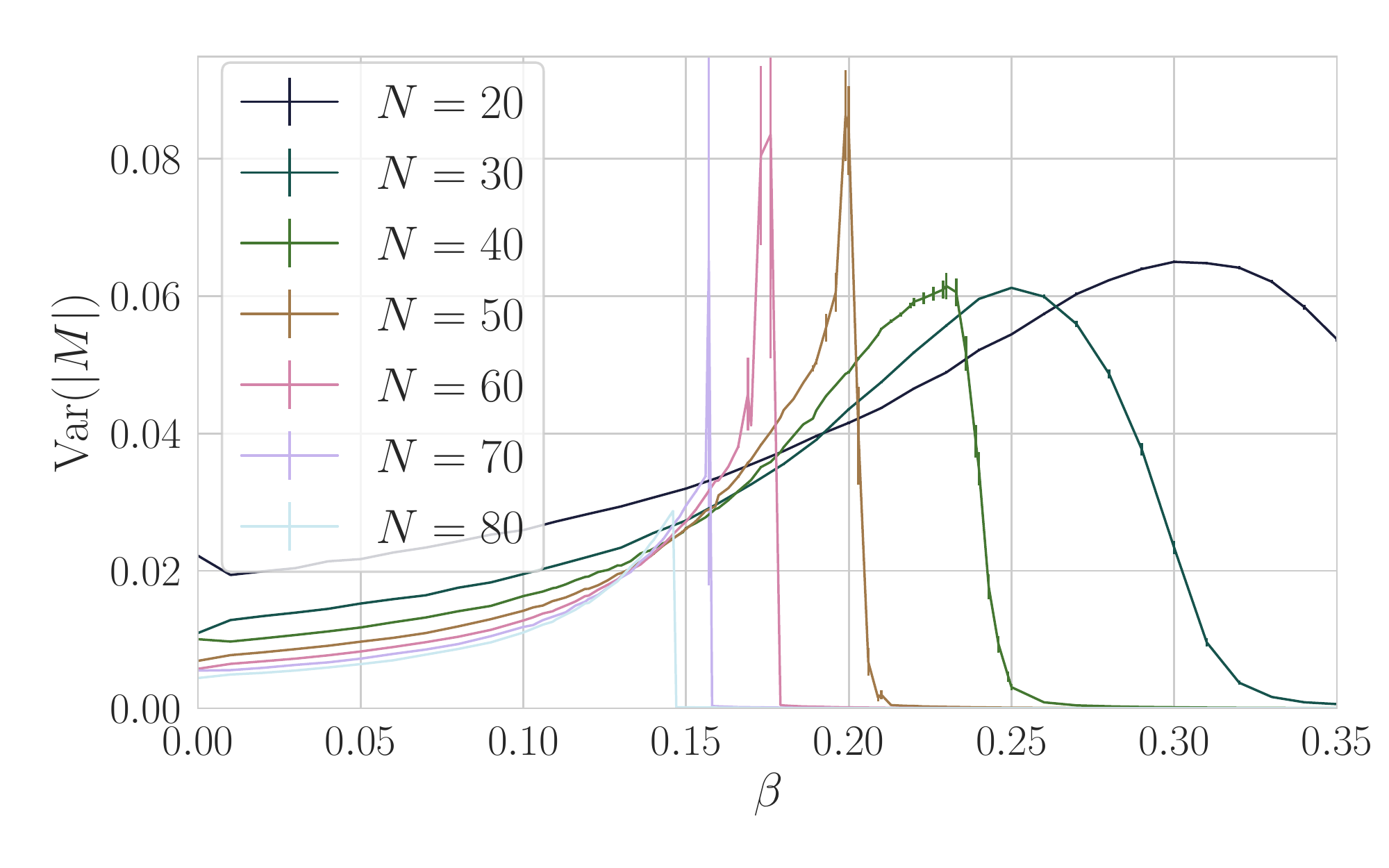}}
  \caption{Plots of the variance for $S$ and $|M|$ at the phase transition for positive $\beta$ and $j=-1$. While the peak in the action becomes higher, the peak in the magnetisation only becomes sharper, without increasing in height. The height of the peaks is plotted on log-scale.}\label{fig:variances}
\end{figure}

\begin{figure}
  \subfloat[$B_{M}$]{
  \includegraphics[width=0.5\textwidth]{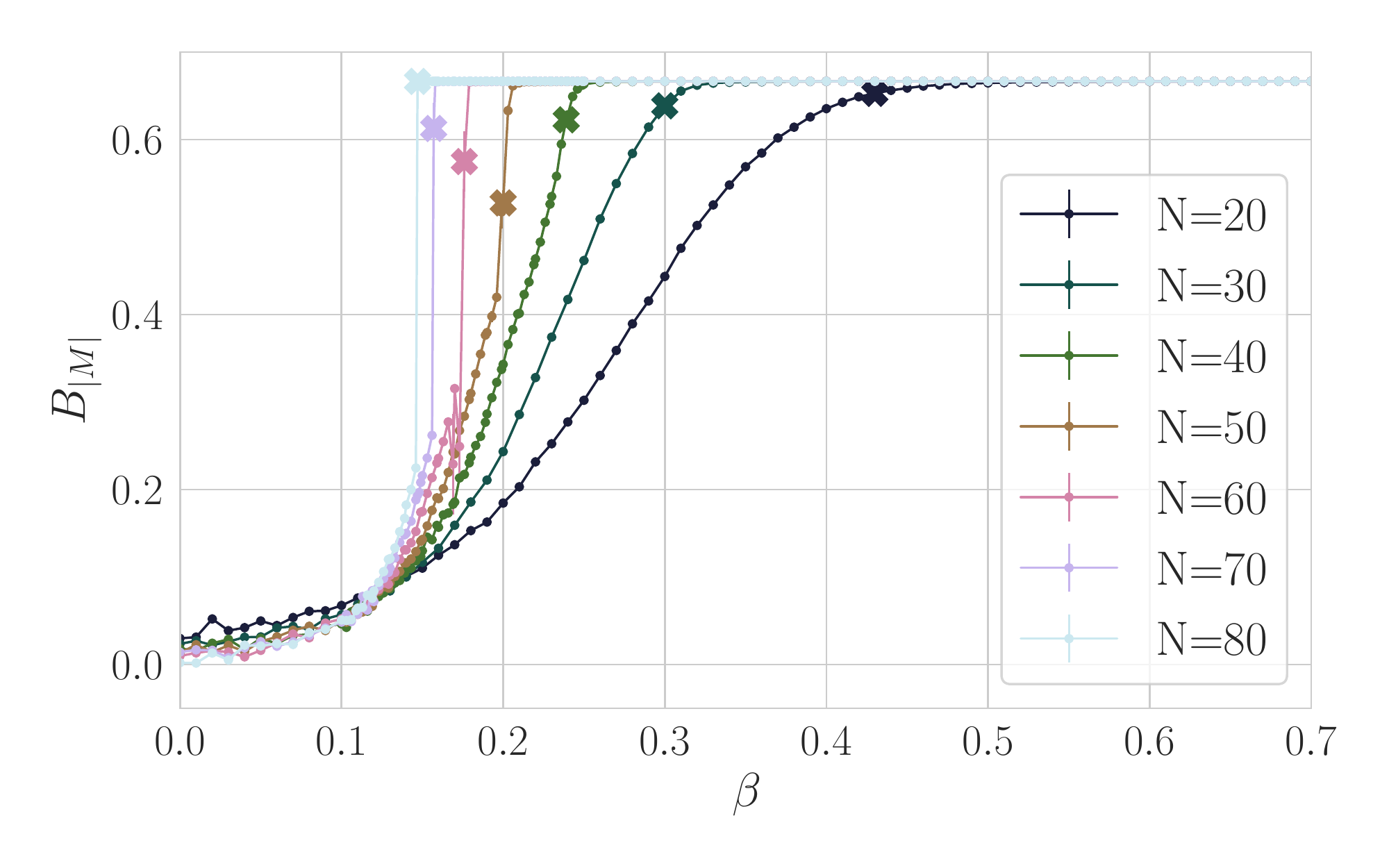}}
  \subfloat[$B_{S}$]{
  \includegraphics[width=0.5\textwidth]{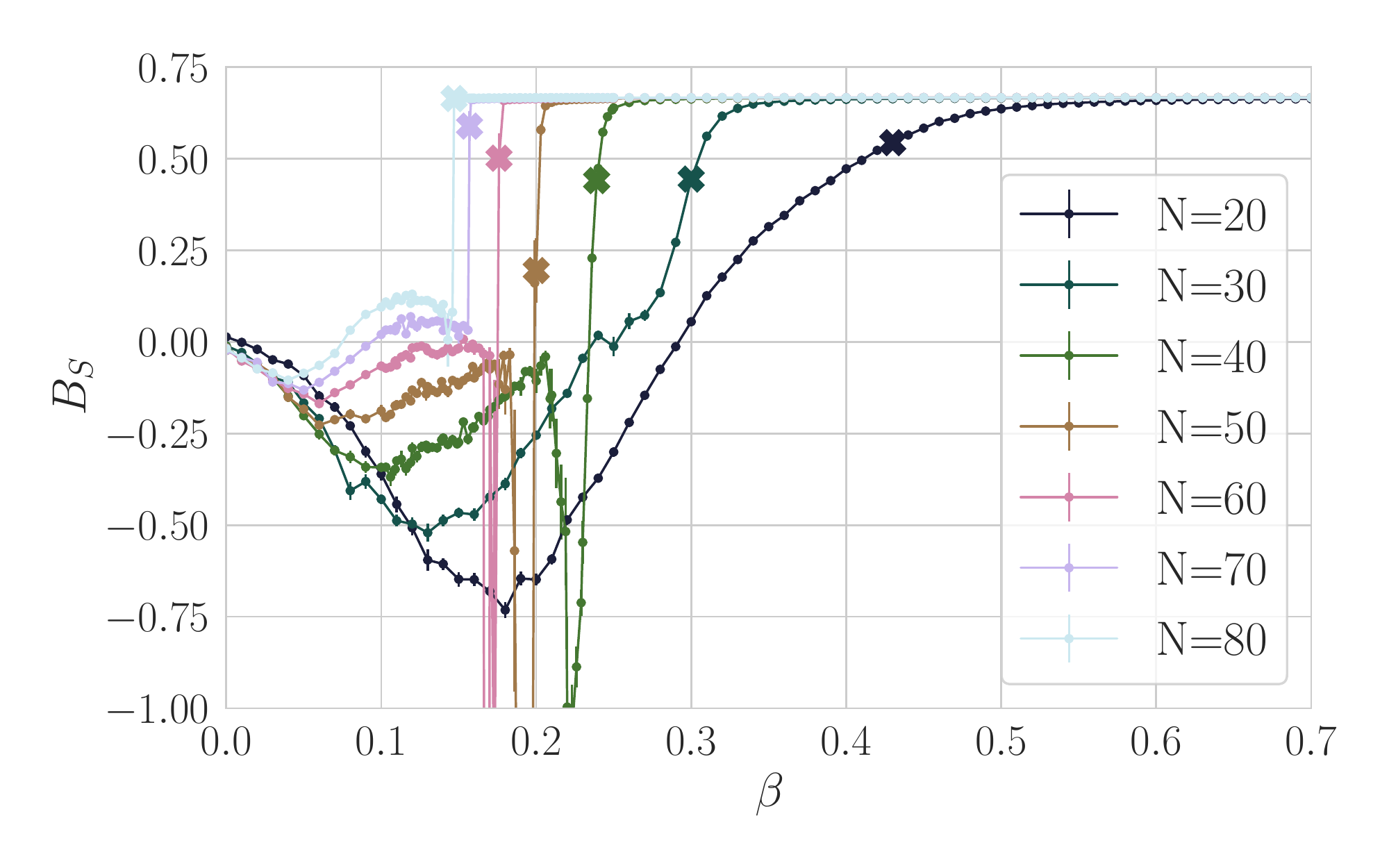}}

  \subfloat[$B_{S_{I}}$]{
  \includegraphics[width=0.5\textwidth]{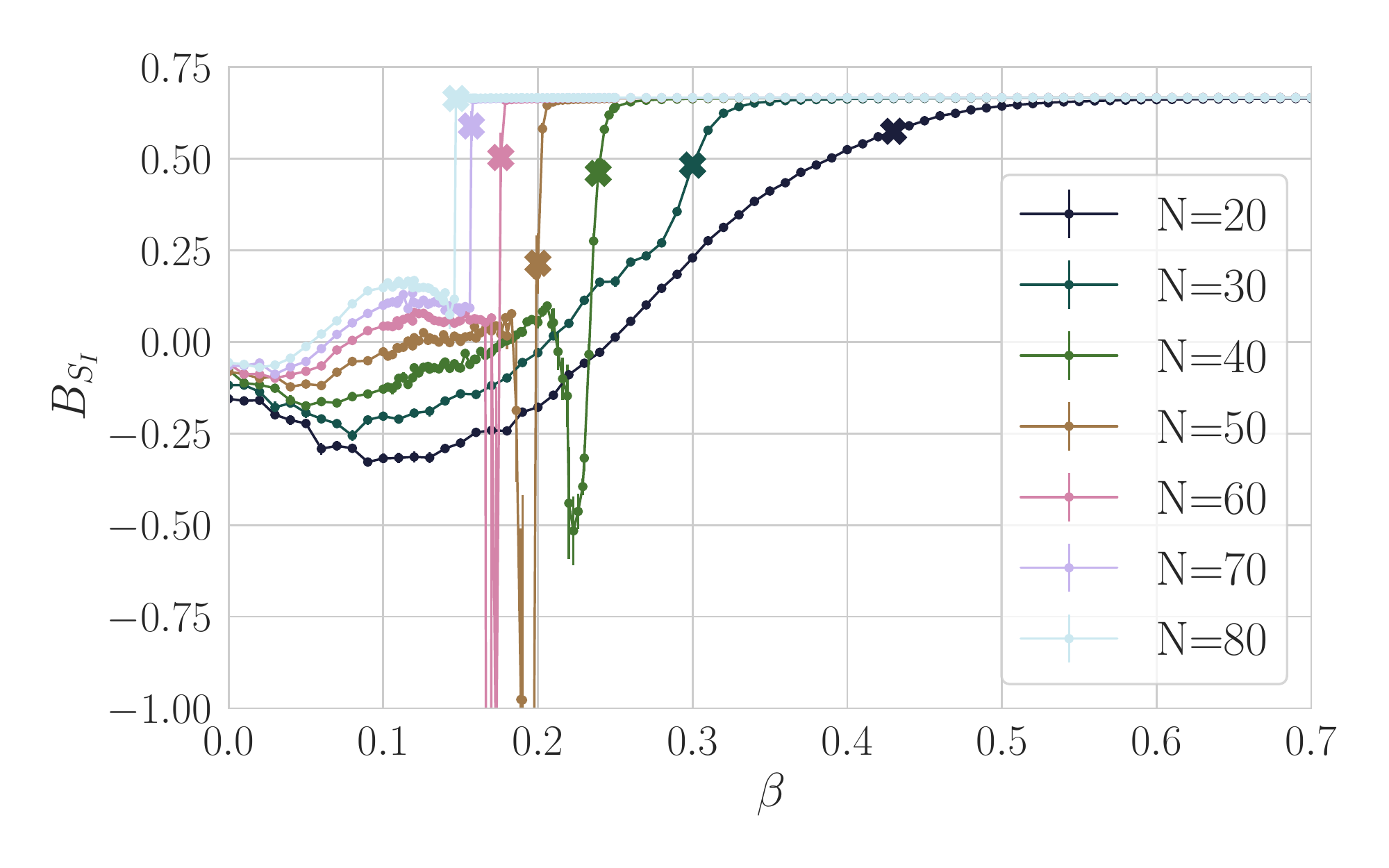}}
  \subfloat[$B_{S_{c}}$]{
  \includegraphics[width=0.5\textwidth]{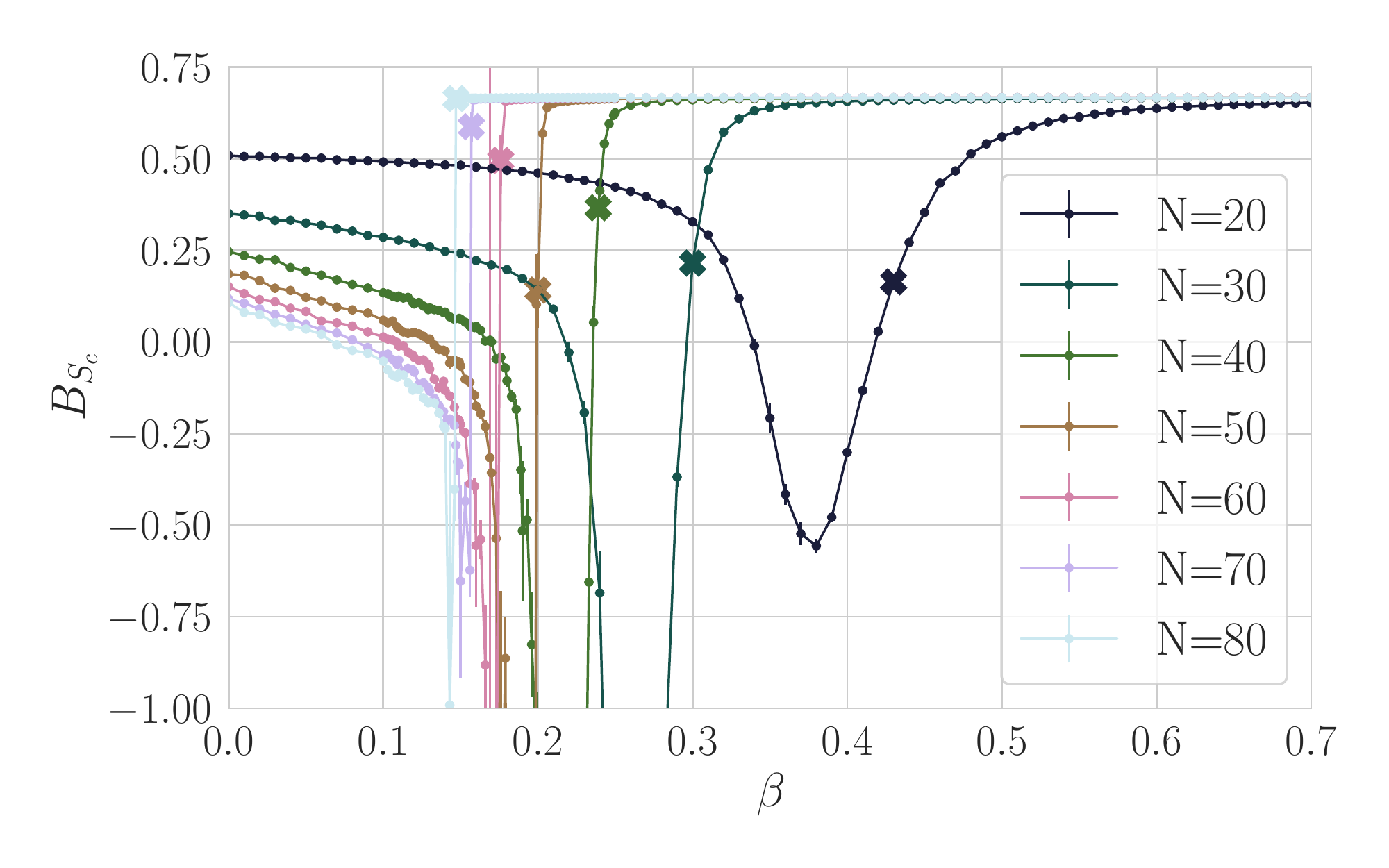}}

  \caption{The fourth order cumulant for the actions and the magnetisation at $j=-1$. The thicker crosses indicate the location of the phase transition, as determined from the peak of $\mathrm{Var}(S)$.}\label{fig:Binder_Coefficient}
\end{figure}

This hints that the phase transition at this point is dominated by the geometry, and that the magnetisation is somewhat following along.
As a supplementary tool we can examine the variance of the observables at the phase transition.
In Figure~\ref{fig:variances} we can see that the variance of the action has a strong peak that becomes higher and sharper as the system size increases.
The magnetisation also has a peak, however this does not become higher, only sharper.
In the pure Ising model the peak of the variance for the magnetisation rises strongly just like the variance for the action.
One possible explanation would be that the transitions in this system are of mixed order, with different phase transition behaviours for matter and geometry.
This hypothesis is also consistent with the plots of the fourth order cumulants, $B_{M},B_{S},B_{S_c}$ and $B_{S_I}$ as defined in equation \eqref{eq:binder} shown in Figure~\ref{fig:Binder_Coefficient}.
While the cumulant for the action shows a clear dip at the phase transition it does not show any such signal for the magnetisation.
The dip in the cumulant of the action is slightly offset from the location where the transition is according to the variance, but this is to be expected, since at finite size the phase transition will not show up exactly at the same spot for different quantities.

The behaviour of the cumulant for the three action variables and the magnetisation on both sides of the phase transition is easy to understand.
Around $\beta=0$ the distribution of both observables is a wide Gaussian, which increases in width as the system size increases.
In the limit of wide distributions\footnote{As can be easily checked analytically using a single Gaussian.} the Binder cumulant goes to $0$, which is what it does here.
For large $\beta$ the distributions become sharply peaked around the energetically optimal states.
In the infinite system size limit, as well as in the infinite $\beta$ limit, we would expect this peak to go towards a delta distribution, the limit of the cumulant for this distribution is $2/3$, as we find in that phase.

At the transition we find smooth change of $B_{|M|}$, similar to that of the magnetisation at the $2$nd order transition in the Blume-Capel model on a fixed lattice~\cite{Tsai_Salinas_1998}.

For the three action observables on the other hand we find a sharp dip down, which as discussed in~\cite{Tsai_Salinas_1998}, is a signal for a first order transition.
However there the cumulant for the action is $0$ away from the phase transition, which is not the case in our model.
It thus remains inconclusive what order this transition is, although there are hints that it might be of mixed order.

\subsection{Order of the two phase transitions at negative $\beta$}

After understanding the phase transition behaviour for $j=-1$ somewhat better we can now turn towards the $j=1$ line.
There are two transitions here, which, as discussed above, become visible in different variables.
The Ising action, as well as the overall action show signals for both phase transitions, while the causal set action only shows the geometric phase transition, and the magnetisation and the relation correlation only show the magnetic phase transition.

\begin{figure}
  \centering
  \includegraphics[width=0.9\textwidth]{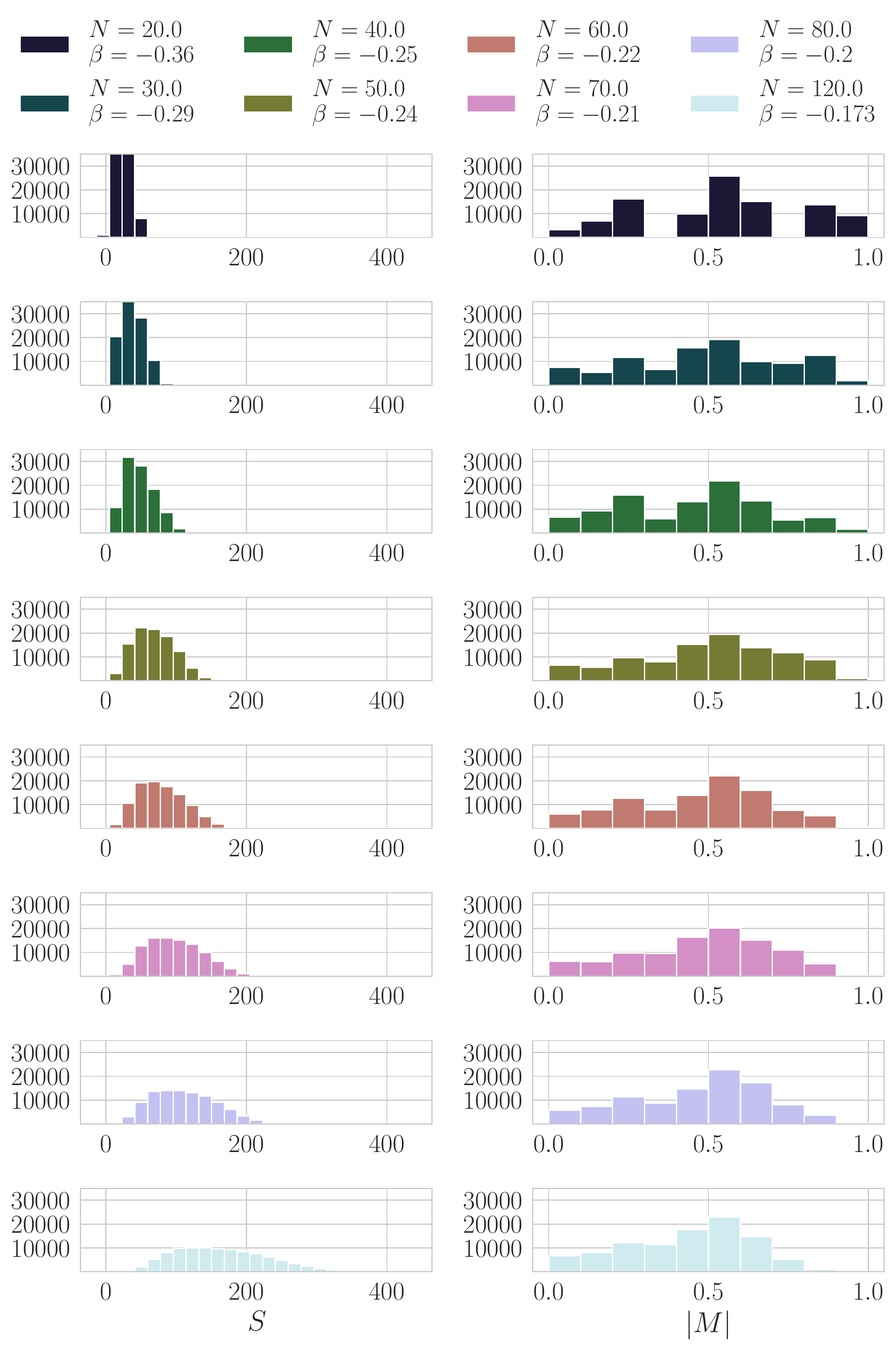}
  \caption{At the first, magnetic phase transition, the action does not change drastically. It is still concentrated around small positive values, although this distribution becomes wider for larger $N$. The magnetisation however is distributed almost flatly over the entire interval $[0,1]$. The height of the peaks is plotted on log-scale.}\label{fig:hist_PT1_j1}
\end{figure}

Figure~\ref{fig:hist_PT1_j1} shows the histograms for the action and the absolute magnetisation at the magnetic phase transition.
Neither of them shows a double peak structure, instead the action shows a wide peak, that increases in width as the system size increases, the value of the action is dominated by the Ising action.
The magnetisation shows values distributed almost flatly in the interval $[0,1]$.
This behaviour could be consistent with a $2$nd order phase transition, but definitely not with a first order transition.
To try and better understand it we can then look at the fourth order cumulant.

\begin{figure}
  \begin{minipage}{0.1\textwidth}
    (a) $B_{M}$
  \end{minipage}
  \begin{minipage}{0.9\textwidth}
  \includegraphics[width=\textwidth]{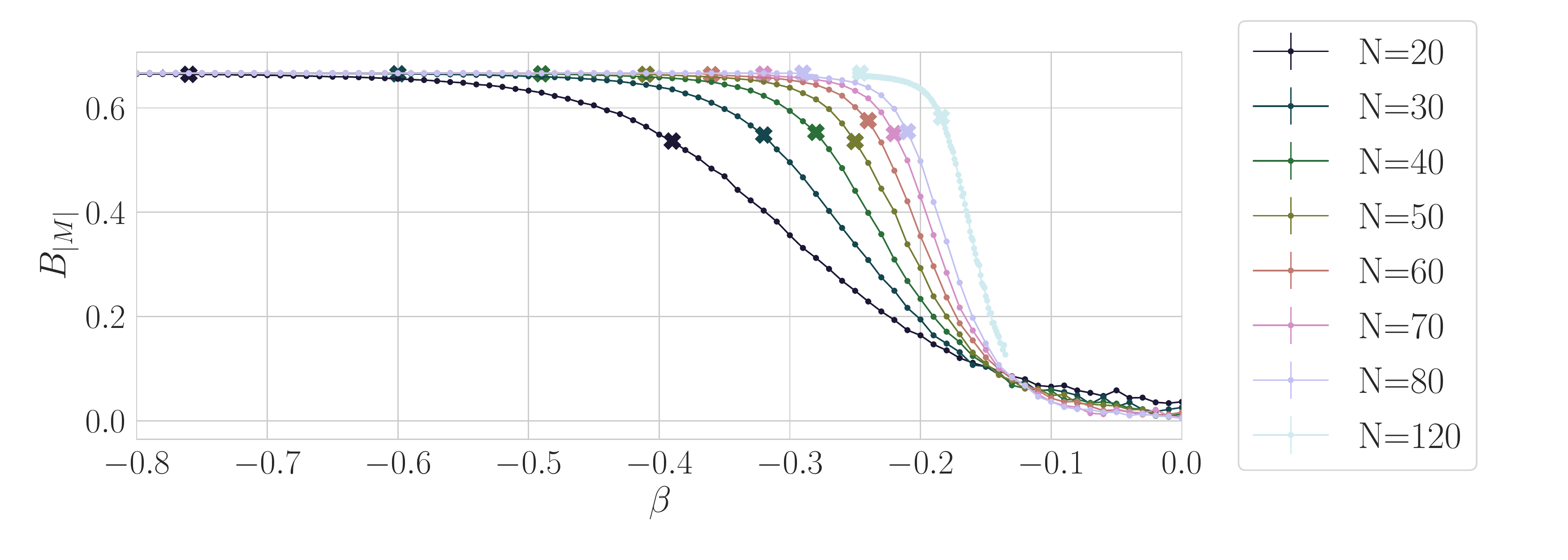}
  \end{minipage}
  \begin{minipage}{0.1\textwidth}
    (b) $B_{S}$
  \end{minipage}
  \begin{minipage}{0.9\textwidth}
    \includegraphics[width=\textwidth]{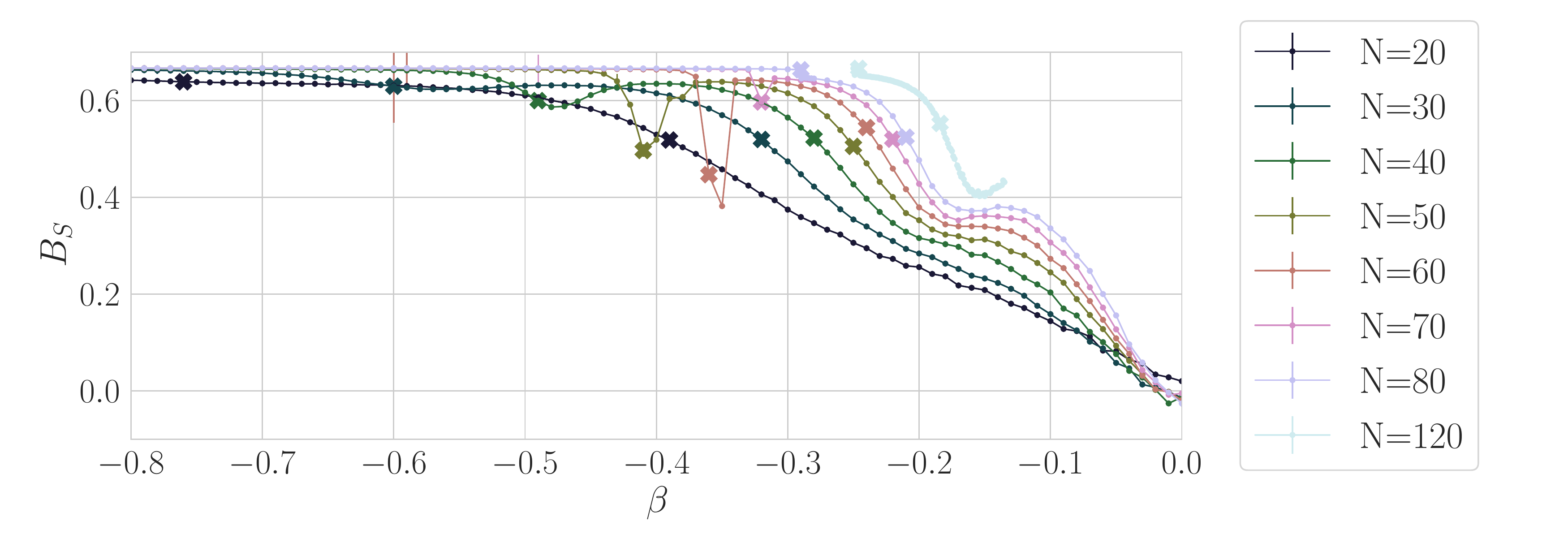}
  \end{minipage}
  \begin{minipage}{0.1\textwidth}
    (c) $B_{S_{I}}$
  \end{minipage}
  \begin{minipage}{0.9\textwidth}
    \includegraphics[width=\textwidth]{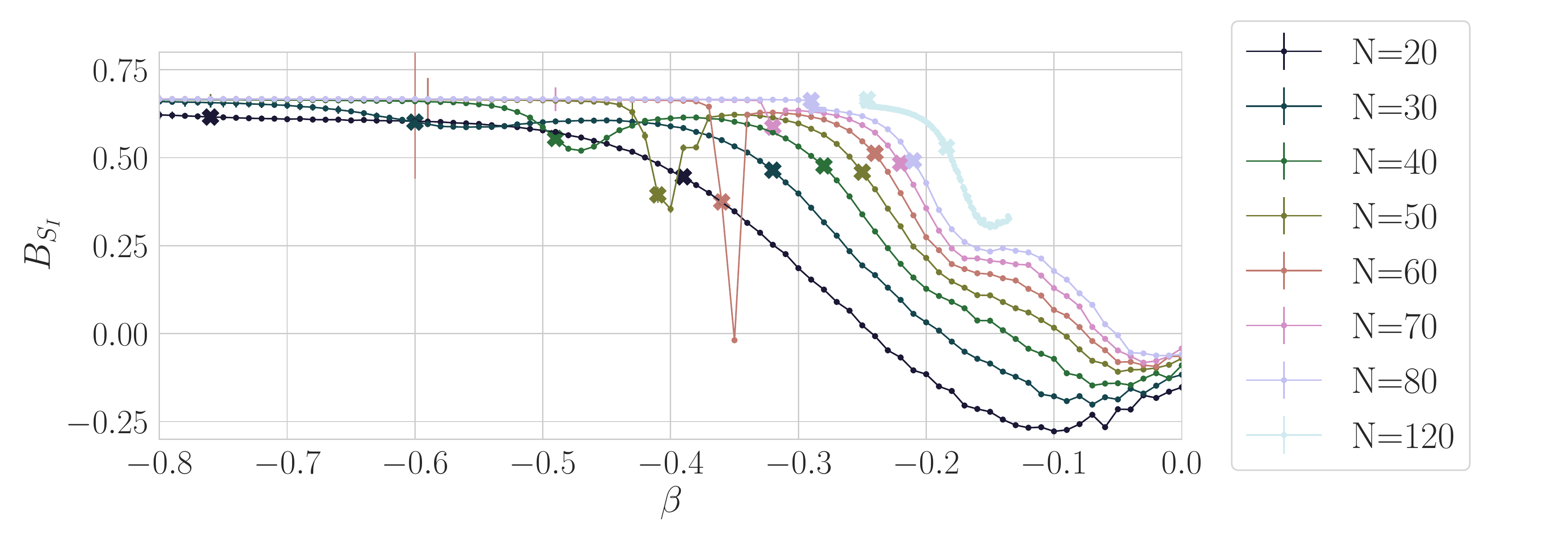}
  \end{minipage}
  \begin{minipage}{0.1\textwidth}
    (d) $B_{S_{c}}$
  \end{minipage}
  \begin{minipage}{0.9\textwidth}
  \includegraphics[width=\textwidth]{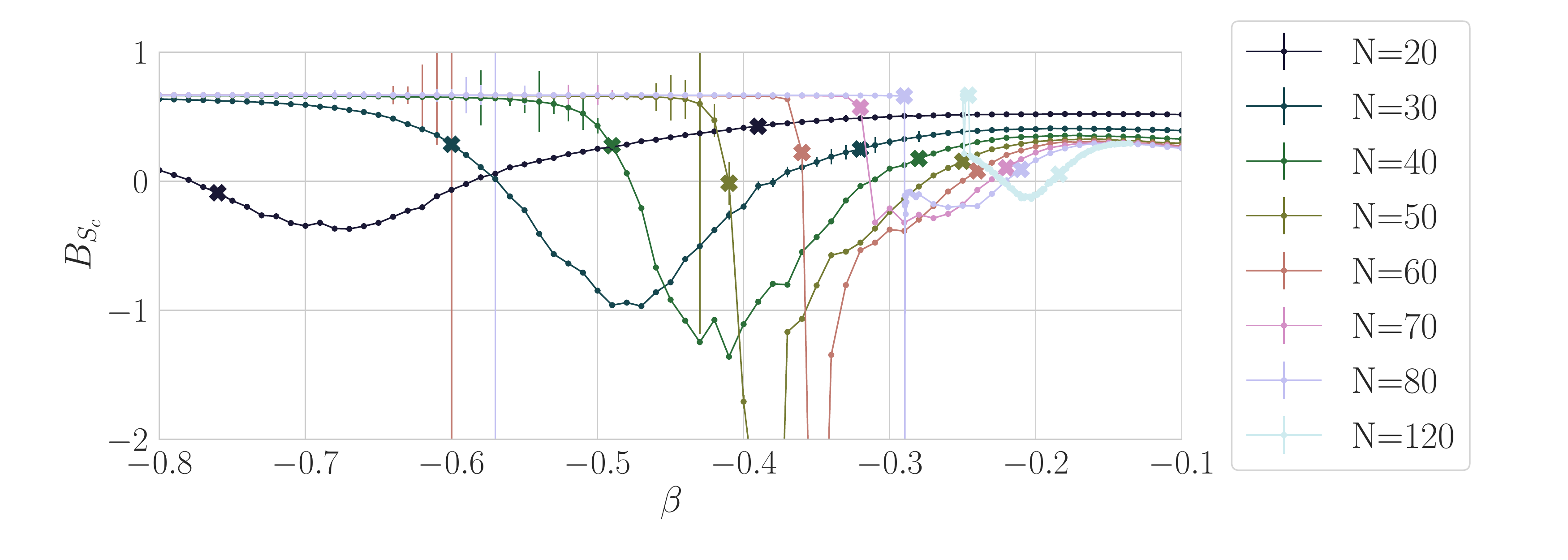}
  \end{minipage}

   \caption{Fourth order cumulant plotted against $\beta$ for both the action and the magnetisation for $j=1$. The thicker crosses indicate the two phase transition points. The rightmost cross of each colour is the magnetic transition,while the second cross indicates the geometric transition, both measured from the peaks of
   $\mathrm{Var}(S)$.}\label{fig:binder_j1_combine}
\end{figure}

 The fourth order cumulant of the magnetisation, shown in Figure \ref{fig:binder_j1_combine} (a), smoothly transitions from $0$ at $\beta=0$ to $2/3$ at very negative $\beta$, which as above could be a signal for a $2$nd order phase transition.
The sharpest ascent of the cumulant is roughly around the magnetic phase transition, as one might expect.
The cumulants for the overall and Ising action observables only show a weak dip at this transition, which becomes more pronounced as the system size increases, thus being best visible for the $N=120$ data.

\begin{figure}
  \centering
  \includegraphics[width=0.9\textwidth]{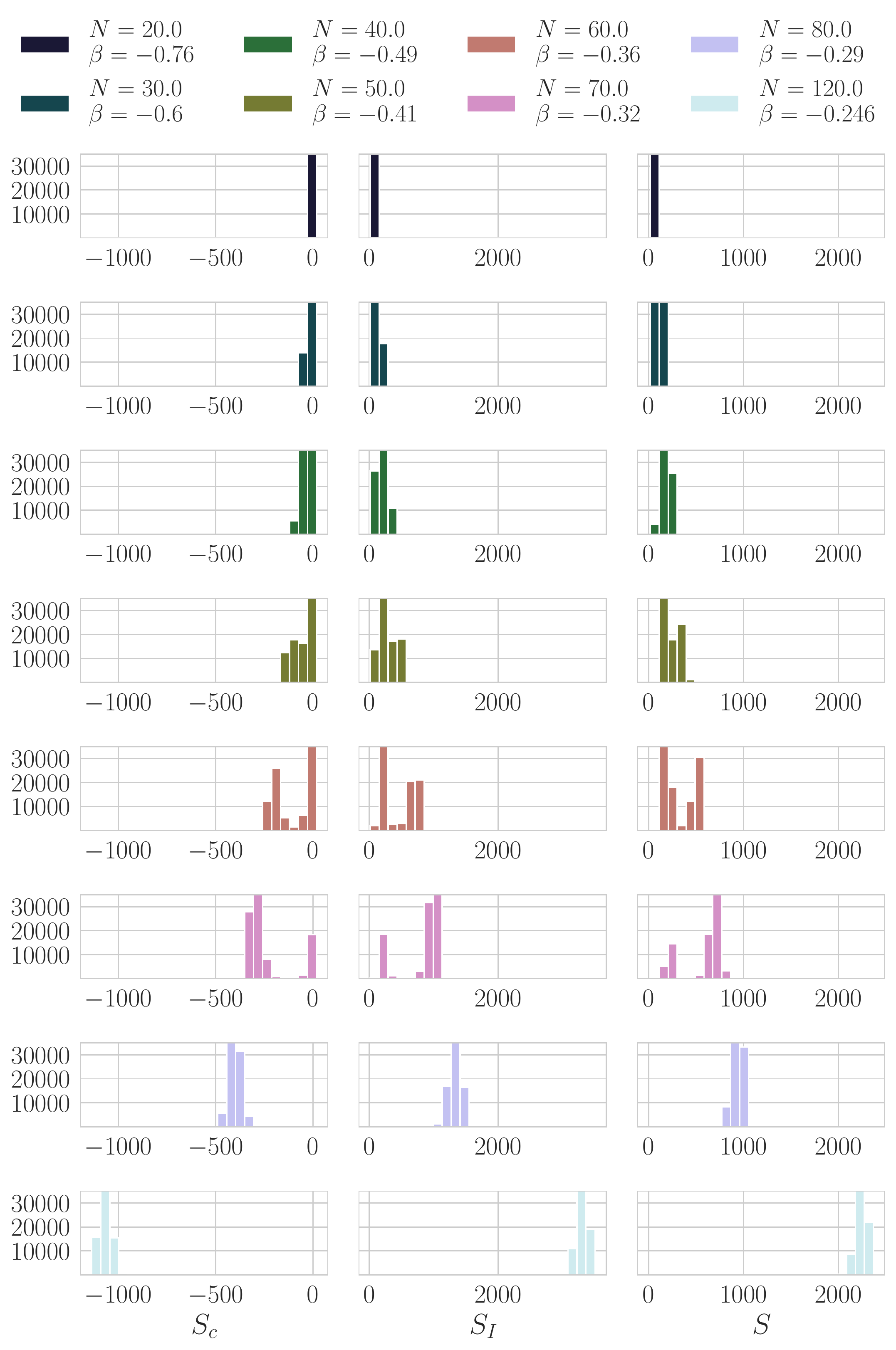}

\caption{The histograms above show how the overall action, and its two parts are distributed at the geometric phase transition. The height of the peaks is plotted on log-scale. }\label{fig:hist_geo_j1}
\end{figure}

For the second, geometric, phase transition we show the histograms in Figure~\ref{fig:hist_geo_j1}.
Since the magnetisation at this transition remains very close to $1$, we thus show histograms for the three action variables.
They show a double peak structure, with the two peaks first becoming distinguishable for $N=50$ and the distance between the peaks increasing until $N=70$.
For $N=80$ and $N=120$ the simulations only have probed one peak, however, as explained before, this is almost certainly a limitation of our simulations and not a feature of the system.
Comparing the magnitudes of the actions it is clear that the Ising action increases more than the causal set action decreases when the crystalline state is taken on, hence it creates the phase transition.
At large negative $\beta$ the action takes on large negative values, which means these states should be energetically disfavoured, however the Ising action takes on positive values of more than twice this value, so the Ising action forces the causal set into a crystalline, link maximizing, state.

At this second phase transition the fourth order cumulants for the three action observables, shown in Figure~\ref{fig:binder_j1_combine} show a small dip for $N=20,30,40,50$, which becomes deeper as $N$ increases.
Since the phase transition becomes sharper as the system size increases, this dip also becomes sharper, and hence harder to resolve.
In finite size systems phase transitions can be slightly shifted between different observables, this is the most likely explanation as to why the dips for $N=60,70,80$ are not visible in $B_{S}$ but visible in $B_{S_c}$ (through a single point in the case of $N=80$).
For $N=120$ we do not have enough resolution to show the dip in any of the observables, however we still think it is justified to assume that this is an effect of the difficulty in probing the system in sufficient detail at this size.

This leads us to hypothesize that the phase transition for the magnetic system is still continuous, while that of the geometry remains of first order.

\section{Scaling of the observables and critical exponents}\label{sec:scaling}
After we determined how the location of the phase transition scales with the systems size $N$ in section~\ref{sec:PTloc}, and examined the order of the phase transition in section~\ref{sec:PTorder} we shall now determine the scaling of the observables with the system size.
This scaling will, in general, be phase dependent.

In~\cite{Glaser:2017sbe} for the pure $2$d orders the scaling for the combination $\beta S$ plotted against $N \beta$ was explored, since this scaling  almost perfectly collapses the phase transitions for different system sizes.
However for the current project it is more convenient to define a reduced $\beta$ as  $\beta_{red}=\frac{\beta-\beta_c}{\beta_c}$, and to plot the observables against this.

\subsection{Negative $j$}
First we will look at the region of positive $\beta$ and negative $j$.
In this region we expect the system to behave similar to the pure $2$d orders with possibly some modifications due to the Ising spins.
The first scalings to look at, are the BD-action $S_c$, the Ising action $S_I$ and the combined action $S=S_c+ S_I$.
A priori nothing requires that the Ising action and the BD action scale in a similar manner with $N$, hence the combined action could show a complicated scaling.
Since the phase transitions in the Ising model and the causal set happen at the same critical temperature for large enough $N$, we plot the action against $\beta_{red}$ using the value $\beta_c$ averaged over all observables.

We expect to find different scaling of the observables in the two phases, hence we split our data at the phase transition and examine the two regions independently.
We know that the expectation value for the BD action at $\beta=0$ is $\langle S_{c} \rangle|_{\beta=0}=4$, and have seen in~\cite{Glaser:2017sbe} that the scaling in the phase connected to this becomes much clearer if one subtracts this offset, the overall action $S$ is a sum so the same offset applies there.

\begin{figure}
\subfloat[$S$ no scaling]{\includegraphics[width=0.5\textwidth]{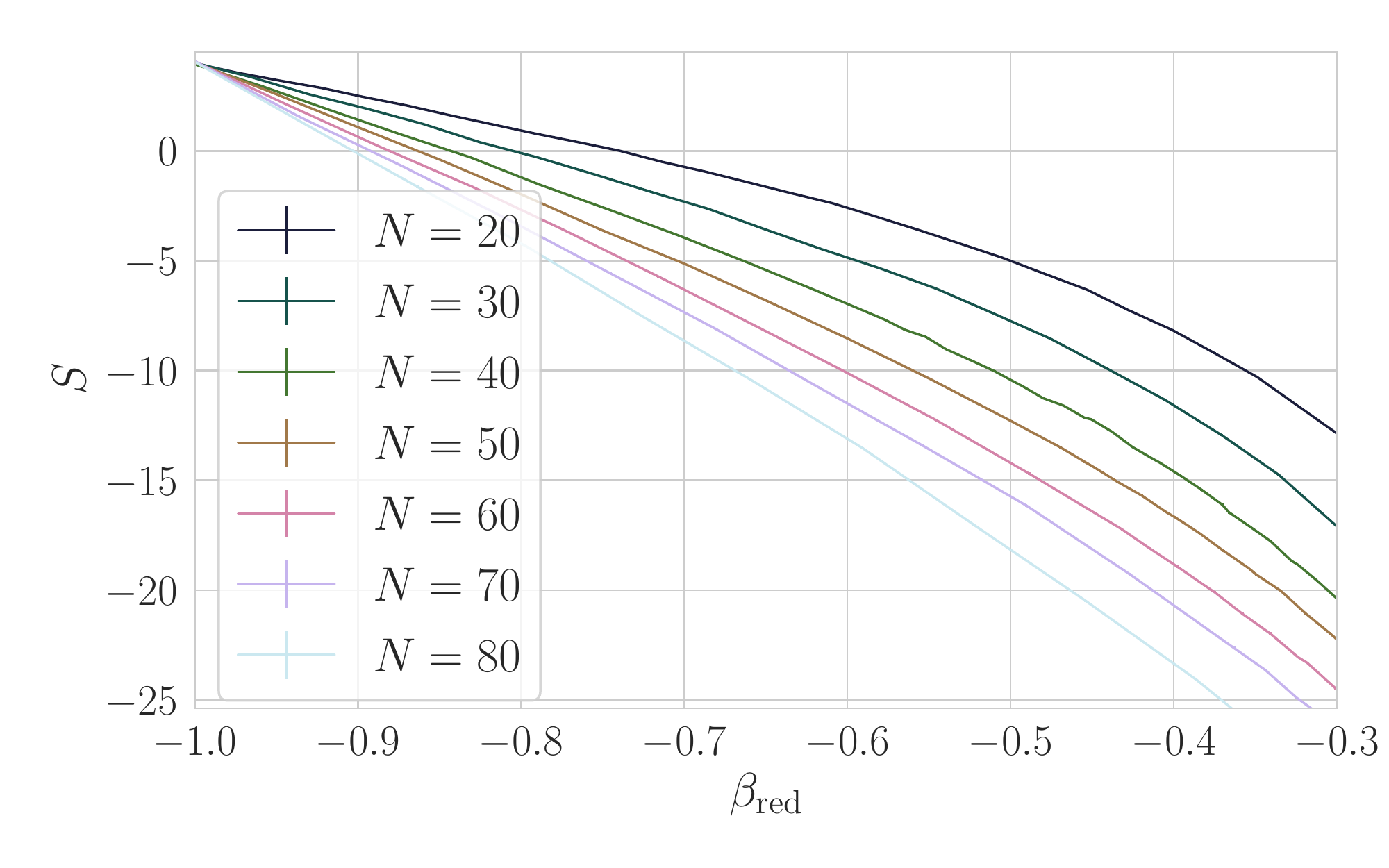}}
\subfloat[$S/N$  rescaled]{\includegraphics[width=0.5\textwidth]{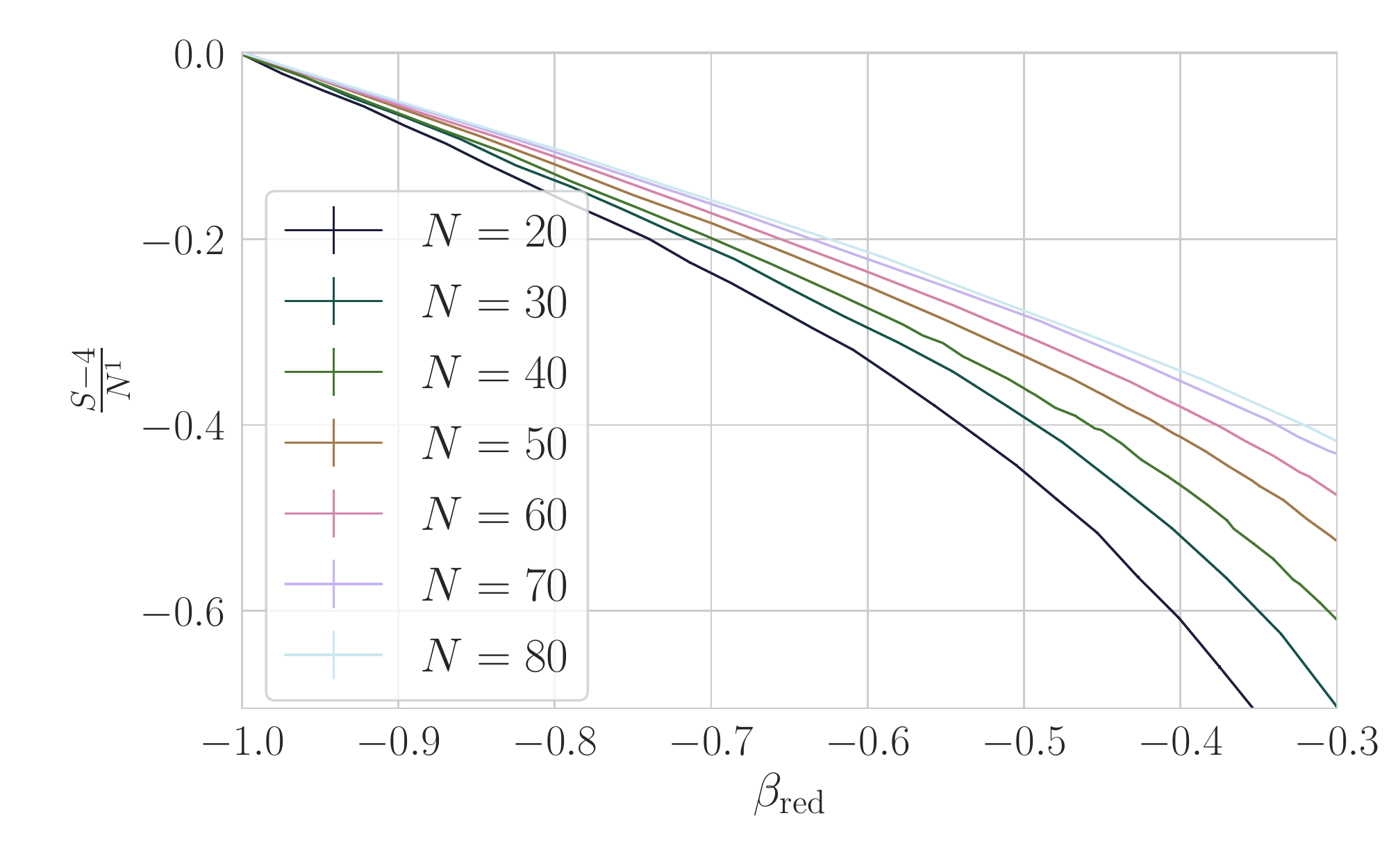}}
\caption{The action in the pre phase-transition region without scaling on the left and rescaled with $1/N$ on the right.\label{fig:scalingActions}}
\end{figure}

Examining $S_c$ and $S_I$ we find that the data collapses when rescaled by a factor of $1/N$, hence in this phase both scale linearly with $N$, and so does $S$, as shown in Figure~\ref{fig:scalingActions}.
This result agrees with those found for the pure $2$d orders in~\cite{Glaser:2017sbe}.
We have also confirmed that the same scaling is valid for $j=-0.5$ which is an indication that this scaling is stable for the entire $j<0,\beta>0$ region.

\begin{figure}
\subfloat[$S$ no scaling]{\includegraphics[width=0.5\textwidth]{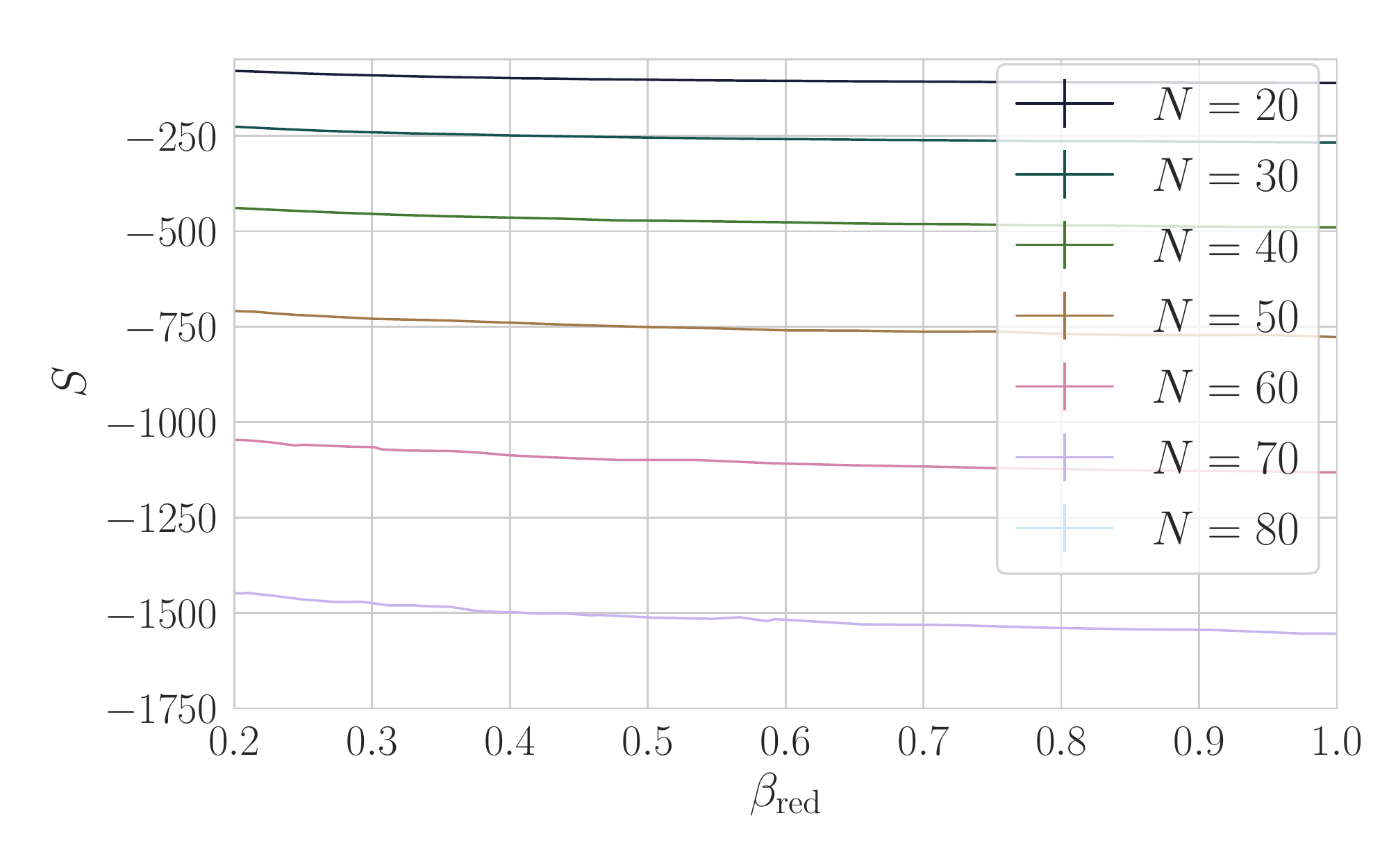}}
\subfloat[$S/N$ rescaled]{\includegraphics[width=0.5\textwidth]{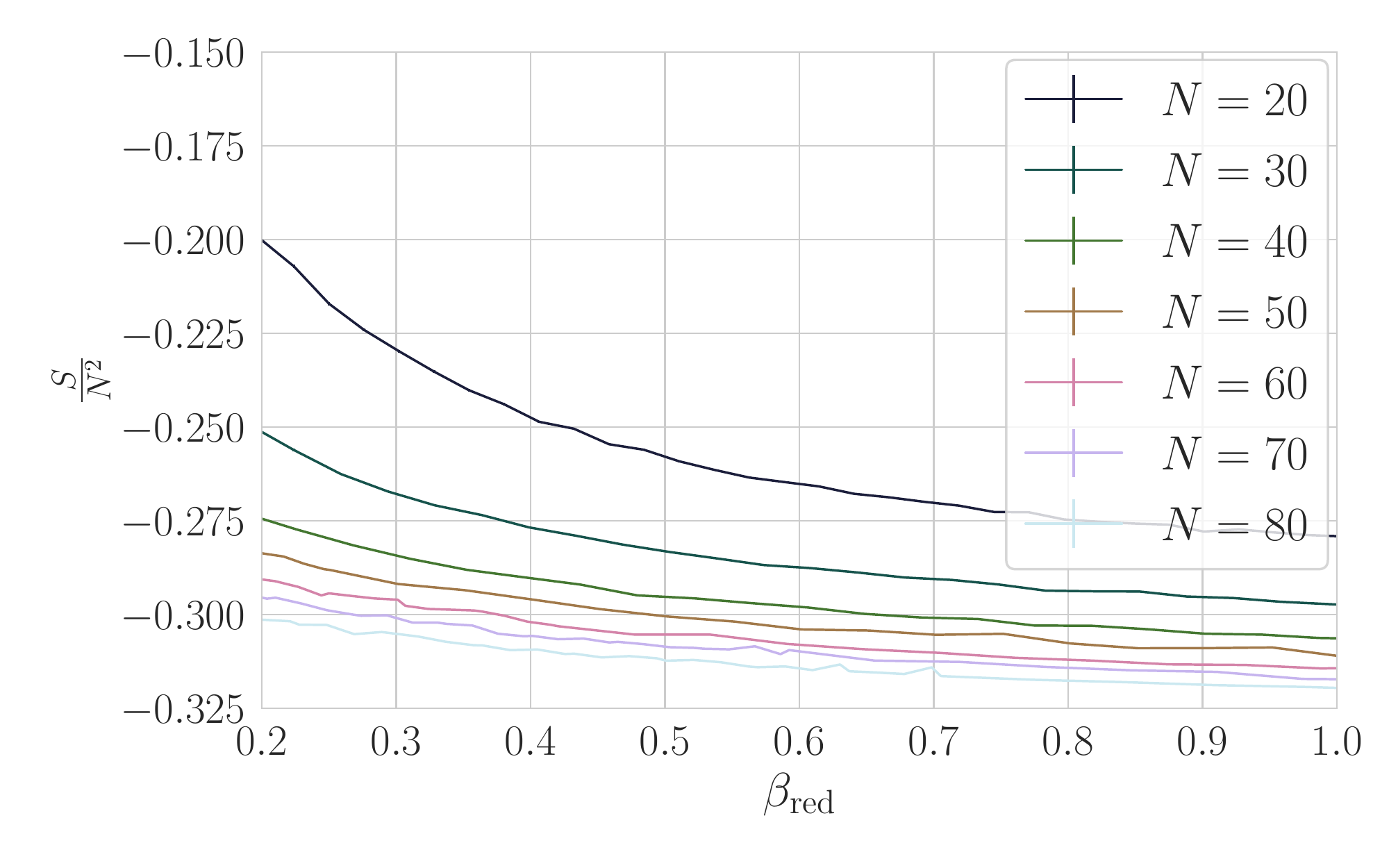}}
\caption{Scaling of the action in the post phase transition region rescaled with $N^2$.\label{fig:scalingActionsPostPT}}
\end{figure}

In the region past the phase transition, both the Ising action and the Benincasa-Dowker action aim to maximize the number of links in the causal set.
Since crystalline orders maximize the number of links, which grows like $N^2$, we expect the actions to be proportional to $N^2$ in this regime, which we can confirm in Figure~\ref{fig:scalingActionsPostPT}.

\begin{figure}
  \centering
\subfloat[at low $\beta$]{\includegraphics[width=0.5\textwidth]{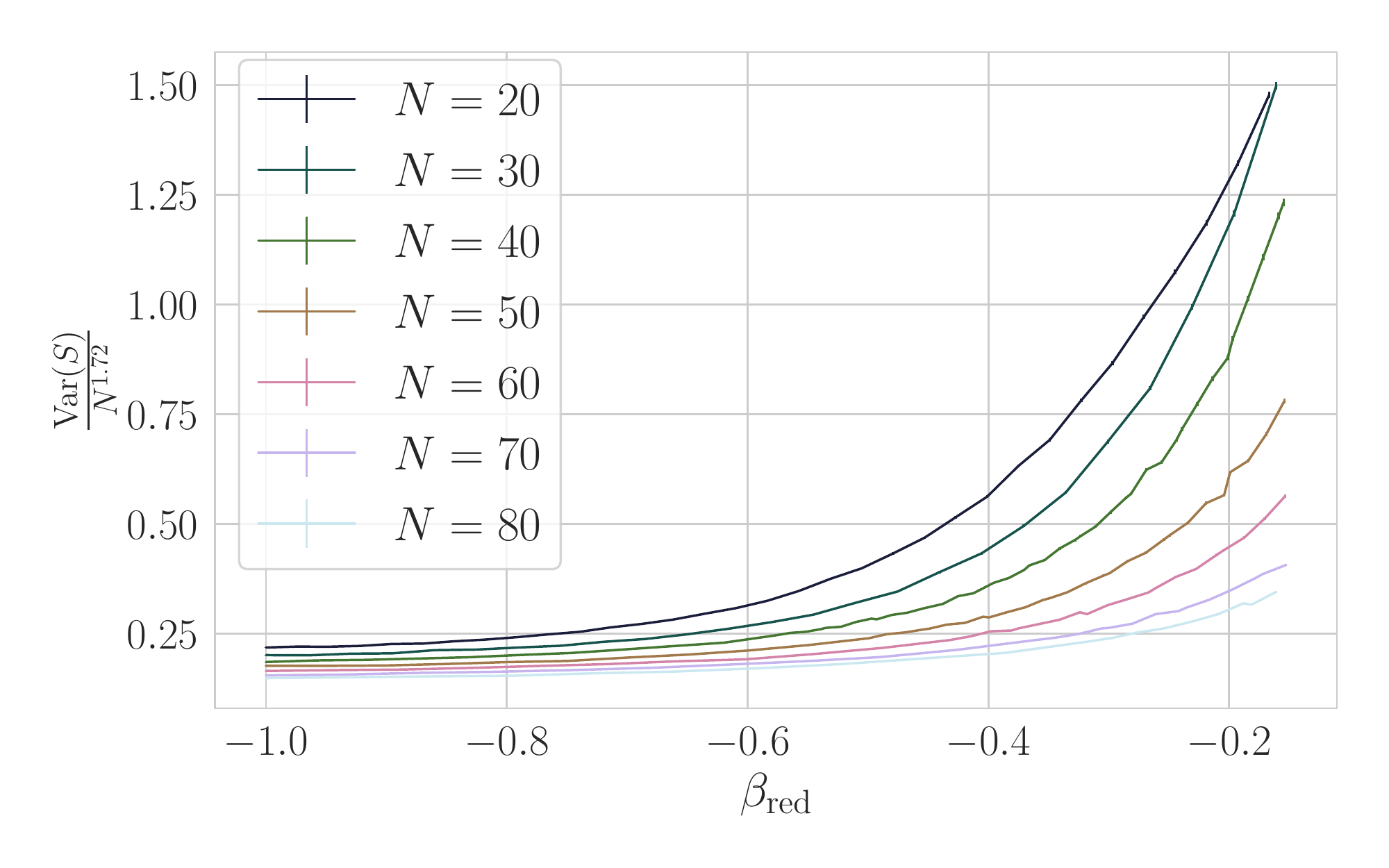}}
\subfloat[at high $\beta$]{\includegraphics[width=0.5\textwidth]{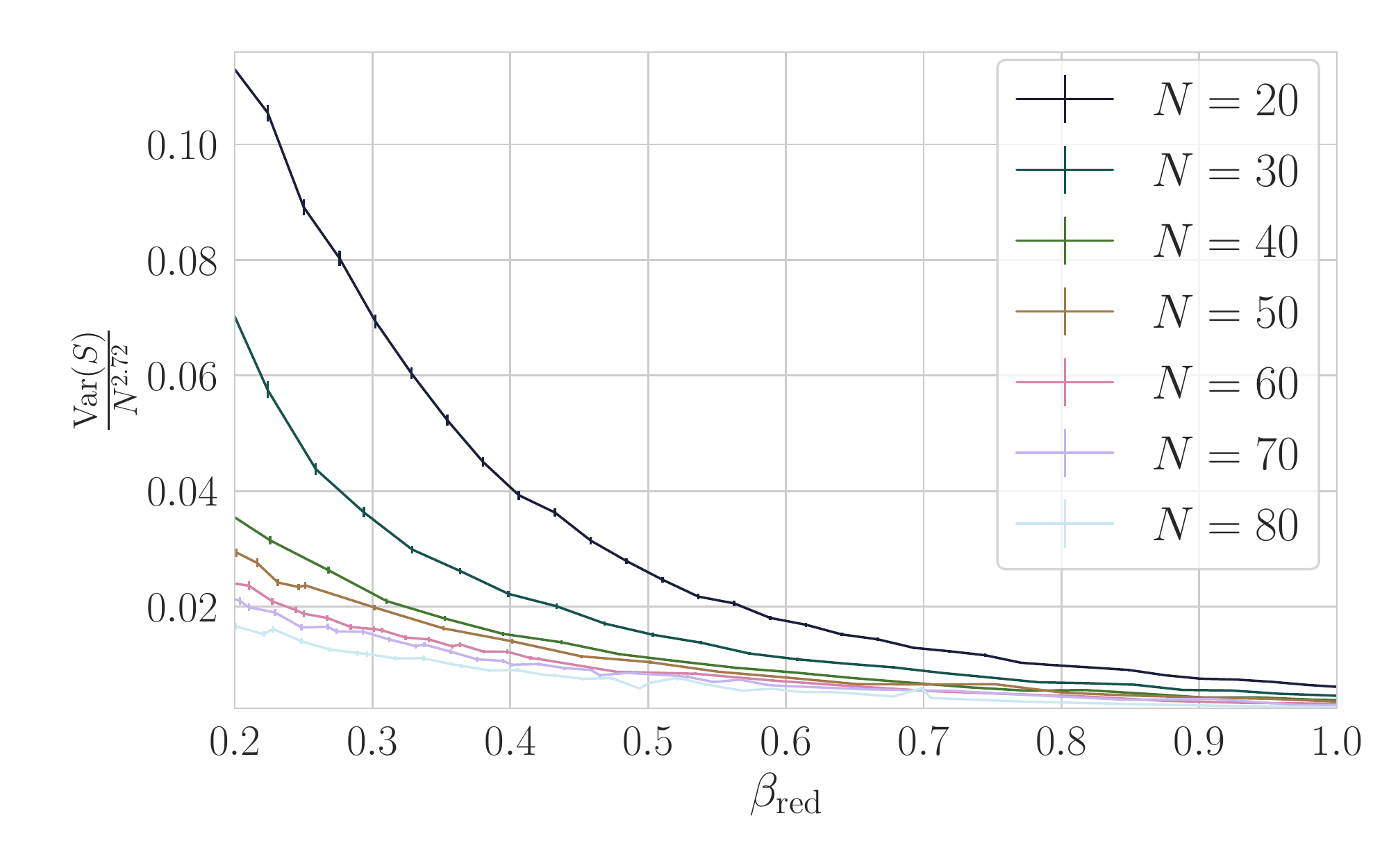}}
  \caption{$\mathrm{Var}(S)$ rescaled with $-\nu+\lambda$, in the left hand region, at low $\beta$ this is $-1.72$ and in the right hand plot, at high $\beta$ it is $-2.72$.}\label{fig:variancejm}
\end{figure}

We have thus determined that the critical exponent $\nu$ is $1$ in the random phase and $2$ in the crystalline, correlated phase.
As explained in the introduction we would expect the variance of $S$ to scale like $\nu-\lambda$.
We can see that this works in Figure~\ref{fig:variancejm} where we have rescaled the actions with $-\nu+\lambda$ to achieve a collapse of the data\footnote{We have also checked that plots of $\beta \av{S}$ and $\beta^2 \mathrm{Var}(S)$, as used in~\cite{Glaser:2017sbe} are consistent for this phase transition, however we decided against using them, since they do not work well for the two phase transitions at negative $\beta$.}

\begin{figure}
  \subfloat[entire range]{\includegraphics[width=\textwidth]{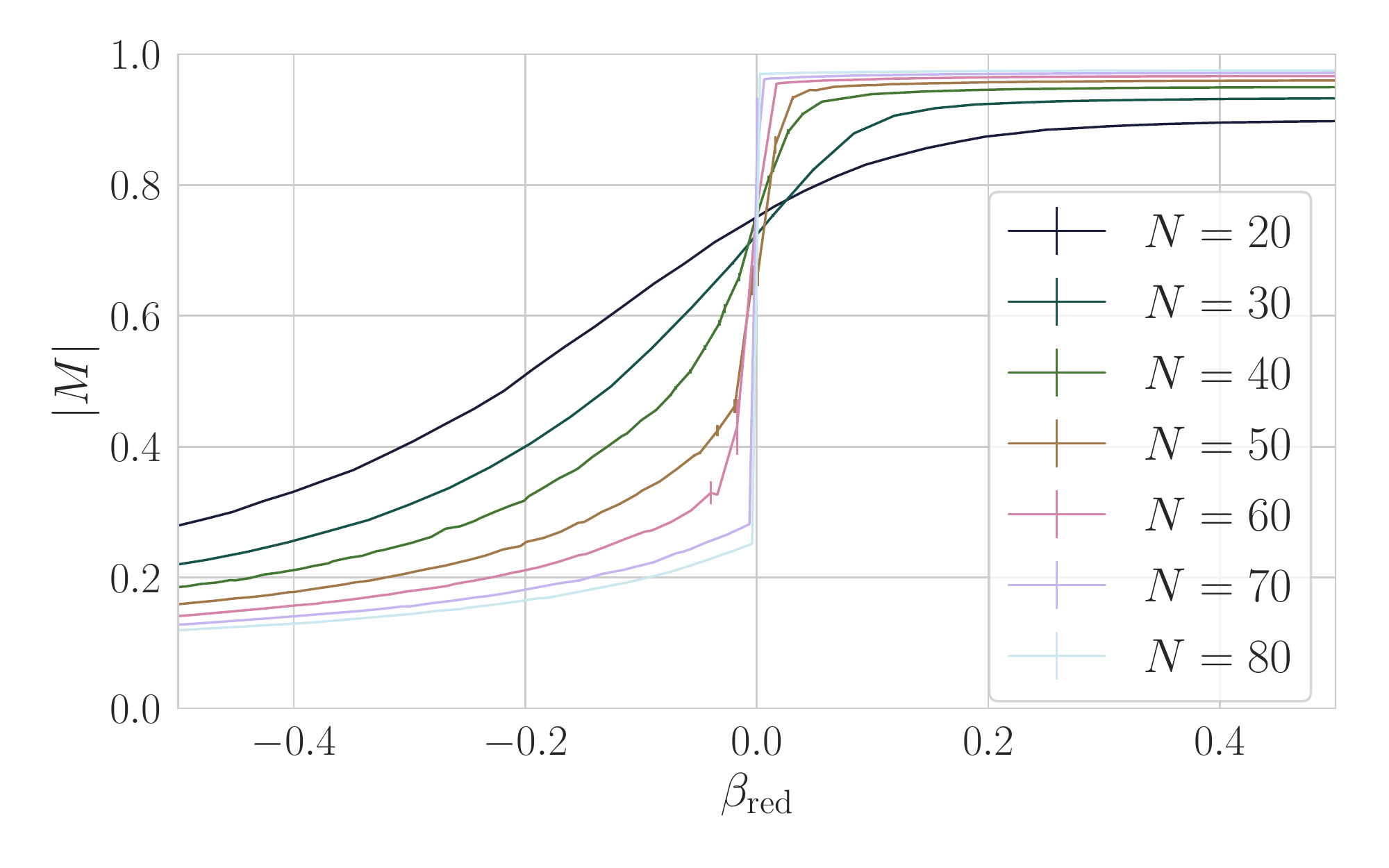}}

  \subfloat[region before phase transitions]{\includegraphics[width=0.5\textwidth]{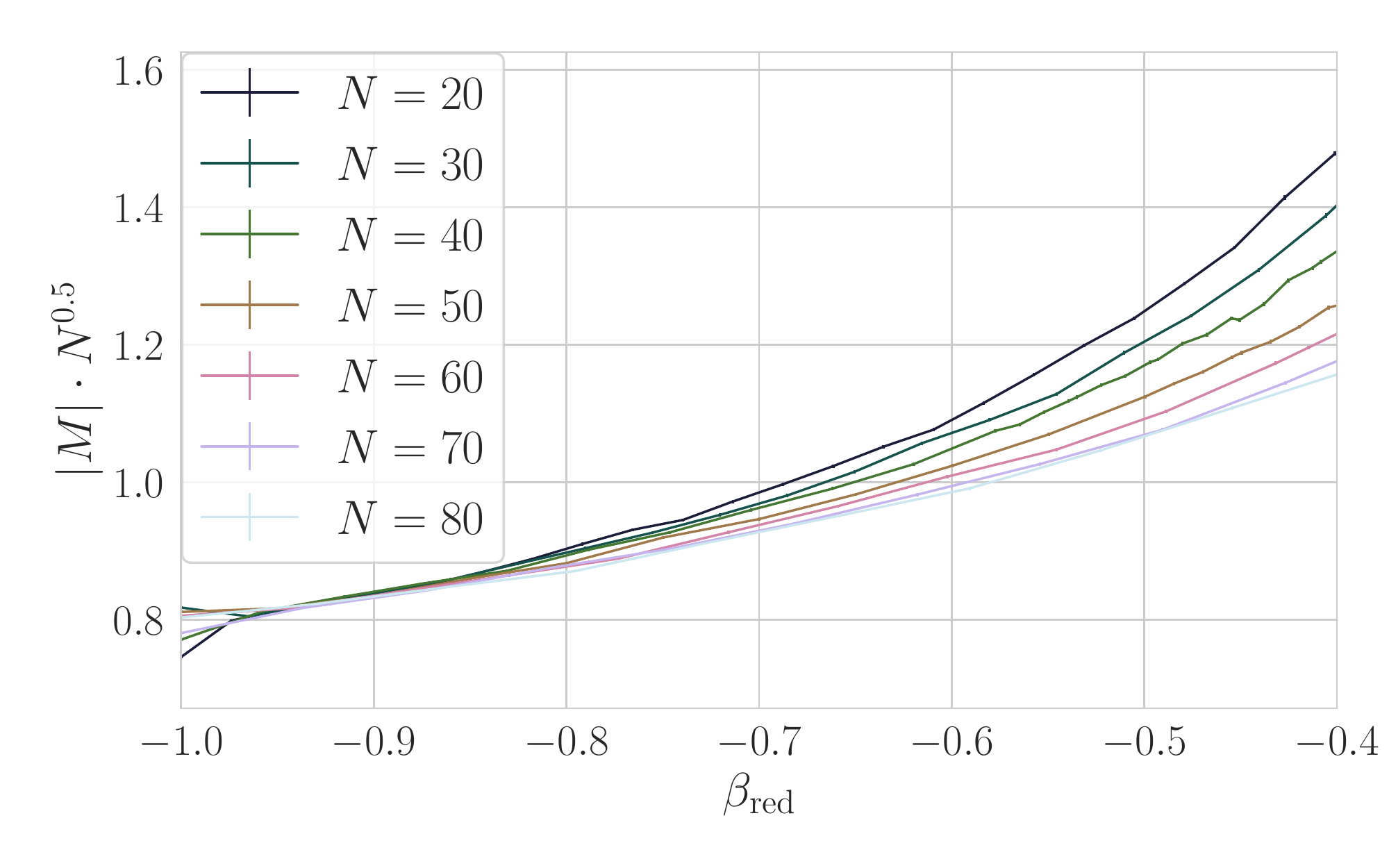}}
  \subfloat[region post phase transitions]{\includegraphics[width=0.5\textwidth]{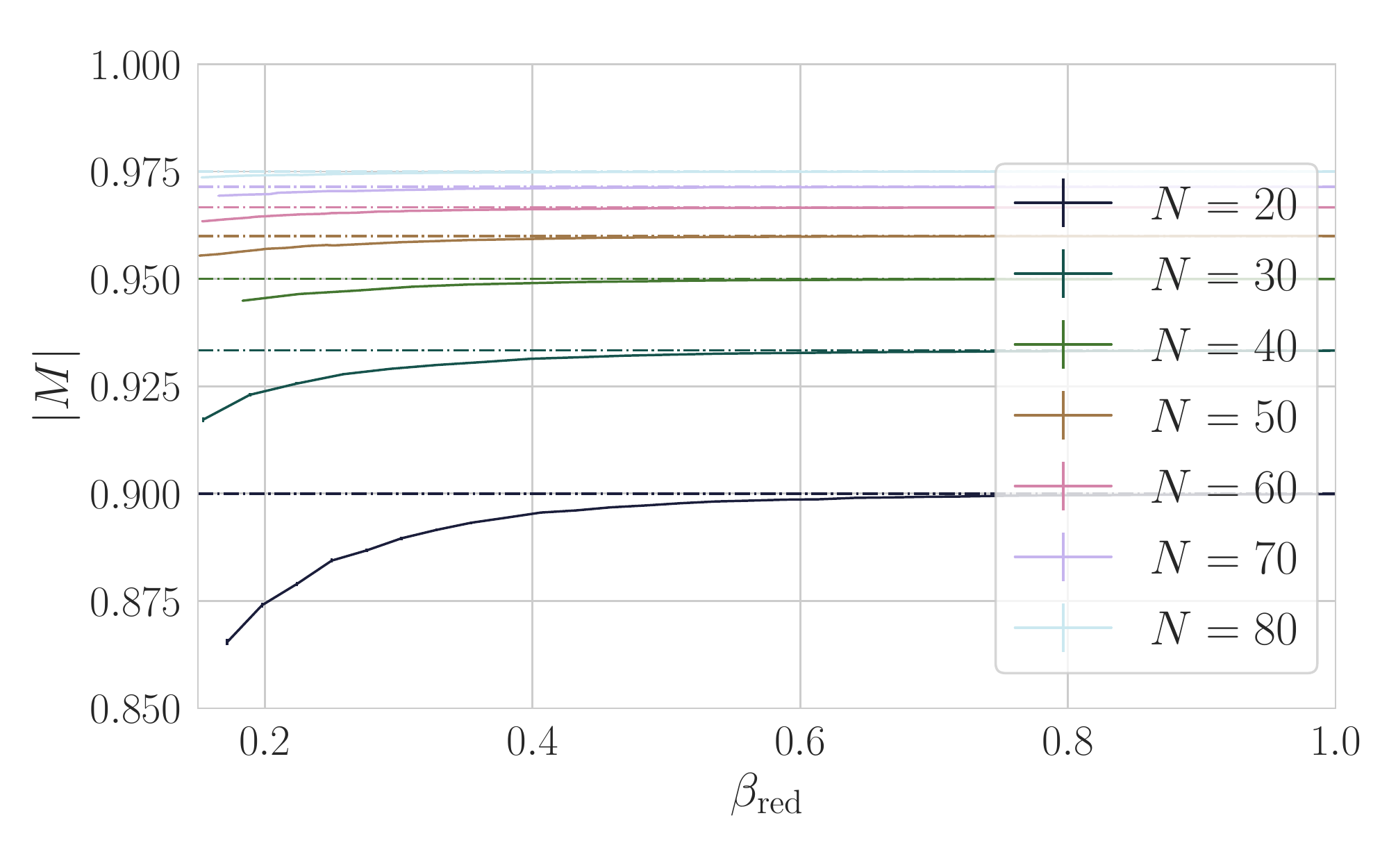}}
  \caption{\label{fig:aMjm1} The absolute magnetisation does not scale with the system size, other than some finite-size effects.}
\end{figure}

The absolute magnetisation does not scale with $N$, as shown in Figure~\ref{fig:aMjm1}~(a).
The value of $|M|$ before the phase transition is close to $0$, the residual value is a finite size effect and scales like $N^{-1/2}$.
The maximal value for the average of $|M|$ in the correlated state of the Ising model is $1-2/N$, as we see in Figure~\ref{fig:aMjm1}~(b), where the dash-dotted lines are the values of $1-2/N$.
This is likely an effect of the residual fluctuations of the spins, which would disappear for much larger $\beta$, assuming the simulations underwent some form of tempering to avoid a freeze out of single states.

\begin{figure}
\centering
  \subfloat[entire range]{\includegraphics[width=0.8\textwidth]{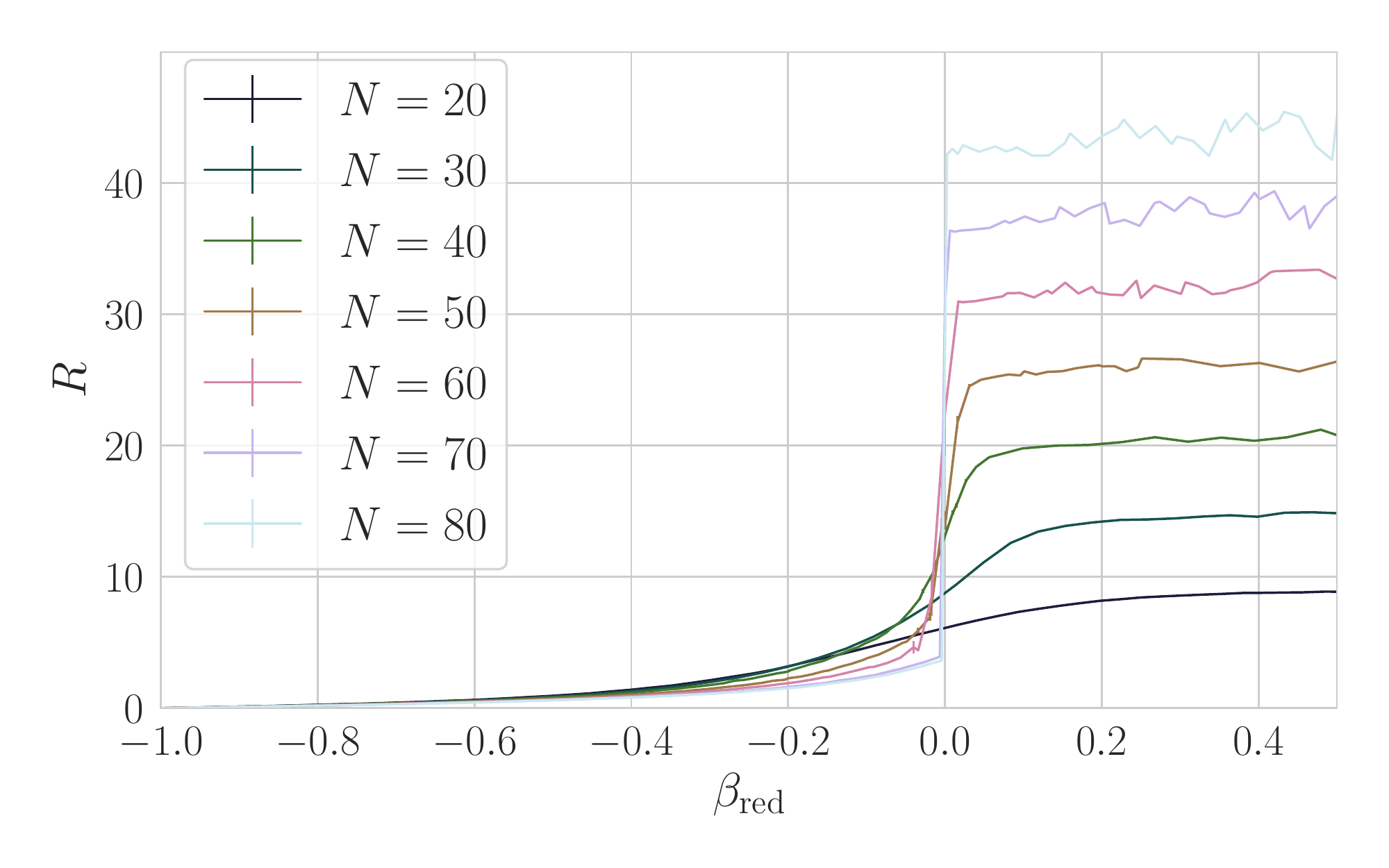}}

  \subfloat[pre phase transition]{\includegraphics[width=0.5\textwidth]{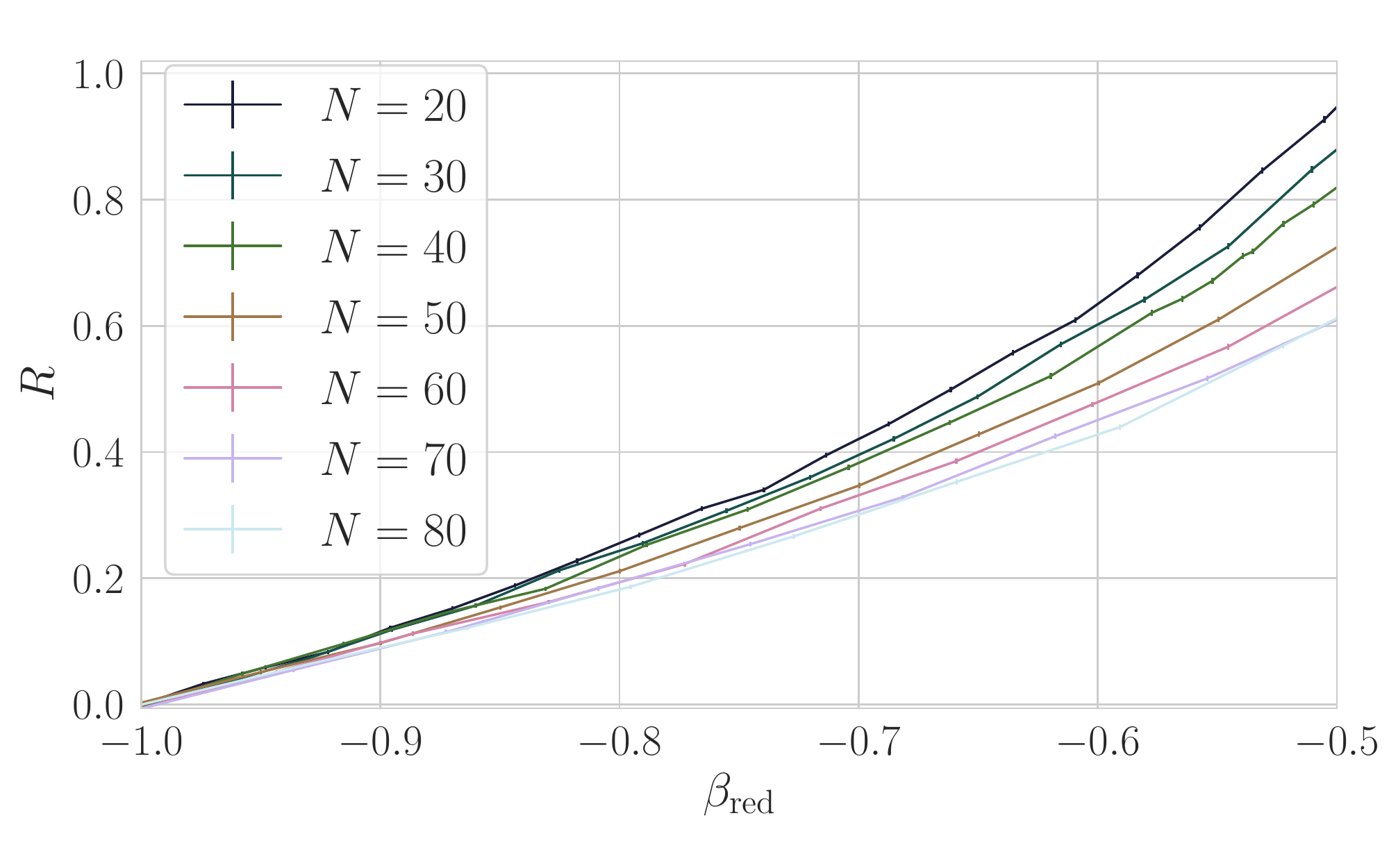}}
\subfloat[post phase transition]{\includegraphics[width=0.5\textwidth]{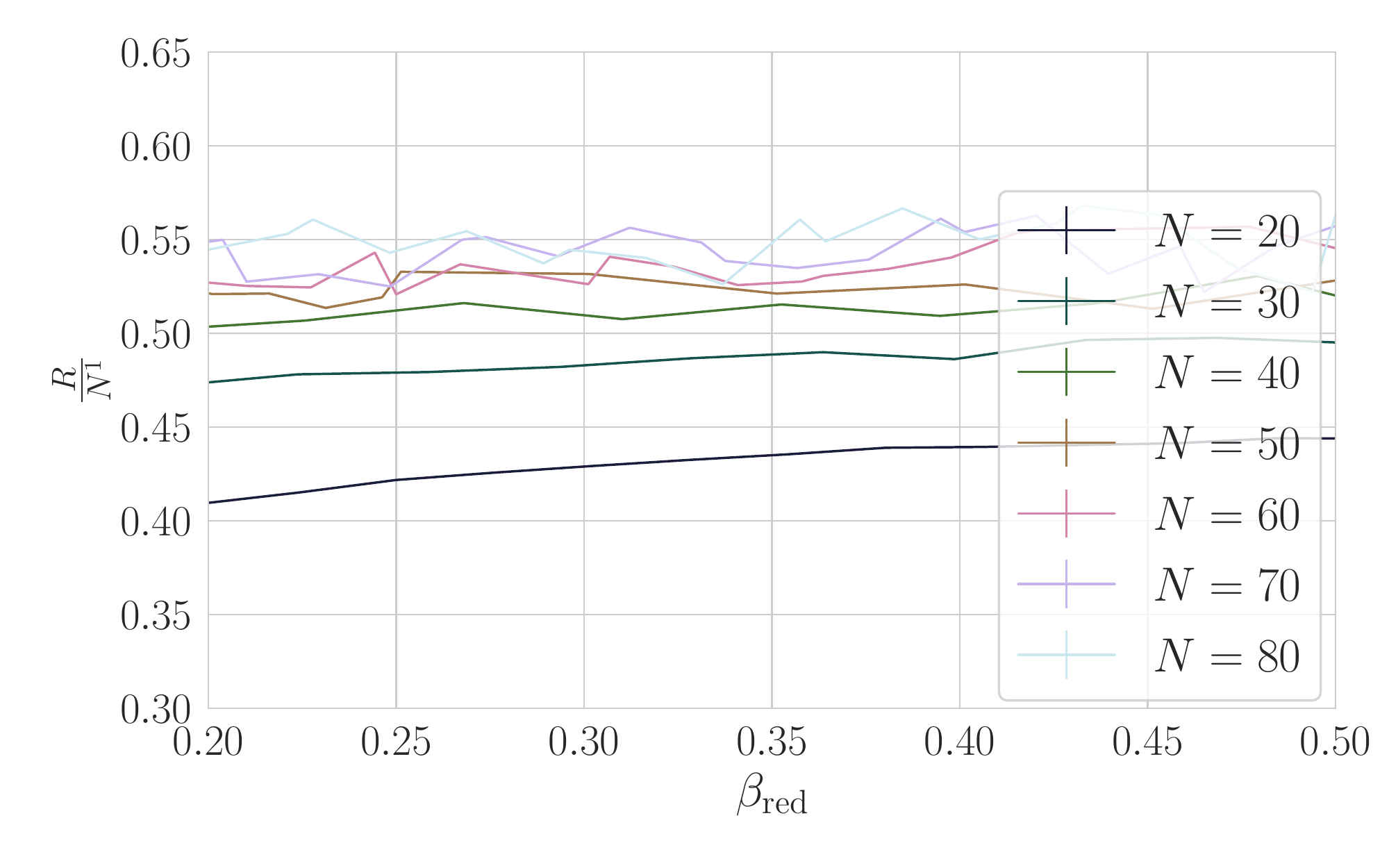}}
\caption{\label{fig:Rscaling_jm1} The relation correlation does not scale with $N$ in the pre phase transition region, while it scales like $N$ in the past phase-transition region.}
\end{figure}

The relation correlation $R$ does not scale with $N$ in the pre phase transition regime, see Figure~\ref{fig:Rscaling_jm1}, although close to the phase transition we start seeing effects that announce the approaching transition.
In the crystalline correlated phase $R$ scales like $N$.
This scaling makes sense considering that the maximal number of relations goes like $N^2$, the observable $R$ is defined with a normalisation of $1/N$, so in the regime where the number of relations is maximized, and all spins are aligned the expected scaling is $R \sim N$.

\subsection{Positive $j$ negative $\beta$}

Understanding the scaling of the system at positive $\beta$ works well, and the results are comparable to old results, as the system only undergoes a single phase transition that behaves similar to the purely geometric transition in~\cite{Glaser:2017sbe}.
However the system behaviour at negative $\beta$ is harder to understand, there are now two phase transitions, and the state of the geometry is changed through the matter interaction.
To explore the scaling of the action we divide the data into three different regions, \I which is the completely random phase, \II which is the phase of random causal sets with correlated Ising spins between the two phase transitions and \III which is  the crystalline correlated phase at large, negative $\beta$.

To explore the scaling we want to work with the reduced $\beta$ again, for \I this is $\beta$ reduced with the first phase transition point, we call this $\brOne$, and for \III we use the second phase transition location calling this $\brTwo$.
For \II there is no unique choice, and we have not been able to determine a clear scaling behaviour for all observables.

\begin{figure}
\subfloat[$S$ no scaling]{\includegraphics[width=0.5\textwidth]{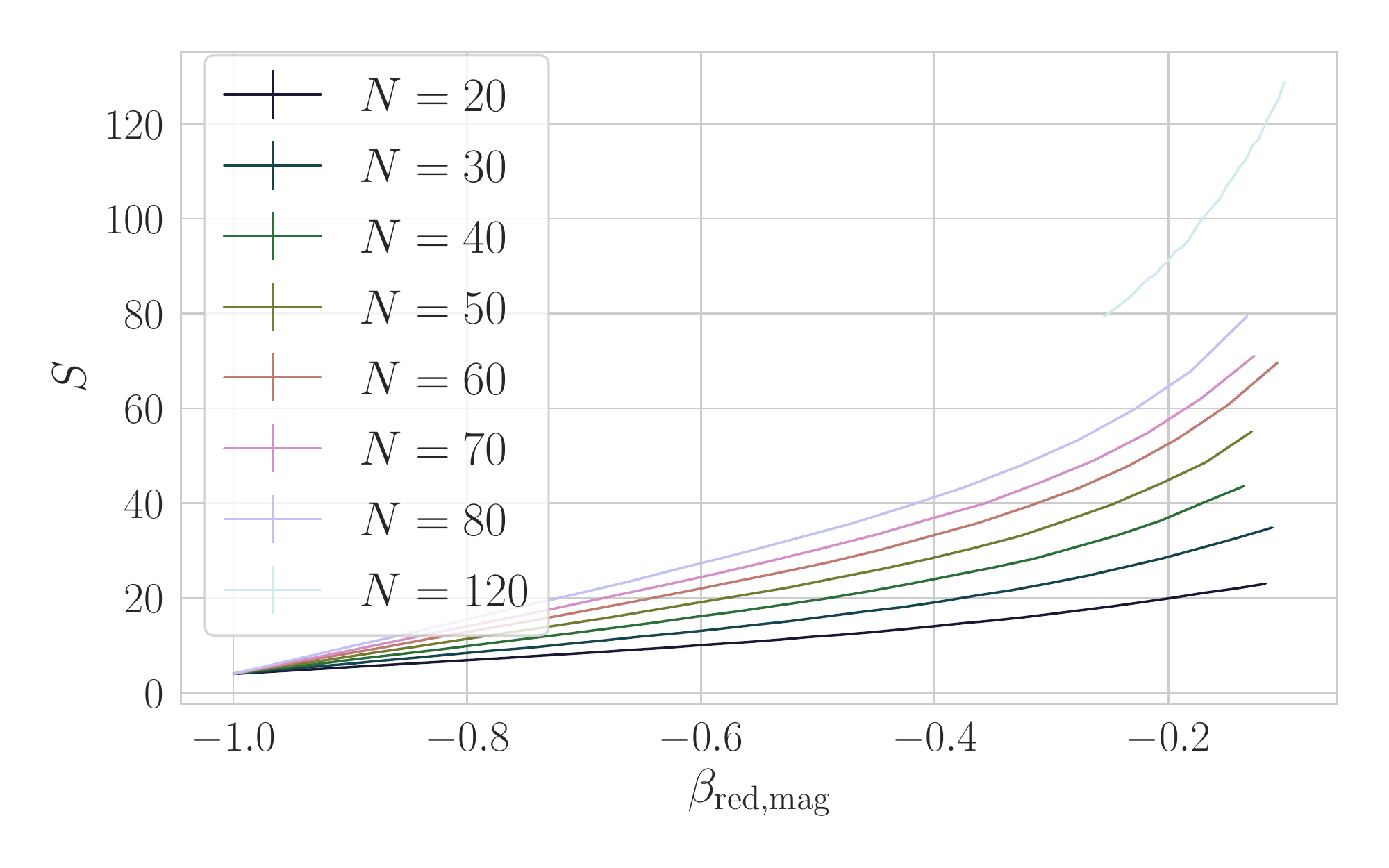}}
\subfloat[$S/N$ rescaled]{\includegraphics[width=0.5\textwidth]{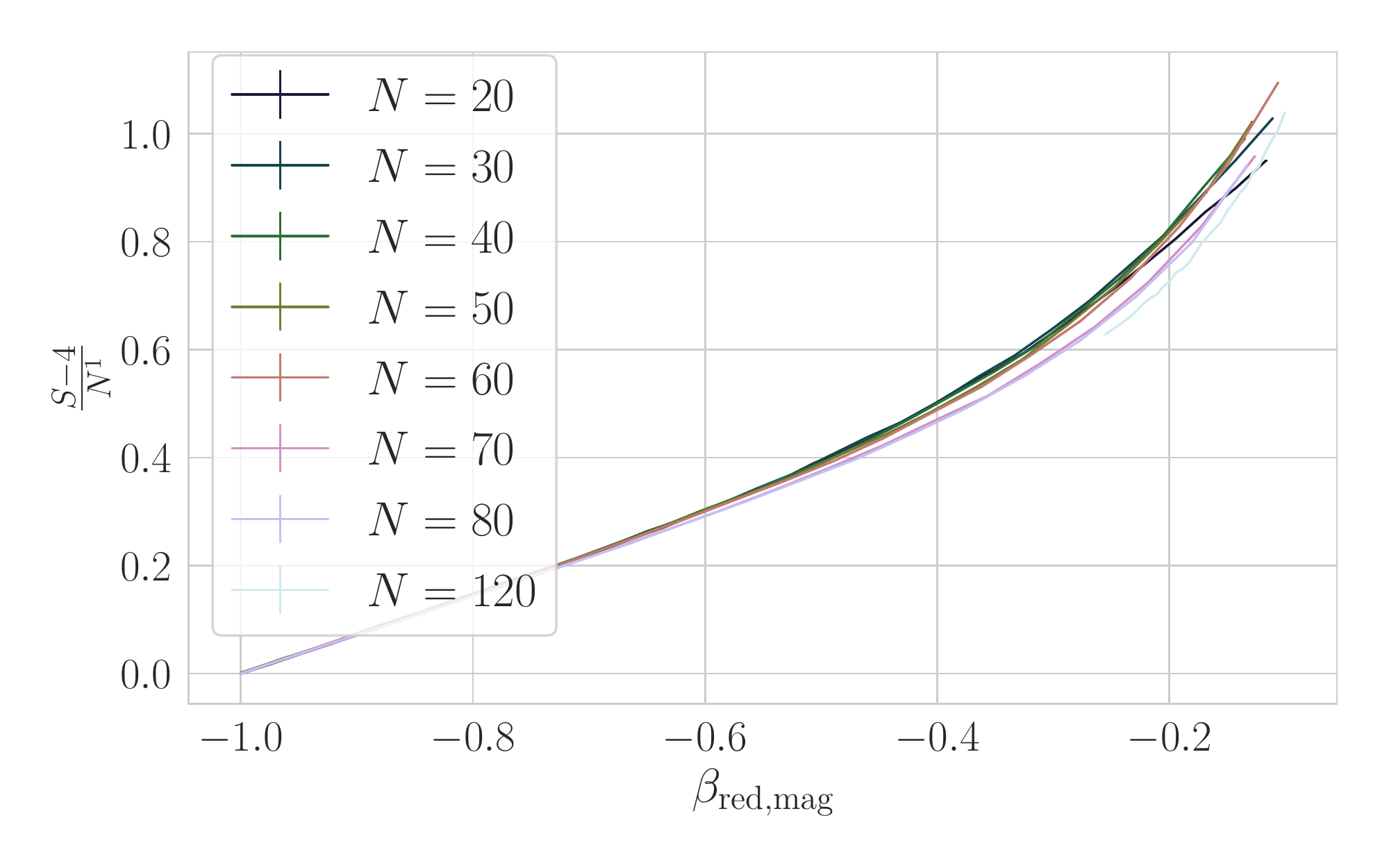}}
  \caption{In \I the actions scale the same way as they do for small positive $\beta$.}\label{fig:Action_reg_I}
\end{figure}

In \I the actions still scale in the same manner as they did for the region of low positive $\beta$, as shown in Figure~\ref{fig:Action_reg_I}.
This makes sense, considering that these regions lie along a continuous line in $\beta$, and that there is no phase transition at $\beta=0$ to divide them.

\begin{figure}
\subfloat[$S$ no scaling]{\includegraphics[width=0.5\textwidth]{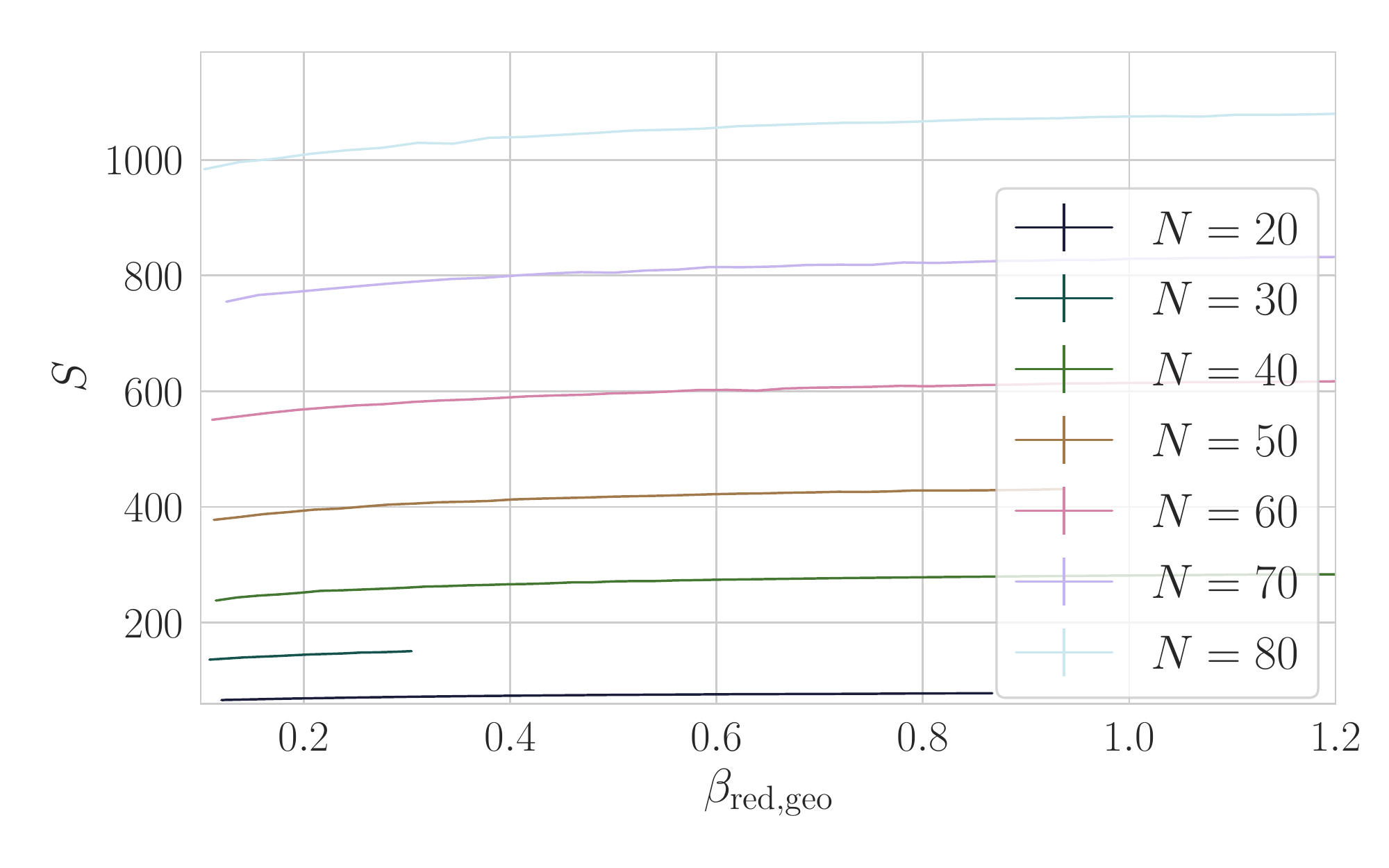}}
\subfloat[$S/N$ rescaled]{\includegraphics[width=0.5\textwidth]{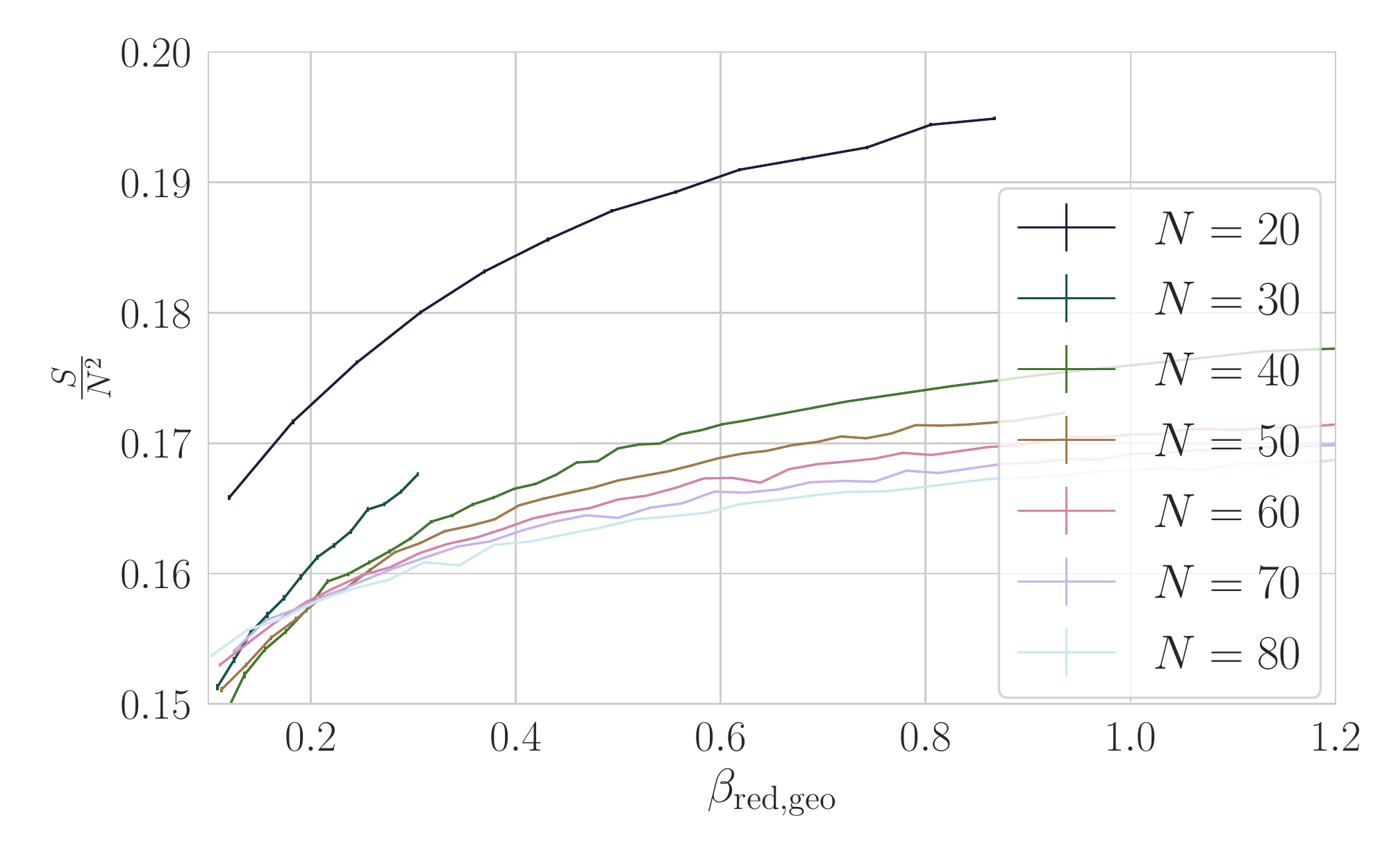}}
  \caption{In \III the actions scale like the number of links $N^2$.}\label{fig:Action_reg_III}
\end{figure}

Next we look at the scaling of the actions in \III, they all scale like $N^2$, which is again the same as the region of crystalline correlated causal sets seen in the last section, shown in Figure~\ref{fig:Action_reg_III}.
This is nice, since it strengthens our qualitative judgement from looking at the partial orders, as e.g.\ in Figure~\ref{fig:N20phases} and Figure~\ref{fig:N20phasesj1}, that these two disjointed regions are examples of the same physical phase, even though once these partial orders arise due to their minimizing the BD-action $S_c$ and the other time they arise because they maximize the number of links, and thus $S_I$.

\begin{figure}
\subfloat[\I]{\includegraphics[width=0.5\textwidth]{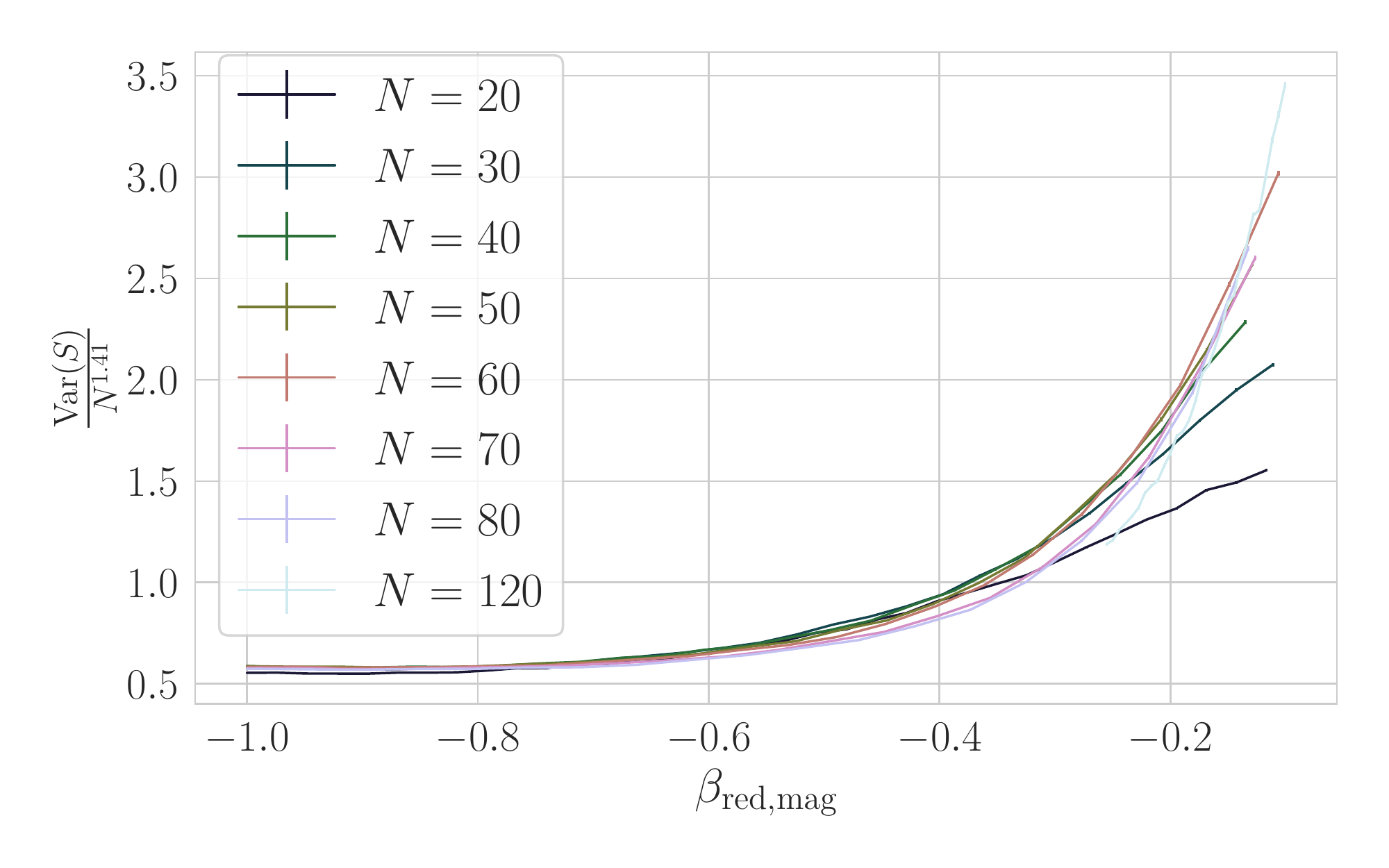}}
\subfloat[\III]{\includegraphics[width=0.5\textwidth]{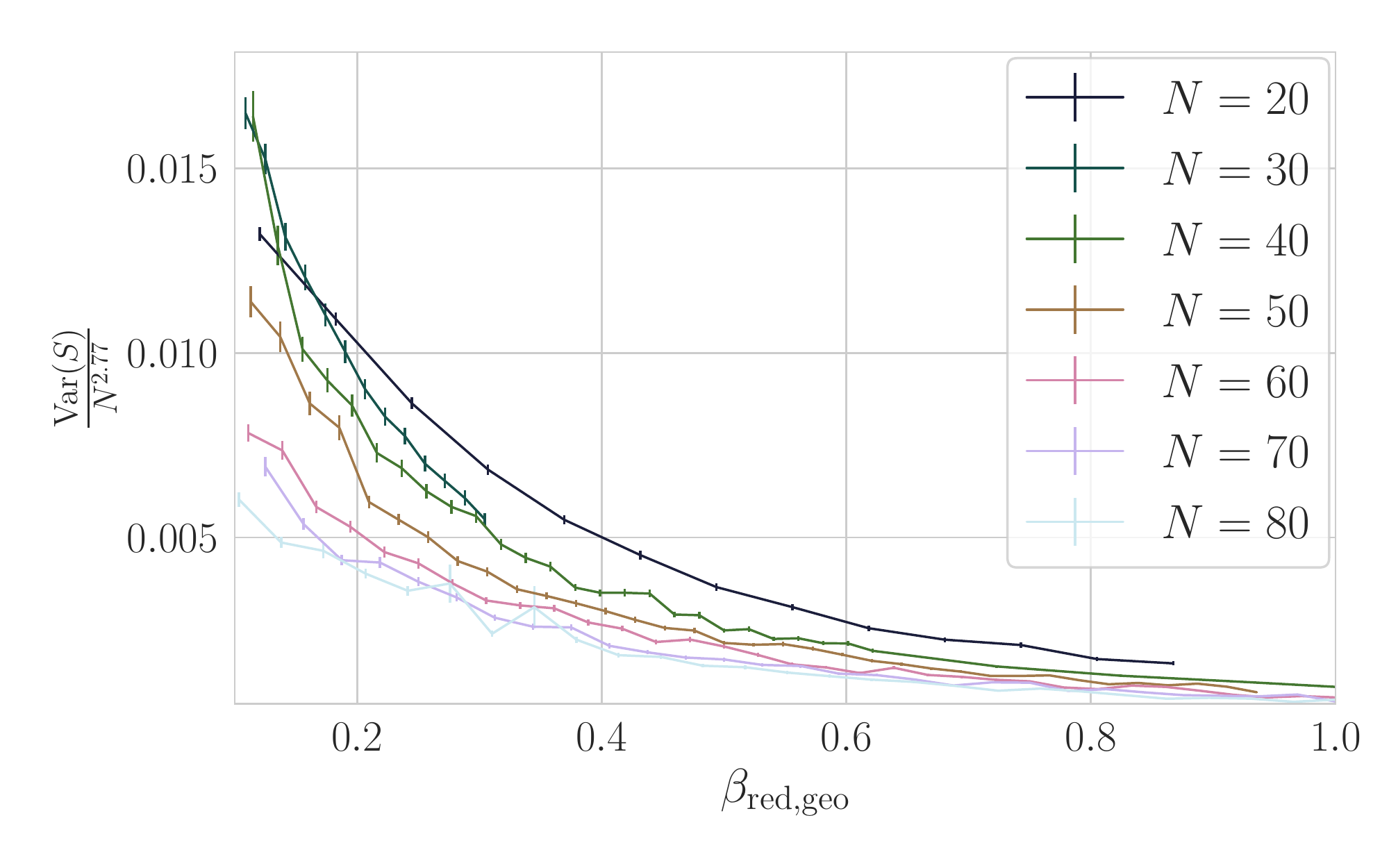}}
  \caption{Scaling of $\mathrm{Var}(S)$ in \I and \III.\@ In \I we rescale with $-\nu+\lambda=-1.41$, while in \III $-\nu+\lambda=-2.77$.}\label{fig:scalingVarj1}
\end{figure}

As before we can confirm that the scaling of the variances is consistent with $\nu - \lambda$, at least for \I and \III where we understand the scaling of the action, we show this in Figure~\ref{fig:scalingVarj1}.

\begin{figure}
\subfloat[Unscaled $|M|$]{\includegraphics[width=0.5 \textwidth]{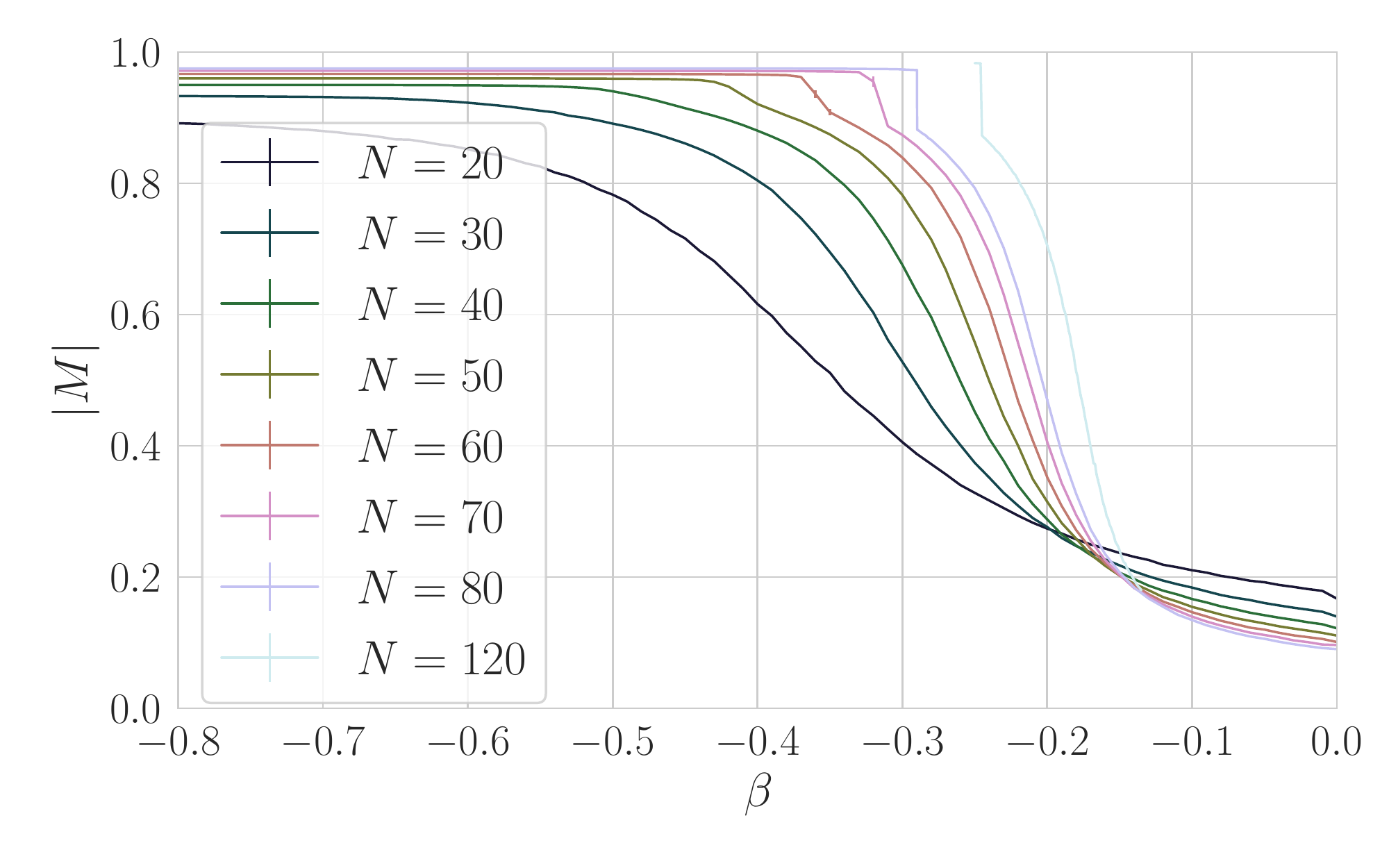}}
\subfloat[\I]{  \includegraphics[width=0.5\textwidth]{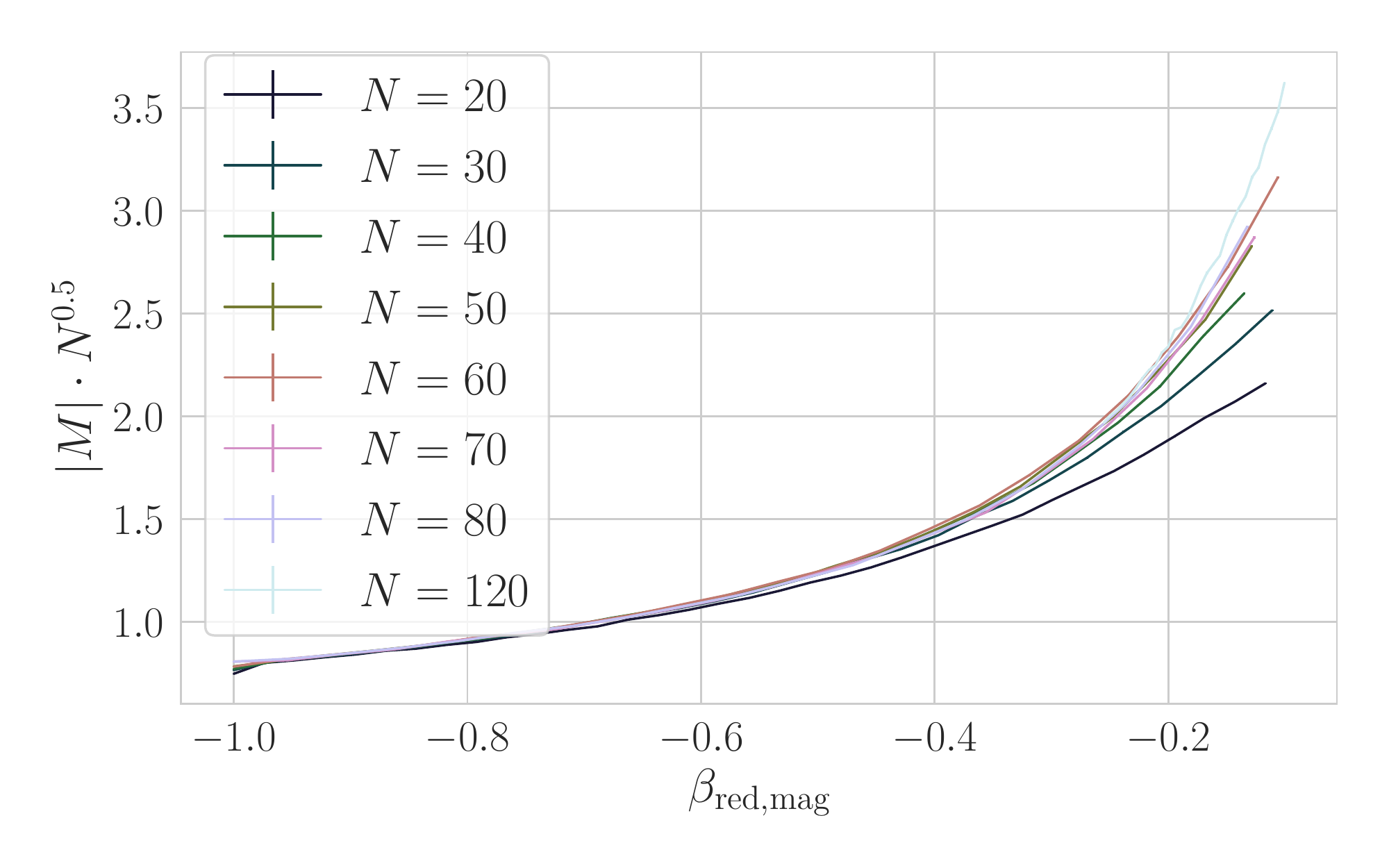}}

\subfloat[\II]{\includegraphics[width=0.5\textwidth]{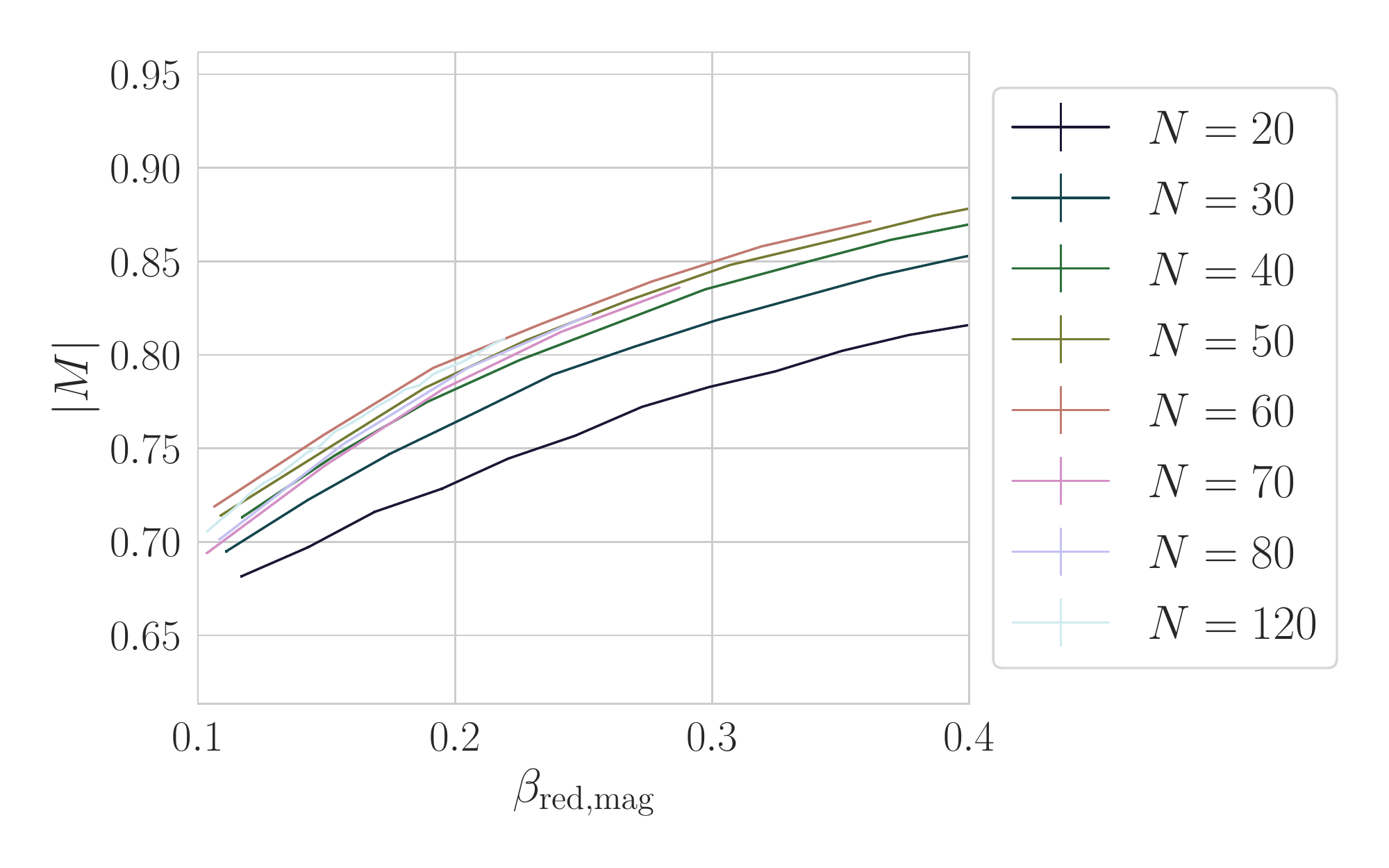}}
\subfloat[\III]{  \includegraphics[width=0.5\textwidth]{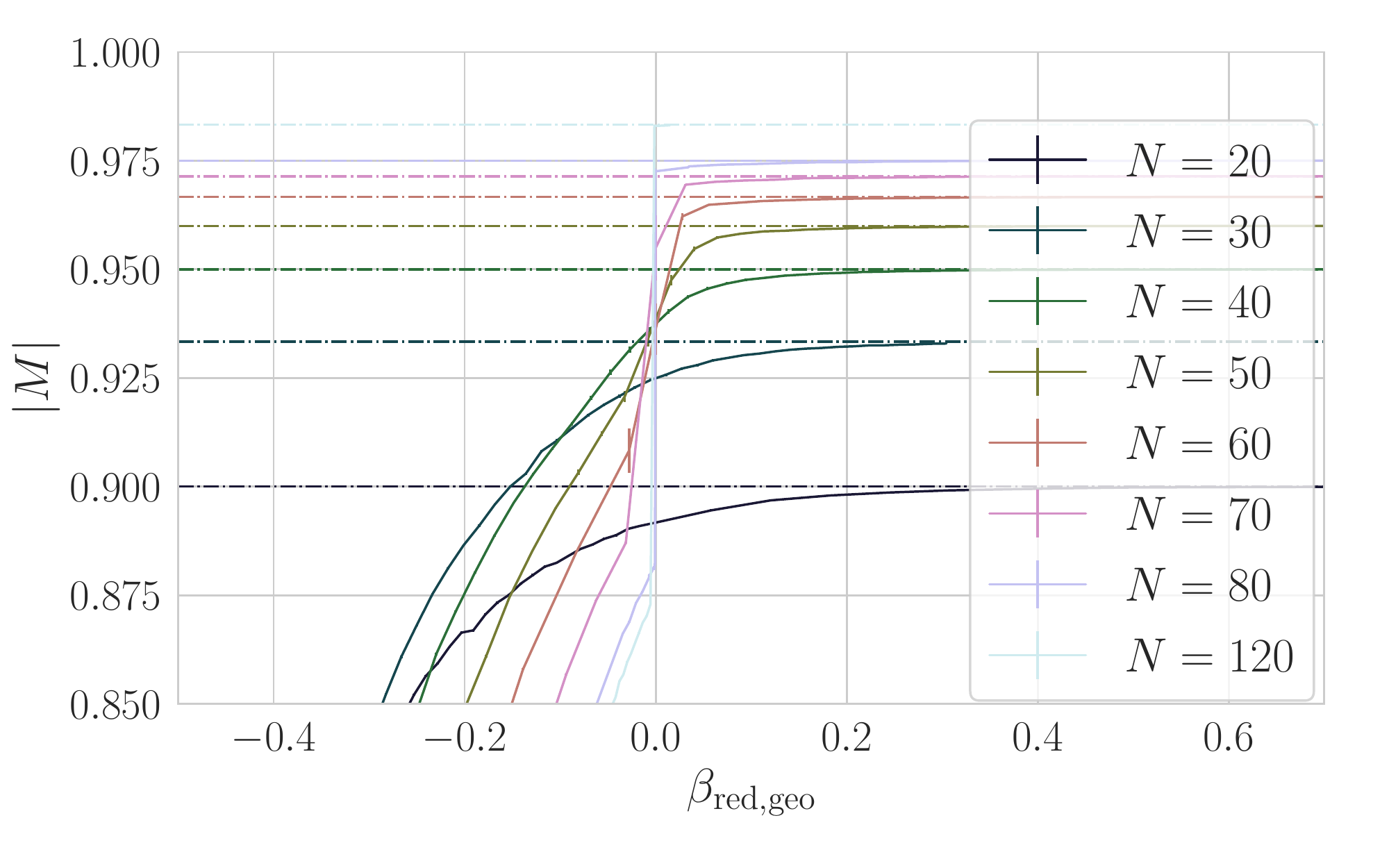}}
  \caption{In subplot (a) we show the unscaled magnetisation, where both phase transitions are clearly visible. In subplot (b) we see the magnetisation rescaled with $\sqrt{N}$ in \I. In (c) we see that the magnetisation does not scale with $N$ in \II, and in (d) we see the finite size effects in \III.}\label{fig:aMj1}
\end{figure}

The absolute magnetisation $|M|$ is show in Figure~\ref{fig:aMj1}, in sub-figure~(a) we see the entire magnetisation without any scaling.
We can see the rise in magnetisation in \II after the first phase transition, and then the steep jump at the second phase transition  to \III.
As for positive $\beta$ the absolute magnetisation goes to $0$ like $1/\sqrt{N}$ as $\beta$ goes to $0$ in the completely random phase (Figure \ref{fig:aMj1} (b)) and to 1 as $1-\frac{2}{N}$ at large negative  $\beta$ in the correlated crystalline phase (Figure \ref{fig:aMj1} (d).

In \II we can plot the absolute magnetisation against $\brOne$, comparing the data for different $N$. In Figure~\ref{fig:aMj1} (c) we see that the magnetisation in this region does not seem to scale with $N$.

\begin{figure}
\subfloat[\I]{  \includegraphics[width=0.5\textwidth]{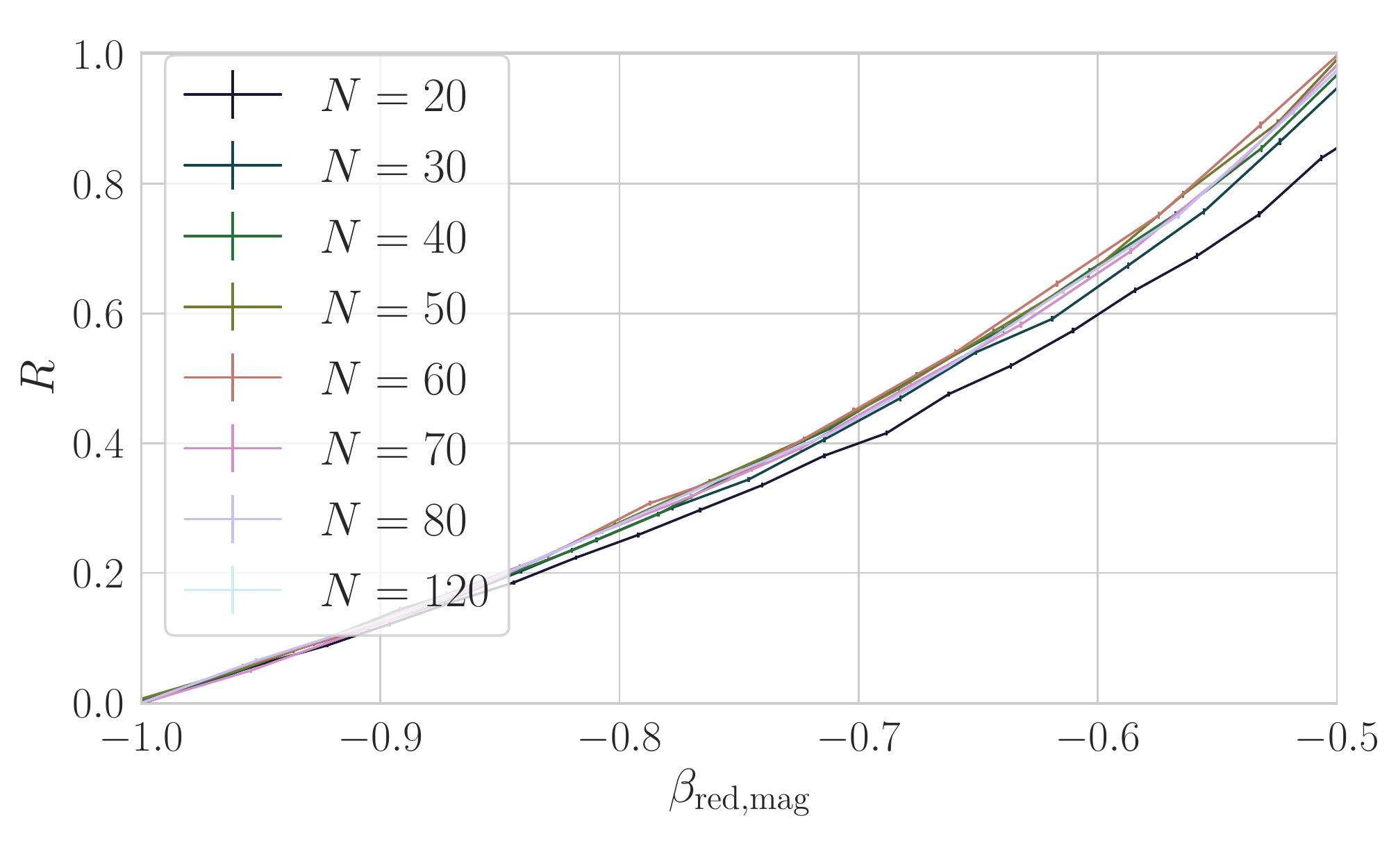}}
\subfloat[\II + \III]{  \includegraphics[width=0.5 \textwidth]{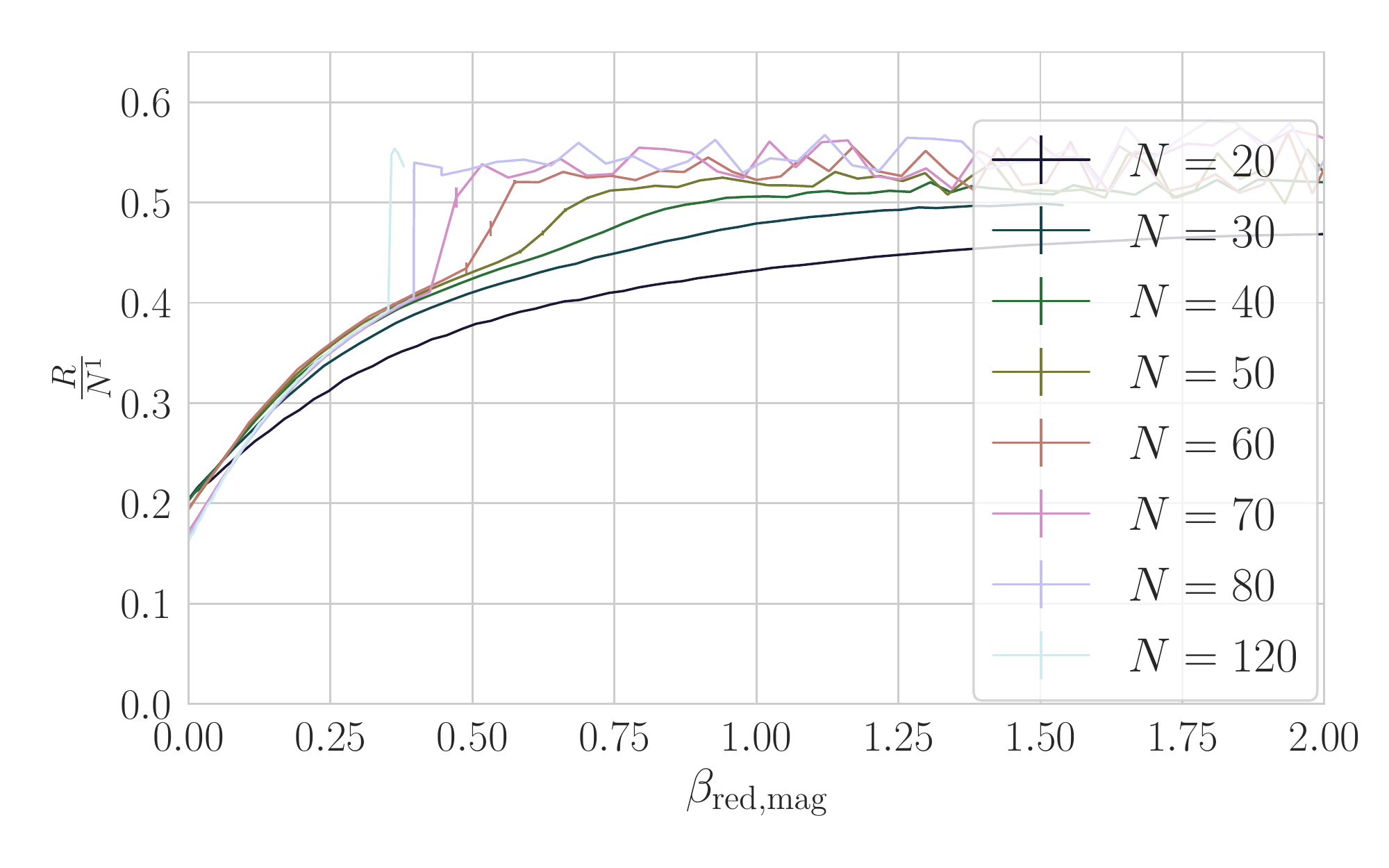}}

  \caption{Scaling of $R$ with the system size. On the left hand side we see that the relation correlation is not scaling with $N$ in \I, while the right hand plot shows that it scales like $N$ in both \II and \III.\@ Both plots are against $\brOne$, since this works for \II, and $R$ in \III is approaches a constant value.}\label{fig:relcor}
\end{figure}

The last scalar observable we examine for scaling behaviour is $R$.
In \I, $R$ does not show scaling with $N$, while in \III it scales like $N$, just as it did in the positive $\beta$ region, we can see this in Figure~\ref{fig:relcor}~(a) and (b) respectively.
Interestingly we find nice scaling for $R$ in \II if we plot using $\brOne$, the scaling then goes like $N$ again, also seen in Figure~\ref{fig:relcor}~(b).
Since $R$ goes to a constant value for large $\brOne, \brTwo$ a separate plot for $\brTwo$ to examine this region is not necessary, instead we can observe the scaling like $N$ on the far right side of Figure~\ref{fig:relcor}.

\section{Discussion}\label{sec:discussion}

In this article we have explored the scaling of the Ising model coupled to the ensemble of the random $2$d orders, weighted with the BD-action.
While the Ising model on fixed causets, as studied as a precursor in~\cite{Glaser:2018jss} and the $2$d orders on their own are  relatively simple and show nice scaling behaviour, their combination becomes considerably more complicated and not everything about the behaviour is immediately obvious.

The Ising model coupled to random geometries has been studied before, in particular analytically in the context of matrix models~\cite{Boulatov_Kazakov_1987}, and numerically for dynamical triangulations~\cite{Ambjorn_Anagnostopoulos_1997}.
In the matrix model description the spin ordering transition is found to be continuous, however it is not immediately clear how the geometry changes.
For a single Ising model coupled to dynamical triangulations the matter does not seem to influence the geometry~\cite{Ambjorn_Anagnostopoulos_1997}.
The difference between these models and the $2$d orders is however that there the number of nearest neighbours of each spin remains fixed, while for the $2$d orders the number of nearest neighbours changes, the $2$d orders are thus non-local.
This is likely why the Ising spins in certain phases strongly influence the geometry of the $2$d orders.
To better understand this behaviour we have examined the scaling with the system size $N$ along the two lines $j=-1, \beta \in [0,0.8]$ and $j=1, \beta\in [-1.4,0]$.
Let us now summarize the results.

Along the line~$j=-1, \beta \in [0,0.8]$ we find one phase transition, from a phase in which both the Ising spins and the geometry are random, into a phase in which the Ising spins are aligned and the causal set is crystalline.
After detailed analysis of the transitions we find:
\begin{itemize}
  \item The phase transition happens at a smaller value of $\beta$ than for the pure $2$d order system.
  \item The critical temperature scales slower than linearly in $N$ and can be best fit as $\beta_c(N)= (3.35 \pm 0.15 )\cdot N^{-0.72 \pm  0.01}$, so $\lambda=-0.72$.
  This is likely due to the Ising model slowing the scaling as compared to that expected for the pure $2$d orders.
  \item This phase transition shows signs for first order behaviour in the observables associated with the geometry, but signs of higher order behaviour in the observables associated with the Ising spins. This leads us to hypothesize that this might be a mixed order phase transition.
  One might wonder whether the two systems are truly coupled, however the location of the phase transition is changed considerably from the uncoupled systems.
  It thus seems that they are influencing each other, which is also supported by the fact that the geometric phase transition at negative $\beta$ is induced by the coupling of the systems.
  \item All three actions here scale linearly in $N$ in the low $\beta$ region where causal sets and spins are random and quadratic in $N$ in the crystalline and correlated region at high $
  beta$, hence $\nu=1$ in the random region and $\nu=2$ in the correlated crystalline region. We can confirm that their variances scale like $\nu-\lambda$ as expected.
\end{itemize}

The other line we explored was~$j=1, \beta\in [-1.4,0]$.
Along this line we find two phase transitions, a magnetic one from a phase of uncorrelated spins and random $2$d orders to a phase of random $2$d orders and correlated spins, and then a geometric one, from random $2$d orders with correlated spins to crystalline $2$d orders with correlated spins.

\begin{itemize}
  \item The phase transitions are best fit with $\beta_{c,mag} = (-1.22 \pm 0.20) \cdot N^{-0.41 \pm 0.04}$ and $\beta_{c,geo}=  (-8.58 \pm 0.34) \cdot N^{-0.77 \pm 0.01}$.
  These lines would meet at $N\sim 500-1000$ taking the uncertainty on the fits into account.
  However looking at the plot in Figure~\ref{fig:betac_j1} we see that the current fit for $\beta_{c,}$ lies somewhat above the last point.
  Physically a merging of the lines seems a remote possibility, since it is the energy of the aligned spins that forces the system into the crystalline state.
  This leads us to expect that even at very large $N$ one will be able to find three distinct phases.
  \item At the magnetic phase transition the fourth order cumulant, and the histograms show behaviour that is mostly consistent with a higher order phase transition, except that the histogram of the action does not peak sharply, instead the distribution grows wider as the system size increases.
  At the geometric phase transition all observations are consistent with a first order phase transition, considering the difficulty we have with resolving the phase transition location at the larger system sizes.
  It thus seems that the matter induces a phase transition of the geometry, however this new geometric phase transition still remains of first order.
  \item The scaling of the action observables is consistent with that observed before. We know that \I is the same phase as the region of low $\beta$ for the other line, so it is reassuring that we find them again to scale linearly in $N$.
  In \III on the other hand, the causal sets are crystalline again, so we find a scaling as $N^2$.
  Unfortunately we have no consistent scaling for the actions for \II, since we can not find a good way how to align the data.
  \item The only scaling in \II that we can clearly identify are $|M|$ and $R$.
  We can plot them against $\brOne$, where we then find that the magnetisation does not scale with the system size at all, while $R$ scales like $N$.
\end{itemize}

So how does this answer the questions we asked at the beginning of this article?
The most crucial question was certainly if the system still shows consistent scaling behaviour, or whether the coupling of matter to the $2$d orders interrupted this crucially.
We find that even with matter, we can still determine scaling exponents for the actions, which agree with those for the pure geometry system.
So the critical exponents here remain the same as found in~\cite{Glaser:2017sbe}, however they do not easily map to any known universality class.

The phase transition behaviour however does become more complicated.
The scaling of the phase transition location changes from a very simple linear scaling in $N$ to more complicated fractional scalings.
In particular the question if the two phase transitions for negative $\beta$ meet or not, while unlikely from a physical perspective, can not be definitely answered from our simulations.
The single phase transition at positive $\beta$ seems to be of mixed order, with a first order transition in the geometry and a continuous transition in the spins.
For negative $\beta$ it seems that the magnetic phase transition remains continuous, while the geometric transition remains of first order.
Within the limitations of our data it thus seems that matter can induce a phase transition in geometry, however it does not lead to a higher order transition.

\section{Conclusion and Outlook}\label{sec:sum}
The $2$d orders are the simplest class of causal sets one can explore in a computer simulation, while the Ising model is one of the simplest matter models to study.

As explored in~\cite{Glaser:2018jss}, when coupled together these two simple models show interesting and complex behaviour.
Trying to understand this behaviour better we looked at the scaling of the system with size, to explore which parts of the behaviour are finite size effects.
One big limitation to these studies is that the computer simulations are still studying relatively small systems, since large systems take more time and computing resources.
Nevertheless we have gathered some intriguing data.
The phase transition in the ordinary Ising model is of second order, so having phase transitions of the geometry induced through the Ising model led to the hope that this might be a higher order phase transition in causal sets.
Higher order phase transitions in quantum geometry systems are of particular interest from the asymptotic safety standpoint~\cite{Niedermaier_Reuter_2006}, since they indicate a finite dimensional critical surface, which is essential to obtain a predictive theory.
Finding such a phase transition once the system is coupled to matter would then be a big indication that `` Matter matters''~\cite{Eichhorn_2017}.
Unfortunately it seems that this system is not so friendly, instead we find indications that the geometric phase transitions of the $2$d orders coupled to the Ising model remain of first order.
This is in contrast to what is found when studying the Ising model coupled to matrix models.
It seems to us that the big difference between the systems is here that the causal sets are non-local and that the nodes change their average valency at the phase transition, while this does not happen in matrix models or triangulations.

Coupling the Ising model to the $2$d orders leads to non-local couplings, turning the Ising model into a long-range system.
The thermodynamics of long range systems is well studied and shows that, for weak long range interactions, their behavior only becomes important around phase transitions, changing the critical exponents of the universality class to become dependent on the form of the long range interaction~\cite{Bouchet_Gupta_Mukamel_2010}.
If something like this is the case here an interesting question for future study would be to see if the critical exponents allow us to determine an approximation to this long range interaction.
It could also be interesting to see how adding interaction terms that are of long range type in an ordinary Ising model behave here, and if one might be able to tune these to recover the behavior of the local Ising model.

\section*{Acknowledgements}
We would like to thank Sumati Surya for comments on an early version of this manuscript.
L.G. has been funded through grant number M 2577 through the Lise Meitner-Programm of the FWF Austria.

\printbibliography

\end{document}